\documentclass[prx, twocolumn, nofootinbib, showpacs, amsmath, amssymb, floatfix, eqsecnum]{revtex4-1}
\usepackage{amsmath}
\usepackage{amssymb}
\usepackage{amsthm}
\usepackage{amsfonts}

\usepackage{graphicx,color,framed,caption}
\usepackage{hyperref}
\usepackage{times}
\usepackage{enumerate}
\usepackage{lipsum}
\usepackage{slashed}
\usepackage{url}
\usepackage{bbm}
\usepackage{wasysym}
\hypersetup{
    colorlinks=true, %set true if you want colored links
    linktoc=all,     %set to all if you want both sections and subsections linked
    linkcolor=blue,  %choose some color if you want links to stand out
}

\def \beq {\begin{equation}}
\def \eeq {\end{equation}}
\def \beqa {\begin{eqnarray}}
\def \eeqa {\end{eqnarray}}
\def \bseq {\begin{subequations}}
\def \eseq {\end{subequations}}

\begin{document}

\title{Topological invariants for SPT entanglers}

\author{Carolyn Zhang}
%\author{Michael Levin}
\affiliation{Department of Physics, Kadanoff Center for Theoretical Physics, University of Chicago, Chicago, Illinois 60637,  USA}
\date{\today}

\begin{abstract}
We develop a framework for classifying locality preserving unitaries (LPUs) with internal, unitary symmetries in $d$ dimensions, based on $(d-1)$ dimensional ``flux insertion operators" which are easily computed from the unitary. Using this framework, we obtain formulas for topological invariants of LPUs that prepare, or entangle, symmetry protected topological phases (SPTs). These formulas serve as edge invariants for Floquet topological phases in $(d+1)$ dimensions that ``pump" $d$-dimensional SPTs. For 1D SPT entanglers and certain higher dimensional SPT entanglers, our formulas are completely closed-form. %In other cases, we still obtain formulas that compute these invariants, but they are not completely closed form, in that they involve restricting operators that are not products of on-site operators.
\end{abstract}

\maketitle
%\tableofcontents
%%%%%%%%%%%%%%%%%%%%%%%%%%%%%%%%%%%%%%%%%%%%%%%%%%%
\section{Introduction}\label{sintroduction}
%%%%%%%%%%%%%%%%%%%%%%%%%%%%%%%%%%%%%%%%%%%%%%%%%%%
In recent years, there has been much progress in not only classifying topological quantum phases, but also in understanding methods for detection and preparation of such phases. Methods for preparing, or ``entangling," a state in a topological phase are deeply connected to the entanglement properties of the phase\cite{chen2010}. For example, it was shown that a state in a long-range entangled phase, such as a 2D topological order, cannot be prepared by any finite depth quantum circuit (FDQC). Instead, such a state can only be prepared either by a circuit whose depth scales with the system size\cite{bravyidepth,hastingslocality,huangcomplexity} or by supplementing an FDQC with measurements and post-selection\cite{natmeasure, aguado, pirolicircuits, natnonabelian, nathierarchy}. On the other hand, short-range entangled phases such as symmetry-protected topological phases (SPTs) can be prepared by FDQCs, but these circuits must contain local gates that break the symmetry\cite{chen2010,sptcohomology}. 

Interestingly, broad classes of SPT phases can be entangled by FDQCs that, though containing gates that break the symmetry, respect the symmetry as a whole\cite{chen2010, nat2018, ellisondisentangle, symmedge}. Sometimes, an SPT cannot be entangled by an FDQC that respects the symmetry as a whole, but can be entangled by a more general locality preserving unitary (LPU) which respects the symmetry\cite{haahnontrivial,haahclifford,moreqca}. When the locality is strict, without exponentially decaying tails, these nontrivial LPUs are also known as quantum cellular automata (QCA)\cite{GNVW, freedmanclassification, freedmangroup}.

LPUs have also recently received attention because they describe the stroboscopic boundary dynamics of many-body localized, periodically driven systems, also known as Floquet systems\cite{chiralbosons,GNVW,fermionic}. LPUs with symmetry describe the boundary dynamics of these kinds of Floquet systems when the drive is constrained to respect the symmetry\cite{cohomology, dynamically,alldimensions,u1floquet}. Nontrivial $G$ symmetric Floquet systems in $d$ spatial dimensions can ``pump" $(d-1)$ dimensional $G$ SPTs to the boundary every period\cite{cohomology, dynamically, bachmannthouless}. For these systems, the stroboscopic boundary dynamics is described by a $G$ symmetric $(d-1)$ dimensional SPT entangler. These kinds of boundary unitaries have been classified and studied in exactly solvable models\cite{dynamically,alldimensions} and matrix product unitaries\cite{mpusymm,mpu,pirolifermionic,xiempu,piroliqca}. 

Although LPUs with various kinds of symmetry have been classified, there exist very few explicit formulas for topological invariants of these LPUs. For bosonic systems in 1D without any symmetry, there is an explicit formula that takes as input an LPU and produces the GNVW index, that classifies LPUs without any symmetry\cite{chiralbosons,GNVW}. In this work, we will provide similar formulas for topological invariants of LPUs with symmetry. These formulas also serve as boundary invariants for many-body localized Floquet systems with symmetry. For simplicity, we will consider only LPUs with strict locality. 

In general, nontrivial $G$ symmetric strict LPUs fall into two classes: those that entangle SPTs and those that do not entangle SPTs. For example, all nontrivial strict LPUs with discrete symmetries in 1D entangle SPTs\cite{cohomology}, while nontrivial strict LPUs with $U(1)$ symmetry in 1D are not related to SPTs\cite{u1floquet}. In this work, we will obtain formulas for topological invariants of $G$ symmetric strict LPUs that are SPT entanglers. To specify that we are restricting to this particular subset of $G$ symmetric strict LPUs, we will refer to them as $G$ symmetric SPT entanglers in the remainder of this work. In particular, we will focus on symmetric entanglers for bosonic ``in-cohomology" SPTs. These SPTs are classified by elements of $H^{d+1}(G,U(1))$, and we will show that for the dimensions and symmetries we consider, the topological invariants computed from our formulas completely specify this element. In short, in this work, we present formulas that take as input a $G$ symmetric SPT entangler and produce topological invariants that completely specify the SPT phase it entangles.

We have two guiding principles. First, of course, our formulas must produce the same result for two equivalent LPUs, that entangle the same SPT phase. Roughly speaking, our formulas must be insensitive to modification of the input unitary by any strictly local, symmetric unitary. This means that they will also be insensitive to $G$ symmetric FDQCs, which are FDQC constructed out of such symmetric local unitaries. Second, we try to make our formulas as closed-form as possible. This means that whenever possible, they only involve the truncation of operators that are products of on-site operators, such as on-site symmetry operators. In particular, when $G$ is unitary, we do not truncate the SPT entangler.

These two guiding principles, along with the fact that we classify $G$ symmetric SPT entanglers rather than SPT states, differentiate the work we present here from previous related work. In particular, most work related to classifying SPTs via their entanglers do not assume that the entanglers respect the symmetry\cite{ogata1d,ogata1dfermionic,sopenko1d,ogata2d, ogata2dfermionic,sopenko2d}. This extra assumption allows us to make our invariants more explicit. Since broad classes of SPTs can be entangled by symmetric entanglers, we do not lose much generality in making this assumption. Ref.~\onlinecite{symmedge} also assumed that the SPT entangler is symmetric as a whole. Using the SPT entangler truncated to a finite disk, they obtained a corresponding ``anomalous edge representation of the symmetry." They then showed how to get the cocycle labeling the SPT phase from the anomalous edge representation of the symmetry. We discuss the relation between our methods and the anomalous representation of the symmetry on the edge in Appendix.~\ref{sboundaryrep}. Unlike Ref.~\onlinecite{symmedge}, we do not truncate the SPT entangler to compute our invariants, when the symmetry is unitary. In some cases, when the entangler is actually a nontrivial QCA (even in the absence of symmetry), it cannot be truncated at all. Furthermore, our invariants are actually gauge invariant quantities: unlike the cocycles computed in Ref.~\onlinecite{symmedge}, which are only defined up to a coboundary, our invariants have no ambiguity.

The rest of this paper is organized as follows. We include in this section a summary of our main results and an illustrative example of an invariant for 1D SPT entanglers. In Sec.~\ref{sspts}, we describe our general framework for classifying SPT entanglers with flux insertion operators. We then apply this framework to SPT entanglers related to 1D SPTs in Sec.~\ref{s1d} and 2D SPTs with discrete, abelian, unitary symmetries in Sec.~\ref{sabelian2d}. We include some results regarding fermionic SPT entanglers in Sec.~\ref{sfermionic}, before concluding with interesting open questions in Sec.~\ref{sdiscussion}. We defer most of the proofs, including the explicit derivations of relations between our invariants and known SPT invariants, to the appendices.
%%%%%%%%%%%%%%%%%%%%%%%%%%%%%%%%%%%%%%%%%%%%%%%%%%%%%%%%%%%%%%%%%%%%%%%%%%%%%%%%%%%%
\subsection{Summary of results}\label{ssummary}
%%%%%%%%%%%%%%%%%%%%%%%%%%%%%%%%%%%%%%%%%%%%%%%%%%%%%%%%%%%%%%%%%%%%%%%%%%%%%%%%%%%%
Our main result is a framework for classifying LPUs with symmetry, which can be applied to both SPT entanglers and LPUs that are not related to SPTs. Using this framework, we obtain topological invariants for various kinds of SPT entanglers. These topological invariants can be divided into two main groups.

Our first group of invariants apply to 1D SPT entanglers with discrete symmetries. When the symmetry is unitary and discrete, we obtain closed form formulas for topological invariants that take as input only a global SPT entangler $U$ and global symmetry operators. These formulas are given in Eq.~(\ref{flowdiscrete}) for abelian symmetries and (\ref{flownonabelian}) for non-abelian symmetries. We also have an invariant for time reversal SPT entanglers, written in Eq.~(\ref{timerev}), but it is not completely closed form because it involves truncating the entangler. These invariants can be easily leveraged to obtain invariants of SPT entanglers in higher dimensions described by decorating domain walls with 1D SPTs, written in Eq.~(\ref{decdomaininv}).

Our second group of invariants apply to 2D SPT entanglers with discrete, unitary, abelian symmetries, beyond those with domain walls decorated by 1D SPTs. The explicit formulas for these invariants are given by Eqs.~(\ref{thetag1def}) and (\ref{thetagigj}), and are not completely closed form in that they involve truncation of certain non-onsite operators. Again, these invariants can be leveraged to obtain invariants of SPT entanglers in higher dimensions described by decorating domain walls with 2D SPTs.

We also obtain a closed form formula, given by Eq.~\ref{flowfermions}, for the $\mathbb{Z}_2$ invariant classifying SPT entanglers with only fermion parity symmetry in 1D. Nontrivial SPT entanglers in this case entangle the Kitaev wire, and differ from the others considered in this work in that they are nontrivial QCAs\cite{fermionic,huangcomplexity}.

%%%%%%%%%%%%%%%%%%%%%%%%%%%%%%%%%%%%%%%%%%%%%%%%%%%%%%%%%%%%%%%%%%%%%%%%%%%%%%%%%%%%
\subsection{Example: $\mathbb{Z}_2\times\mathbb{Z}_2$ SPT entangler in 1D}\label{sexample}
%%%%%%%%%%%%%%%%%%%%%%%%%%%%%%%%%%%%%%%%%%%%%%%%%%%%%%%%%%%%%%%%%%%%%%%%%%%%%%%%%%%%
To give a flavor of the kinds of formulas for topological invariants studied in this work, we begin with a simple example. In this example, we present a set of formulas that compute topological invariants for SPT entanglers in 1D with $\mathbb{Z}_2\times\mathbb{Z}_2$ symmetry. The classification of 1D SPTs with this symmetry is $\mathbb{Z}_2$: there is one trivial phase and one nontrivial phase. We will show that our closed form formulas compute a set of $U(1)$ phases $\{c(g,h)\}=\left\{\frac{\omega(g,h)}{\omega(h,g)}\right\}$, where $\omega(g,h)\in H^2(\mathbb{Z}_2\times\mathbb{Z}_2,U(1))$ labels the SPT phase\footnote{See Appendix~\ref{sgroupcohomology} for a review of group cohomology}. The set of phases $\{c(g,h)\}$ for all $g,h\in \mathbb{Z}_2\times\mathbb{Z}_2$ completely defines the SPT phase. 

The physical setup consists of a finite, periodic 1D chain with an even number of spin-1/2's. The two global $\mathbb{Z}_2$ symmetries, generated by unitary operators $U_{g_1}$ and $U_{g_2}$, are spin flips on all the even sites and all the odd sites respectively:
\begin{equation}
U_{g_1}=\prod_{r\text{ even}}\sigma^x_{r}\qquad U_{g_2}=\prod_{r\text{ odd}}\sigma^x_{r},
\end{equation}
where $\sigma^x_r$ is the Pauli $x$ operator on site $r$. An example of a symmetric, gapped Hamiltonian with a trivial ground state is given by
\begin{equation}
H_0=-\sum_r\sigma^x_r.
\end{equation}

The ground state of $H_0$ is the state with all the spin-1/2's in the $+1$ eigenstate of $\sigma^x_r$. A $\mathbb{Z}_2\times\mathbb{Z}_2$ symmetric SPT entangler is given by
\begin{equation}\label{uz2z2}
U=\prod_re^{\frac{i\pi}{4}(-1)^r\sigma^z_r\sigma^z_{r+1}}.
\end{equation}

Notice that $U$ is symmetric under both global $\mathbb{Z}_2$ symmetries, but its individual gates $e^{\frac{i\pi}{4}(-1)^r\sigma^z_r\sigma^z_{r+1}}$ are not symmetric under either symmetry. This is expected, because for $U$ to be a nontrivial SPT entangler, it must contain gates that break the symmetry. 

To confirm that $U$ indeed entangles the $\mathbb{Z}_2\times\mathbb{Z}_2$ SPT, notice that $U$ transforms $H_{0}$ into $H_{\mathrm{SPT}}$, whose ground state is the well-known cluster state:
\begin{equation}
H_{\mathrm{SPT}}=U^\dagger H_{0}U=-\sum_r\sigma^z_{r-1}\sigma^x_r\sigma^z_{r+1}.
\end{equation}

Our formula for $c(g,h)$ takes as input $U$ and two restricted symmetry operators $U_{A,g}$ and $U_{B,h}$, where $g,h\in\mathbb{Z}_2\times\mathbb{Z}_2$. $U_{A,g}$ and $U_{B,h}$ are restrictions of global symmetry operators $U_g$ and $U_h$ to intervals $A$ and $B$ respectively. Because each of the global symmetry operators is a product of on-site operators, these restrictions can be done unambiguously. In particular,
\begin{equation}
U_{A,g_1}=\prod_{\substack{r\in A\\ r\text{ even}}}\sigma^x_r\qquad U_{B,g_2}=\prod_{\substack{r\in B\\ r\text{ odd}}}\sigma^x_r.
\end{equation}
\begin{figure}[tb]
   \centering
   \includegraphics[width=.9\columnwidth]{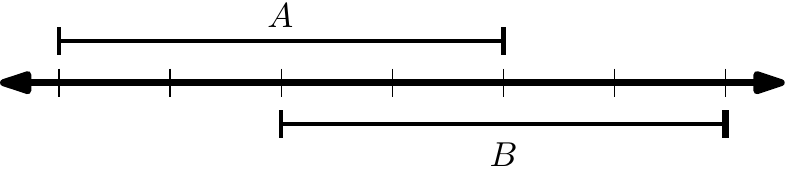} % requires the graphicx package
   \caption{$c(g,h)$ in Eq.~(\ref{example}) is defined using symmetry operators for sufficiently large, overlapping intervals $A$ and $B$ on a 1D lattice.}
   \label{fig:ABintervals}
\end{figure}

It is important that we choose $A$ and $B$ to be sufficiently overlapping intervals in 1D, as illustrated in Fig.~\ref{fig:ABintervals}. For concreteness, let $A=[0,2a]$ and $B=[a,3a]$ where $a$ is an odd integer and $a\gg 1$ (we will later precisely define the relevant length scales). Then $c(g,h)$ is given by
\begin{equation}\label{example}
c(g,h)=\overline{\mathrm{Tr}}\left(U^\dagger U_{A,g} U U_{B,h}U^\dagger U_{A,g}^\dagger U U_{B,h}^\dagger\right),
\end{equation} 
where $\overline{\mathrm{Tr}}$ is a trace normalized by the dimension of the total Hilbert space, so that $\overline{\mathrm{Tr}}(\mathbbm{1})=1$. 

Let us check that Eq.~(\ref{example}) produces the correct $c(g,h)$ for $U=\mathbbm{1}$ and for the SPT entangler in Eq.~(\ref{uz2z2}). It is easy to see that, because $U_{A,g}$ and $U_{B,h}$ commute, $c(g,h)=1$ for all $g,h\in \mathbb{Z}_2\times\mathbb{Z}_2$ if $U=\mathbbm{1}$. On the other hand, if $U$ is the SPT entangler defined in (\ref{uz2z2}), then $U$ attaches a $\sigma^z_r$ of the opposite sublattice near the left and right endpoints of a restricted symmetry operator. For example, $U^\dagger U_{A,g}U=\sigma^z_{-1}U_{A,g}\sigma^z_{2a+1}$. It follows that $c(g_1,g_2)=-1=c(g_2,g_1)$ and $c(g,h)=1$ otherwise, which matches with the set $\{c(g,h)\}$ defining the 1D $\mathbb{Z}_2\times\mathbb{Z}_2$ SPT.%, confirming that the gauge invariant phases defining the SPT can be easily computed from the SPT entangler. 

Notice that Eq.~(\ref{example}) satisfies our two guiding principles. First, it is insensitive to modifications of $U$ by local, symmetric unitaries. If $U\to U_rU$ where $U_r$ is fully supported in $A$ or $\overline{A}$, then we can commute $U_r$ through $U_{A,g}$ and $U_{\overline{A},g}^{\dagger}$ to cancel with its inverse. If $U_r$ is fully supported deep in $B$ or $\overline{B}$, then we can use $U_rU=U(U^\dagger U_rU)$ and then commute $U^\dagger U_rU$ through $U_{B,h}$ and $U_{B,h}^\dagger$. For sufficiently large and overlapping $A$ and $B$, any local unitary is supported deep inside $A,\overline{A},B,$ or $\overline{B}$, so Eq.~(\ref{example}) is insensitive to $U\to U_rU$ for any local, symmetry unitary. This ensures that Eq.~(\ref{example}) produces a topological invariant. Second, it is completely closed form in that it only takes as input the global SPT entangler $U$ and restrictions of $U_g$ and $U_h$, which are products of on-site operators. Formulas like Eq.~(\ref{example}) are the main result of this paper. 
%%%%%%%%%%%%%%%%%%%%%%%%%%%%%%%%%%%%%%%%%%%%%%%%%%%%%%%%%%%%%%%%%%%%%%%%%%%%%%%%%%%%
\section{Framework for classifying LPUs with symmetry}\label{sspts}
%%%%%%%%%%%%%%%%%%%%%%%%%%%%%%%%%%%%%%%%%%%%%%%%%%%%%%%%%%%%%%%%%%%%%%%%%%%%%%%%%%%%
In this section, we will present our framework for classifying LPUs with symmetry. This framework is based on a set of $(d-1)$ dimensional operators that we call flux insertion operators, that form an anomalous representation of the symmetry. These operators are useful for our purposes because they can be easily computed from the SPT entangler when the symmetry is on-site, and completely classify the entangler. 
%%%%%%%%%%%%%%%%%%%%%%%%%%%%%%%%%%%%%%%%%%%%%%%%%%%%%%%%%%%%%%%%%%%%%%%%%%%%%%%%%%%%
\subsection{Preliminaries: SPT phases and SPT entanglers}\label{sdefs}
%%%%%%%%%%%%%%%%%%%%%%%%%%%%%%%%%%%%%%%%%%%%%%%%%%%%%%%%%%%%%%%%%%%%%%%%%%%%%%%%%%%%
For the most part, we will consider only bosonic systems in this work. Specifically, we consider a lattice of bosonic spins on a general $d$ dimensional lattice $\Lambda$ with a symmetry $G$, which may contain antiunitary elements such as time reversal. The action of $G$ on the lattice spins is given by $\{U_gK_g\}$, where $K_g$ is complex conjugation for antiunitary elements and $K_g=1$ for unitary elements. 

\begin{figure}[tb]
   \centering
   \includegraphics[width=.9\columnwidth]{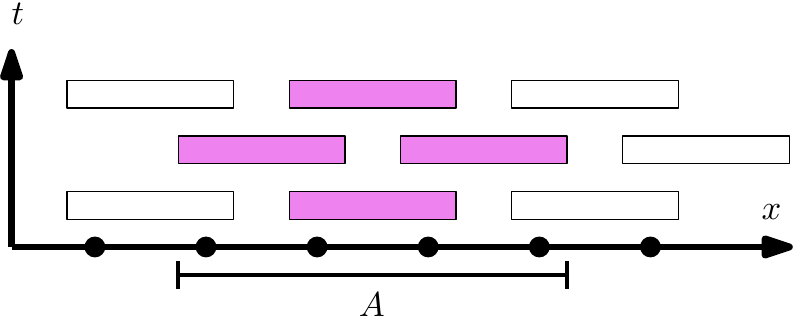} % requires the graphicx package
   \caption{A depth 3 FDQC in 1D consists of 3 layers where each layer is a product of commuting local unitary operators. Here, the $x$-axis is the spatial dimension and the $y$-axis is time. Each black circle represents a lattice site and each rectangle is a local unitary operator. To restrict the FDQC to the region $A$, we simply delete all the local unitaries with support outside of $A$. The restricted FDQC is a product of all the colored rectangles.}
   \label{fig:fdqc}
\end{figure}

To define SPT phases and SPT entanglers, it is useful to  first define what we mean by FDQC. An FDQC is a unitary $U$ that can be written as a finite product of layers $U_n,n\in[1,N]$, where each layer $U_n$ is a product of commuting local unitary operators (``gates"):
\begin{equation}\label{FDLU}
U=U_NU_{N-1}\cdots U_1\qquad U_n=\prod_rU_{n,r}.
\end{equation}

Here, each gate $U_{n,r}$ is strictly supported within a bounded distance $\lambda$ of the site $r\in\Lambda$. By definition, $N$ must be finite, in that it does not grow with the system size. The generic form of a 1D FDQC is illustrated in Fig.~\ref{fig:fdqc}. FDQCs can be used to approximate, by Trotter decomposition, finite time evolution by any local Hamiltonian. 

An important property of FDQCs is that they can be \emph{restricted}. To restrict an FDQC to a region $A$, we simply remove all the gates with support outside of $A$. For example, a restriction of a 1D FDQC to an interval $A$ is shown in Fig.~\ref{fig:fdqc}. 

A $G$-symmetric FDQC is an FDQC in which every local gate $U_{n,r}$ is symmetric. In other words, every $U_{n,r}$ commutes with every element of $\{U_gK_g\}$. They describe finite time evolution with a local Hamiltonian that respects the symmetry at all points in time.

One can also consider a more general unitary operator which we call an LPU, that simply maps local operators to nearby local operators. Recall that in this work we only consider strict LPUs, which map strictly local operators to nearby strictly local operators, without exponentially decaying tails. We can associate with any LPU an ``operator spreading length" $\xi$, which is the maximum distance it can spread a local operator. Specifically, for any operator $O_r$ supported on site $r$, $U^\dagger O_rU$ is supported within a disk of radius $\xi$ centered at $r$. It is easy to see that for FDQCs, which form a subset of LPUs, $\xi=2N\lambda$. A $G$-symmetric LPU respects the symmetry as a whole, but may not have a decomposition into an FDQC built out of local, symmetric gates.

A $G$-symmetric strict LPU $U$ is an SPT entangler if it satisfies 
\begin{equation}
U|\psi_{0}\rangle=|\psi_{\mathrm{SPT}}\rangle,
\end{equation}
where $|\psi_0\rangle$ is a symmetric product state of the form $|\psi_{0,r}\rangle^{\otimes |\Lambda|}$, satisfying $U_gK_g|\psi_{0,r}\rangle=|\psi_{0,r}\rangle$ for all $g\in G$. Here, $|\psi_{\mathrm{SPT}}\rangle$ is a (possibly trivial) SPT state. In this paper, we say that $|\psi_{\mathrm{SPT}}\rangle$ is in a nontrivial SPT phase if $U$ is locality preserving, but there is an obstruction to making $U$ a $G$-symmetric FDQC. Note that this is different from the more standard definition that $|\psi_{\mathrm{SPT}}\rangle$ is in a nontrivial SPT phase if it can be connected to $|\psi_0\rangle$ by an FDQC, but only if the FDQC contains gates that break the symmetry. In this paper, we will consider $U$ that are more general symmetric LPUs, which may not be FDQCs even after forgetting about the symmetry. This is because for some SPT phases, the entangler can only be made symmetric if it is a nontrivial LPU\cite{haahnontrivial,haahclifford,moreqca}.

 %Two states are in the same SPT phase if they differ by a $G$ symmetric FDQC:
%\begin{equation}\label{stateequiv}
%|\psi'\rangle\sim|\psi\rangle:|\psi'\rangle=G\text{ FDQC }\cdot|\psi\rangle
%\end{equation}
We define two SPT entanglers as equivalent if they differ by a $G$ symmetric FDQC:
\begin{equation}\label{Usim}
U'\sim U: U'=G\text{ FDQC }\cdot U.
\end{equation}

This means that the SPT states they entangle are equivalent, because 
\begin{equation}
|\psi_{\mathrm{SPT}}'\rangle=G\text{ FDQC }\cdot|\psi_{\mathrm{SPT}}\rangle,
\end{equation}
which is the usual definition of equivalence for SPT states\cite{chen2010}. Note that the converse does not necessarily hold: two equivalent SPT states, that differ by a $G$ symmetric FDQC, may have inequivalent entanglers.

%%%%%%%%%%%%%%%%%%%%%%%%%%%%%%%%%%%%%%%%%%%%%%%%%%%%%%%%%%%%%%%%%%%%%%%%%%%%%%%%%%%%
\subsubsection{Flux insertion}\label{sanomalous}
%%%%%%%%%%%%%%%%%%%%%%%%%%%%%%%%%%%%%%%%%%%%%%%%%%%%%%%%%%%%%%%%%%%%%%%%%%%%%%%%%%%%
\begin{figure}[tb]
   \centering
   \includegraphics[width=.9\columnwidth]{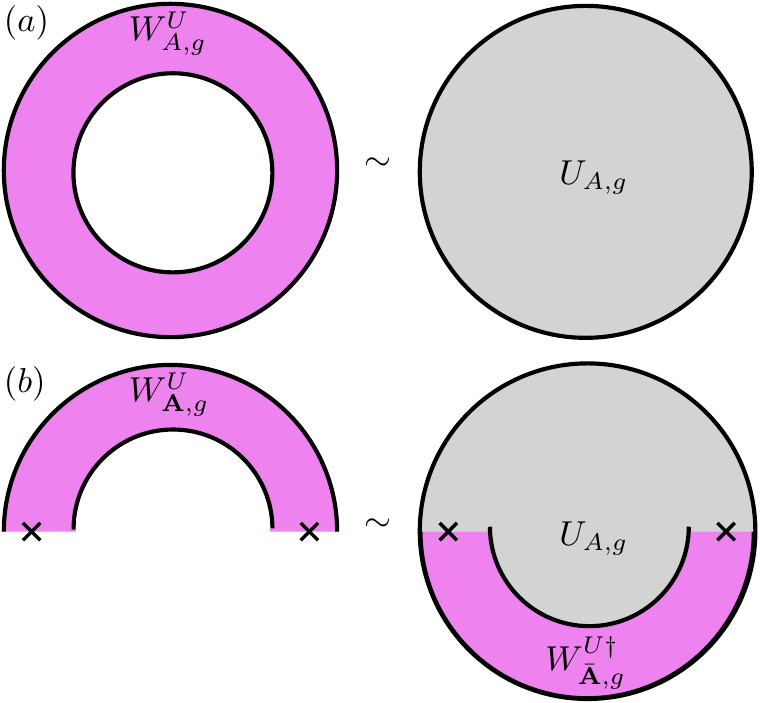} % requires the graphicx package
   \caption{For illustration purposes, we specialize here to 2D. (a) A closed flux insertion operator $W_{A,g}^U$ (pink annulus) acts, on the SPT state $|\psi_{\mathrm{SPT}}\rangle=U|\psi_0\rangle$, as the symmetry transformation restricted to a 2D patch $A$ (grey disk). (b) When $W_{A,g}^U$ is an FDQC, we can restrict it to $W_{\mathbf{A},g}^U$, which is supported on an open 1D interval $\mathbf{A}$. This operator inserts symmetry flux at the boundary of $\mathbf{A}$ (black crosses) and has the same action on $|\psi_{\mathrm{SPT}}\rangle$ as the symmetry defect operator $D_{\mathbf{A},g}^U=W_{\overline{\mathbf{A}},g}^{U\dagger}U_{A,g}$. $D_{\mathbf{A},g}^U$ is a symmetry defect operator because it acts like the symmetry transformation near the top boundary of $A$ (which we denote by $\mathbf{A}$), but leaves $|\psi_{\mathrm{SPT}}\rangle$ invariant near the lower boundary of $A$ (which we denote by $\overline{\mathbf{A}}$).}
   \label{fig:symmflux}
\end{figure}
One way to detect the anomaly of an SPT is by inserting symmetry flux and measuring the degrees of freedom bound to the flux. For example, in an integer quantum Hall state, the Hall conductance is computed from the quantized $U(1)$ charge bound to $2\pi$ flux insertion\cite{laughlin}. 

When $G$ is a unitary, on-site symmetry, we can insert flux using flux insertion operators. A ``closed" flux insertion operator $W_{A,g}^U$ is defined as follows. For any region $A\subset\Lambda$ deep in the bulk of the SPT, $W_{A,g}^U$ is a $(d-1)$ dimensional operator supported near the boundary of $A$ that has the same action on the SPT state as $U_{A,g}=\prod_{r\in A}U_{r,g}$, where $U_{r,g}$ is the on-site representation of the symmetry:
\begin{equation}\label{WAgU}
W_{A,g}^U|\psi_{\mathrm{SPT}}\rangle=U_{A,g}|\psi_{\mathrm{SPT}}\rangle.
\end{equation}

A \emph{closed} flux insertion operator does not insert any symmetry flux, but is useful for defining an \emph{open} flux insertion operator, which does insert symmetry flux. Assuming that $W_{A,g}^U$ is a FDQC, we can restrict $W_{A,g}^U$ to a region $\mathbf{A}\subset\partial A$, where $\partial A$ is the support of $W_{A,g}^U$. The resulting operator, which we denote by $W_{\mathbf{A},g}^U$, is strictly supported on a $(d-1)$ dimensional manifold $\mathbf{A}$. We can also define $W_{\overline{\mathbf{A}},g}^U=W_{A,g}^UW_{\mathbf{A},g}^{U\dagger}$, which is roughly supported on $\overline{\mathbf{A}}=\partial A\setminus\mathbf{A}$.\footnote{Strictly speaking, $W_{\overline{\mathbf{A}},g}^U$ is supported on a slightly larger region and overlaps slightly with $\mathbf{A}$} Using these definitions, we have
\begin{equation}
W_{\mathbf{A},g}^U|\psi_{\mathrm{SPT}}\rangle=W_{\overline{\mathbf{A}},g}^{U\dagger}U_{A,g}|\psi_{\mathrm{SPT}}\rangle.
\end{equation}

Here, $D_{A,g}^U=W_{\overline{\mathbf{A}},g}^{U\dagger}U_{A,g}$ is a symmetry defect operator: it acts as $U_{A,g}$ near $\mathbf{A}$ and the interior of $A$, and $W_{\overline{\mathbf{A}},g}^{U\dagger}$ dresses the operator so that it leaves the ground state invariant near $\overline{\mathbf{A}}$. $W_{\mathbf{A},g}^U$ inserts symmetry flux through the boundaries of the restriction, as illustrated in Fig.~\ref{fig:symmflux}. For example, in Fig.~\ref{fig:symmflux}, $W_{\mathbf{A},g}^U$ still acts like $U_{A,g}$ on $|\psi_{\mathrm{SPT}}\rangle$ near the upper boundary of $A$, but it leaves the ground state unchanged in the lower boundary of $A$. This means that it creates an extrinsic defect line along the upper boundary of $A$, which terminates at symmetry fluxes.

%%%%%%%%%%%%%%%%%%%%%%%%%%%%%%%%%%%%%%%%%%%%%%%%%%%%%%%%%%%%%%%%%%%%%%%%%%%%%%%%%%%%
\subsection{Flux insertion operators from LPUs}\label{sfluxinsert}
%%%%%%%%%%%%%%%%%%%%%%%%%%%%%%%%%%%%%%%%%%%%%%%%%%%%%%%%%%%%%%%%%%%%%%%%%%%%%%%%%%%%
We will now introduce our main tool for studying SPT entanglers, which is a particular choice of flux insertion operators. Note that in the study of SPT phases, \emph{any} $W_{A,g}^U$ that has the action defined by (\ref{WAgU}) on the SPT ground state is a valid flux insertion operator. However, we can define a particular $W_{A,g}^U$ that satisfies (\ref{WAgU}) that is easy to compute using the SPT entangler. This definition uses the SPT entangler $U$, the symmetry operator $U_{A,g}$, and a slightly smaller symmetry operator $U_{A_{\mathrm{in},g}}$, as follows:
\begin{equation}\label{WAgdef}
W_{A,g}^U=U_{A,g}UU_{A_{\mathrm{in}},g}^\dagger U^\dagger.
\end{equation}

Specifically, $A_{\mathrm{in}}$ is a subset of $A$ containing all the points lying deeper than $\xi$ within $A$: $A_{\mathrm{in}}=\{r\in A:\mathrm{dist}(r,\overline{A})>\xi\}$. Since $U$ is $G$-symmetric and locality preserving, it can only modify $U_{A_{\mathrm{in}},g}$ within $\xi$ of the boundary of $A_{\mathrm{in}}$, which is a strip of width $2\xi$ inside $A$. Denoting this strip by $\partial A$, we have
\begin{equation}\label{uuau}
UU_{A_{\mathrm{in}},g}U^\dagger=U_{A_{\mathrm{in}},g}U_{\partial A,g},
\end{equation}
where $U_{\partial A,g}$ is an operator fully supported in $\partial A$. %Because we chose $A_{\mathrm{in}}$ to be a set containing only points that are deeper than $\xi$ inside $A$, $U_{\partial A_{\mathrm{in}}}$ is fully supported in $A$. Likewise, $U^\dagger U_{A,g}U=U_{A,g}U_{\partial A}$ where $U_{\partial A}$ is fully supported outside of $A_{\mathrm{in}}$. As we will show, these specifications on $A$ and $A_{\mathrm{in}}$ ensures that $\{W_{A,g}^U\}$ forms a representation of $G$.

Let us now check that $W_{A,g}^U$ is a (closed) flux insertion operator. It is easy to see that $W_{A,g}^U$ is supported near the boundary of $A$, in $\partial A$. To check that $W_{A,g}^U$ has the same action on the SPT state as $U_{A,g}$, note that
\begin{align}
\begin{split}
W_{A,g}^U|\psi_{\mathrm{SPT}}\rangle&=U_{A,g}UU_{A_{\mathrm{in}},g}^\dagger|\psi_0\rangle\\
&=U_{A,g}U|\psi_0\rangle\\
&=U_{A,g}|\psi_{\mathrm{SPT}}\rangle.
\end{split}
\end{align}

To get the second line, we used the fact that $|\psi_0\rangle$ is invariant under $U_{A_{\mathrm{in}},g}^\dagger$.  
\begin{figure}[tb]
   \centering
   \includegraphics[width=.7\columnwidth]{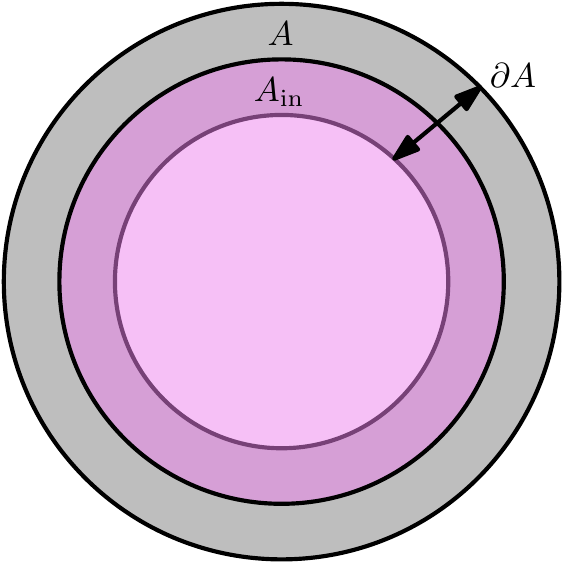} % requires the graphicx package
   \caption{$W_{A,g}^U$ is defined using $U_g$ restricted to two regions $A$ (large disk) and $A_{\mathrm{in}}$ (middle disk), which contains points deeper than $\xi$ inside $A$. $W_{A,g}^U$ is fully supported on $\partial A$, which is a strip of width $2\xi$ inside $A$.}
   \label{fig:edgerep}
\end{figure}
%%%%%%%%%%%%%%%%%%%%%%%%%%%%%%%%%%%%%%%%%%%%%%%%%%%%%%%%%%%%%%%%%%%%%%%%%%%%%%%%%%%%
\subsubsection{Properties of $\{W_{A,g}^U\}$}\label{sboundprop}
%%%%%%%%%%%%%%%%%%%%%%%%%%%%%%%%%%%%%%%%%%%%%%%%%%%%%%%%%%%%%%%%%%%%%%%%%%%%%%%%%%%%
The set of operators $\{W_{A,g}^U\}$ is easy to compute because it only involves restricting the global symmetry operator $U_g$, which can be done unambiguously for unitary, on-site symmetries. It has several important properties:
\begin{enumerate}
\item{Every element in $\{W_{A,g}^U\}$ is a $(d-1)$ dimensional strict LPU.}
\item{$\{W_{A,g}^U\}$ forms a representation of $G$.}
\item{$W_{A,g}^U$ satisfies
\begin{equation}
U_h^\dagger W_{A,g}^UU_h=W_{A,h^{-1}gh}^U.
\end{equation}}
\item{$U'\sim U$ according to Eq.~(\ref{Usim})\footnote{Stricly speaking, we will only prove that $U'\sim U$ up to multiplication by lower dimensional $G$ symmetric QCA. However, we conjecture that this stronger statement holds, as we discuss in appendix~\ref{sproofbd}.} if and only if $\{W_{A,g}^{U'}\}\sim \{W_{A,g}^{U}\}$ for any $A\subset\Lambda$, where 
\begin{equation}\label{wequiv}
\{W_{A,g}^{U'}\}\sim \{W_{A,g}^{U}\}: W_{A,g}^{U'}=V^{\dagger}W_{A,g}^{U}V
\end{equation}
for every $g\in G$, where $V$ is a $G$ symmetric FDQC fully supported within $\xi$ of the boundary of $A_{\mathrm{in}}$.}
\end{enumerate}

The first three properties help us classify different possible $\{W_{A,g}^U\}$ while the fourth property justifies using $\{W_{A,g}^U\}$ to classify SPT entanglers and, more generally, $G$ symmetric LPUs. 

We will now prove each of the four properties. 
\\

\noindent\emph{Proof of Property 1}: We will first show that $W_{A,g}^U$ is supported on a $(d-1)$ dimensional manifold, matching the description of a flux insertion operator in Sec.~\ref{sanomalous}. This follows directly from the definition of $W_{A,g}^U$ in (\ref{WAgU}) together with (\ref{uuau}). Note that (\ref{uuau}) relies on $U$ being $G$-symmetric and locality preserving. Next, $W_{A,g}^U$ is a strict LPU because it is the product of four strict LPUs: $U_{A,g}$ and $U_{A_{\mathrm{in}},g}$ are obviously strict LPUs, and $U$ is also a strict LPU. $U_{A,g}$ and $U_{A_{\mathrm{in}},g}$ both have operator spreading length zero because they are products of on-site operators, while $U$ and $U^\dagger$ both have operator spreading length $\xi$. Therefore, $W_{A,g}^U$ is a strict LPU with operator spreading length $2\xi$.
\\ 

\noindent\emph{Proof of Property 2}: For $\{W_{A,g}^U\}$ to form a representation of $G$, $\{W_{A,g}^U\}$ must satisfy $W_{A,g}^UW_{A,h}^U=W_{A,gh}^U$. By definition,
\begin{equation}
W_{A,g}^UW_{A,h}^U=U_{A,g}UU_{A_{\mathrm{in}},g}^\dagger U^\dagger U_{A,h}UU_{A_{\mathrm{in}},h}^\dagger U^\dagger.
\end{equation}

Notice that $U$ only modifies $U_{A,h}$ by an operator $U_{\partial A',h}$ supported within $\xi$ of the boundary of $A$: $U^\dagger U_{A,h}U=U_{A,h}U_{\partial A',g}$. Since $U_{\partial A',h}$ is supported outside of $A_{\mathrm{in}}$, $U^\dagger U_{A,h}U$ acts as $U_{h}$ within $A_{\mathrm{in}}$. Therefore, to pull it through $U_{A_{\mathrm{in}},g}^\dagger$, we conjugate $g$ by $h$:
\begin{align}
\begin{split}
W_{A,g}^UW_{A,h}^U&=U_{A,g}UU^\dagger U_{A,h}UU_{A_{\mathrm{in}},h^{-1}gh}^\dagger U_{A_{\mathrm{in}},h}^\dagger U^\dagger\\
&=U_{A,g}U_{A,h}UU_{A_{\mathrm{in}},h^{-1}gh}^\dagger U_{A_{\mathrm{in}},h}^\dagger U^\dagger.
\end{split}
\end{align}

Simplifying further, we get
\begin{align}
\begin{split}
W_{A,g}^UW_{A,h}^U&=U_{A,gh}UU_{A_{\mathrm{in}},gh}^\dagger U^\dagger\\
&=W_{A,gh}^U.
\end{split}
\end{align}

One important implication of Property 2 is that, in order for $\{W_{A,g}^U\}$ to form a representation of a finite group $G$, it must have finite order: $\left(W_{A,g}^U\right)^{|g|}=\mathbbm{1}$. The only nontrivial bosonic QCA in the absence of symmetry in 1D and 2D are translations\cite{GNVW,freedmanclassification}, which have order proportional to the system size. This means that for systems of spatial dimension up to three, $\{W_{A,g}^U\}$ must all be FDQCs (if we ignore symmetry). When this is the case, we can always truncate $W_{A,g}^U$ to insert symmetry flux.
\\

\noindent\emph{Proof of Property 3}: We use the fact that $U$ commutes with global symmetry operators $U_h$ to obtain
\begin{align}
\begin{split}
U_h^\dagger W_{A,g}^UU_h&=U_h^\dagger U_{A,g}U_h UU_h^\dagger U_{A_{\mathrm{in}},g}^\dagger U_h\\
&=U_{A,h^{-1}gh}UU_{A_{\mathrm{in}},h^{-1}gh}^\dagger U^\dagger\\
&=W^U_{A,h^{-1}gh}.
\end{split}
\end{align}

In particular, if $G$ is abelian, then $W_{A,g}^U$ commutes with all global symmetry operators. 
\\

\noindent\emph{Proof of Property 4}: We will sketch the idea of the ``only if" direction here; the precise version and the proof for the ``if" direction are more complicated so we defer them to Appendix.~\ref{sproofbd}. Note that the ``only if" direction is sufficient for using $\{W_{A,g}^U\}$ to classify LPUs with symmetry, in that if $\{W_{A,g}^{U'}\}$ is not equivalent to $\{W_{A,g}^U\}$, then $U'$ is not equivalent to $U$. The other direction ensures that $\{W_{A,g}^U\}$ \emph{completely} classifies symmetric LPUs. 

The rough idea of the ``only if" direction is that if we modify $U'=YU$ where $Y$ is a $G$ symmetric FDQC, then by definition, $W_{A,g}^{U'}$ is given by
\begin{equation}
W_{A,g}^{U'}=U_{A,g}YUU_{A_{\mathrm{in}},g}^\dagger U^\dagger Y^\dagger.
\end{equation}

Because $Y$ is a $G$ symmetric FDQC, we can commute the gates fully supported deep inside $A_{\mathrm{in}}$ and far outside $A_{\mathrm{in}}$ through $UU_{A_{\mathrm{in}},g}^\dagger U^\dagger$. Let us denote the product of the remaining gates by $\tilde{Y}$, so that $W_{A,g}^{U'}=U_{A,g}\tilde{Y}UU_{A_{\mathrm{in}},g}^\dagger U^\dagger\tilde{Y}^\dagger$. In Appendix~\ref{sproofbd}, we show that $\tilde{Y}$ is guaranteed to be fully supported within $\xi$ of the boundary of $A_{\mathrm{in}}$. This means that we can commute it through $U_{A,g}$:
\begin{equation}
W_{A,g}^{U'}=\tilde{Y}W_{A,g}^U\tilde{Y}^\dagger,
\end{equation}
where $\tilde{Y}$ is, as desired, a $G$ symmetric FDQC fully supported within $\xi$ of the boundary of $A_{\mathrm{in}}$. 

%%%%%%%%%%%%%%%%%%%%%%%%%%%%%%%%%%%%%%%%%%%%%%%%%%%%%%%%%%%%%%%%%%%%%%%%%%%%%%%%%%%%
\subsubsection{Using $\{W_{A,g}^U\}$ to classify LPUs}\label{sframework}
%%%%%%%%%%%%%%%%%%%%%%%%%%%%%%%%%%%%%%%%%%%%%%%%%%%%%%%%%%%%%%%%%%%%%%%%%%%%%%%%%%%%
We showed that $\{W_{A,g}^U\}$ forms a $(d-1)$ dimensional representation of $G$. However, this representation may be anomalous, in that there may be an obstruction to making $\{W_{A,g}^U\}$ equivalent, according to Property 4, to $\{W_{A,g}^{\mathbbm{1}}\}$. Our framework for obtaining topological invariants for LPUs with symmetry is computing $\{W_{A,g}^U\}$ and then detecting these different kinds of obstructions. Some kinds of obstructions are not related to SPT invariants; these are related to $G$ symmetric LPUs that are not SPT entanglers. We will focus on the obstructions that can be directly related to known SPT invariants. 

In fact, as we show in Appendix~\ref{sboundaryrep}, $\{W_{A,g}^U\}$ carries the same anomaly as the boundary representation of the symmetry described in Ref.~\onlinecite{symmedge}. We use $\{W_{A,g}^U\}$ rather than the boundary representation because it is more explicit, and can be obtained without truncating the SPT entangler. Moreover, anomalies are sometimes easier to detect using $\{W_{A,g}^U\}$ rather than the boundary representation because $\{W_{A,g}^U\}$ satisfies additional properties described in Sec.~(\ref{sboundprop}). In particular, Properties 3 and 4 do not apply to the boundary representation of the symmetry.

With the above approach in mind, we will now derive formulas for topological invariants for $G$ symmetric SPT entanglers in various dimensions. We begin with 1D SPT entanglers.

%%%%%%%%%%%%%%%%%%%%%%%%%%%%%%%%%%%%%%%%%%%%%%%%%%%%%%%%%%%%%%%%%%%%%%%%%%%%%%%%%%%%
\section{1D SPT entanglers}\label{s1d}
%%%%%%%%%%%%%%%%%%%%%%%%%%%%%%%%%%%%%%%%%%%%%%%%%%%%%%%%%%%%%%%%%%%%%%%%%%%%%%%%%%%%
In this section, we will present formulas for topological invariants for 1D bosonic SPT entanglers with discrete symmetries. We will first focus on abelian, unitary symmetries, and present the simple generalization to non-abelian symmetries in Sec.~\ref{snonabelian}. Our invariants for SPT entanglers with antiunitary time-reversal symmetry is less closed form; we present it in Sec.~\ref{str}. The topological invariants, in the abelian case, simply compute $\{c(g,h)\}=\left\{\frac{\omega(g,h)}{\omega(h,g)}\right\}$. 

%%%%%%%%%%%%%%%%%%%%%%%%%%%%%%%%%%%%%%%%%%%%%%%%%%%%%%%%%%%%%%%%%%%%%%%%%%%%%%%%%%%%
\subsection{1D SPT entanglers with abelian symmetries}\label{sabelian}
%%%%%%%%%%%%%%%%%%%%%%%%%%%%%%%%%%%%%%%%%%%%%%%%%%%%%%%%%%%%%%%%%%%%%%%%%%%%%%%%%%%%
Consider a 1D bosonic spin chain, where $A$ is a finite 1D interval. The boundary of $A$ consists of two disconnected points, so $W_{A,g}^U$ is a product of two local operators:
\begin{align}
\begin{split}\label{Uproj}
W_{A,g}^U&=U_{A,g} UU_{A_{\mathrm{in}},g}^\dagger U^\dagger\\
&=L_{A,g}^U\otimes R_{A,g}^U.
\end{split}
\end{align}

Since $G$ is abelian, Property 3 says that $W_{A,g}^U$ commutes with the global symmetry operator $U_h$ for every $h\in G$. Notice that if $W_{A,g}^U=V^\dagger W_{A,g}^{\mathbbm{1}}V$, where $V$ satisfies the definition in Property 4, then $L_{A,g}^U=V^\dagger L_{A,g}^{\mathbbm{1}}V$ and $R_{A,g}^U=V^\dagger R_{A,g}^{\mathbbm{1}}V$ would both commute with $U_h$. Therefore, if $L_{A,g}^U$ and $R_{A,g}^U$ fail to individually commute with $U_h$, then $\{W_{A,g}^U\}$ is \emph{not} equivalent to $\{W_{A,g}^{\mathbbm{1}}\}$, so there must be an obstruction to making $U$ an $G$-symmetric FDQC. 

Physically, this means that nontrivial SPT entanglers ``decorate" the endpoints of a symmetry operator with charge of other global symmetries. Because $L_{A,g}^U$ and $R_{A,g}^U$ are far separated, in order for $L_{A,g}^U\otimes R_{A,g}^U$ to commute with $U_{h}$, the commutator of $R_{A,g}^U$ with $U_h$ must be a phase and the commutator of $L_{A,g}^U$ with $U_h$ must be the opposite phase. To measure the phase, we can compute $\mathrm{Tr}(R_{A,g}^UU_hR_{A,g}^{U\dagger} U_h^\dagger)$, but this involves the extra step of truncating $W_{A,g}^U$ to isolate $R_{A,g}^U$. Instead, we compute the commutator of $W_{A,g}^U$ with $U_{B,h}$, where $B$ is an interval that includes the support of $R_{A,g}^U$, but not the support of $L_{A,g}^U$. This gives
\begin{equation}\label{cghw}
c(g,h)=\overline{\mathrm{Tr}}\left(W_{A,g}^UU_{B,h}W_{A,g}^{U\dagger}U_{B,h}^\dagger\right),
\end{equation}
where $\overline{\mathrm{Tr}}$ refers to a trace that is normalized such that $\overline{\mathrm{Tr}}(\mathbbm{1})=1$. 

Eq.~(\ref{cghw}) is already completely closed form, but we can simplify it even further by using the explicit form of $W_{A,g}^U$ from (\ref{WAgdef}). To generalize more easily to non-abelian symmetries, it is convenient to use a different representation $\{\mathcal{W}_{A,g}^U\}$, given by
\begin{equation}
\mathcal{W}_{A,g}^U=U^\dagger W_{A,g}^UU=U_{A_{\mathrm{in}},g}^\dagger U^\dagger U_{A,g}U.
\end{equation}

This representation carries the same anomaly as $\{W_{A,g}^U\}$ because it is obtained from $\{W_{A,g}^U\}$ by conjugation by an LPU. Furthermore, defining $\mathcal{W}_{A,g}^U=\mathcal{L}_{A,g}^U\otimes\mathcal{R}_{A,g}^U$, we see that $\mathcal{R}_{A,g}^U=U^\dagger R_{A,g}^UU$ has the same commutator with $U_h$ as $R_{A,g}^U$, because $U$ commutes with $U_h$. $\mathcal{R}_{A,g}^U$ is fully supported to the right of $A_{\mathrm{in}}$, and since $B$ only needs to contain the full support of $\mathcal{R}_{A,g}^U$, we can choose $A_{\mathrm{in}}$ and $B$ to be adjacent and disjoint. Replacing $W_{A,g}^U$ in (\ref{cghw}) by $\mathcal{W}_{A,g}^U$, we get
\begin{equation}
c(g,h)=\overline{\mathrm{Tr}}\left(U_{A_{\mathrm{in}},g}^\dagger U^\dagger U_{A,g}UU_{B,h}U^\dagger U_{A,g}^\dagger UU_{A_{\mathrm{in}},g}U_{B,h}^\dagger\right).
\end{equation}

Since we chose $U_{A_{\mathrm{in}},g}$ and $U_{B,h}$ to have disjoint support, we can commute $U_{A_{\mathrm{in}},g}$ through $U_{B,h}$ to obtain 
\begin{equation}\label{flowdiscrete}
c(g,h)=\overline{\mathrm{Tr}}\left(U^\dagger U_{A,g} U U_{B,h}U^\dagger U_{A,g}^\dagger U U_{B,h}^{\dagger}\right).
\end{equation}

Eq.~(\ref{flowdiscrete}) is the main result of this section, and the generalization of (\ref{example}) to general abelian groups. Notice that when $G$ is abelian, we can always commute $U_{A_{\mathrm{in}},g}$ through $U_{B,h}$, regardless of their support. However, when $G$ is non-abelian, we can only do this if the two operators are supported on disjoint intervals, as they are here.

To check that the invariant given by Eq.~(\ref{flowdiscrete}) is invariant under modification of $U$ by any $G$ symmetric FDQC, we can simply check that $c(g,h)$ is invariant under $U\to UU_r$ or $U\to U_rU$ where $U_r$ is a local, $G$ symmetric unitary anywhere in the system. We already checked this in Sec.~\ref{sexample}. Since an FDQC is built out of such $U_r$ operators, this means that $c(g,h)$ is invariant under modification of $U$ by any $G$ symmetric FDQC, as long as $A$ and $B$ are sufficiently large.

The proof follows the same line of argument as the proof of Theorem 1.1 from Ref.~\onlinecite{flow}. We sketch the proof again here. Suppose that $U=YU$ where $Y=\prod_{n=1}^NU_n$ is a $G$-symmetric FDQC, with $U_n=\prod_rU_{n,r}$ a product of disjoint $G$-symmetric local unitaries. We can first remove all the gates $U_{1,r}$ in $U_1$ fully supported deep in $A$ or $\overline{A}$ by commuting $\left(\prod_{n=2}^NU_n\right)^\dagger U_{1,r}\left(\prod_{n=2}^NU_n\right)$ through $U_{A,g}$ and all the gates $U_{1,r'}$ in $U_1$ fully supported deep in $B$ or $\overline{B}$ (i.e. the rest of the gates in $U_1$) by commuting $U^\dagger U_{1,r'} U$ through $U_{B,h}$. This is possible as long the endpoints of (overlapping) $A$ and $B$ are all separated by distances greater than $\mathrm{max}(\xi+2\lambda,2N\lambda)$, where $\xi$ is the operator spreading length of $U$ and $\lambda$ is the radius of a single gate in $Y$. These length scales ensure that all operators $\left(\prod_{n=2}^{N}U_n\right)^\dagger U_{1,r} \left(\prod_{n=2}^{N}U_n\right)$ and $U^\dagger U_{1,r'} U$ are fully supported in $A,\overline{A},B,$ or $\overline{B}$. Assuming that $A$ and $B$ are sufficiently large and overlapping, we can proceed in the same way to remove all the gates in $U_2$, then $U_3$, up to $U_N$. This completely removes $Y$.

This concludes the proof that $c(g,h)$ defined in (\ref{flowdiscrete}) is a topological invariant. Specifically, it is invariant under $U\to YU$ for any $G$-symmetric FDQC $Y$, as long as $A$ and $B$ are sufficiently large and overlapping.
%
%We choose $A$ and $B$ to be sufficiently overlapping, so that every point in the system lies deep inside at least one of four intervals: $A,B,\overline{A}$, or $\overline{B}$. Since $U_r$ is $G$ symmetric, it commutes with $U_{B,h}$ if it is fully supported in $B$ or $\overline{B}$. We can easily see from (\ref{flowdiscrete}) that if we modify $U\to U U_r$, then we can remove $U_r$ by either commuting $U_r$ through $U_{A,g}$ or $UU_rU^\dagger$ through $U_{B,h}$. Similarly, if $U\to U_rU$, then we can remove $U_r$ by commuting it through $U_{A,g}$ or commuting $U^\dagger U_rU$ through $U_{B,h}$. Since $c(g,h)$ in (\ref{flowdiscrete}) is invariant under modification of $U$ with any $G$ symmetric FDQC, $c(g,h)$ is a topological invariant of $G$ symmetric SPT entanglers. 
%%%%%%%%%%%%%%%%%%%%%%%%%%%%%%%%%%%%%%%%%%%%%%%%%%%%%%%%%%%%%%%%%%%%%%%%%%%%%%%%%%%%
\subsection{Relation to SPT invariants}\label{sspt1d}
%%%%%%%%%%%%%%%%%%%%%%%%%%%%%%%%%%%%%%%%%%%%%%%%%%%%%%%%%%%%%%%%%%%%%%%%%%%%%%%%%%%%
The obstruction to making the left and right parts of $W_{A,g}^U$ individually commute with $U_h$ is directly related to the projective representation defining the 1D SPT phase entangled by $U$, when $G$ is abelian. More generally, not restricting to abelian groups, $W_{A,g}^U=L_{A,g}^U\otimes R_{A,g}^U$ forms a linear representation of $G$ while $L_{A,g}^U$ and $R_{A,g}^U$ individually can form opposite projective representations of $G$:
\begin{align}
\begin{split}
L_{A,g}^UL_{A,h}^U&=\omega(g,h)^{-1}L_{A,gh}^U\\
R_{A,g}^UR_{A,h}^U&=\omega(g,h)R_{A,gh}^U.
\end{split}
\end{align}

The function $\omega(g,h):G\times G\to U(1)$ has an ambiguity in that we can attach a phase $\beta(g)$ to each $R_{A,g}^U$ and $\beta^{-1}(g)$ to each $L_{A,g}^U$, which changes $\omega(g,h)$ by a coboundary: $\omega(g,h)\to \omega(g,h)\beta(g)\beta(h)\beta^{-1}(gh)$. When $G$ is abelian, $R_{A,gh}^U=R_{A,hg}^U$, so we can define $\omega(g,h)$ by a set of gauge-invariant phases $\{c(g,h)\}$, given by 
\begin{equation}\label{cghdef}
c(g,h)=\frac{\omega(g,h)}{\omega(h,g)}=\overline{\mathrm{Tr}}\left(R_{A,g}^UR_{A,h}^UR_{A,g}^{U\dagger}R_{A,h}^{U\dagger}\right).
\end{equation}

$c(g,h)$ is clearly gauge invariant because any phase attached to $R_{A,g}^U$ or $R_{A,h}^U$ is canceled by the opposite phase attached to $R_{A,g}^{U\dagger}$ or $R_{A,h}^{U\dagger}$. Using the fact that $L_{A,g}^U\otimes R_{A,g}^U$ is an ordinary representation of $G$, it is easy to show that $L_{A,g}^U$ has the opposite set of phases: $L_{A,g}^UL_{A,h}^UL_{A,g}^{U\dagger}L_{A,h}^{U\dagger}=c(g,h)^{-1}$. The set of phases $\{c(g,h)\}$ for every pair of group elements $g,h\in G$ completely defines the projective representation of $R_{A,g}^U$ when $G$ is abelian, and therefore completely classifies 1D bosonic entanglers with unitary, discrete, abelian on-site symmetries. %When $G$ is non-abelian, we can also specify the cocycle defining the SPT by computing all the gauge invariant phases, but these are not given simply by $\{c(g,h)\}$. We disc Sec.~\ref{snonabelian}).

It is not obvious that the invariant defined in (\ref{flowdiscrete}) is $c(g,h)$ defined in (\ref{cghdef}). We prove that these two quantities are equal in Appendix~\ref{sproofs1D}.
%%%%%%%%%%%%%%%%%%%%%%%%%%%%%%%%%%%%%%%%%%%%%%%%%%%%%%%%%%%%%%%%%%%%%%%%%%%%%%%%%%%%
\subsection{1D SPT entanglers with discrete, non-abelian symmetries}\label{snonabelian}
%%%%%%%%%%%%%%%%%%%%%%%%%%%%%%%%%%%%%%%%%%%%%%%%%%%%%%%%%%%%%%%%%%%%%%%%%%%%%%%%%%%%
SPTs with non-abelian, unitary, on-site symmetries are also classified by projective representations. In this section, we will show how specify the projective representation from quatntities computed using the SPT entangler, using formulas similar to (\ref{flowdiscrete}).% to extract a set of gauge-invariant phases, which generalize $\{c(g,h)\}$ to non-abelian symmetries, from an SPT entangler . These phases completely specify the projective representation.  

According to Schur's theorem, 1D projective representations (also known as Schur multipliers) of on-site symmetries are completely specified by all the gauge-invariant phases $\{e^{i\phi(\gamma_n)}\}$\cite{pollmann2012}. To obtain these gauge-invariant phases, we consider all products of commutators $\gamma_n$, of the form 
\begin{equation}
\gamma_n=\prod_ig_ih_ig_i^{-1}h_i^{-1},
\end{equation}
satisfying the property that multiplying the elements on the right hand side in the group gives the identity. Notice that this is a natural generalization of the abelian case, where we consider elements of the form $ghg^{-1}h^{-1}$ which multiply to identity in the group. The fact that multiplication in the group gives identity means that, if instead we multiply the projective representations, we get a phase:
\begin{equation}\label{ephigamma}
e^{i\phi(\gamma_n)}=\overline{\mathrm{Tr}}\left[\prod_i\left(R_{A,g_i}^UR_{A,h_i}^UR_{A,g_i}^{U\dagger} R_{A,h_i}^{U\dagger}\right)\right].
\end{equation}

Phases of this form are gauge invariant because every group element on the right hand side appears an equal number of times as its inverse. Therefore, any phase attached to $R_{A,g_i}^U$ is canceled by the opposite phase attached to $R_{A,g_i}^{U\dagger}$. Phases of the form (\ref{ephigamma}) naturally generalize $\{c(g,h)\}$ to non-abelian symmetries. We show in Appendix~\ref{sproofnonabelian} that we can write $e^{i\phi(\gamma_n)}$ as
\begin{equation}\label{flownonabelian}
e^{i\phi(\gamma_n)}=\overline{\mathrm{Tr}}\left[\prod_i\left(U^\dagger U_{A,g_i}U U_{B,h_i}U^\dagger U_{A,g_i}^{\dagger} U U_{B,h_i}^{\dagger}\right)\right],
\end{equation}
where $A$ and $B$ are overlapping intervals as in Eq.~(\ref{flowdiscrete}). %The set of phases $\{e^{i\phi(\gamma)}\}$ for all $\gamma\in\left([F,F]\cap R\right)/[R,F]$ completely determines the projective representation associated with $U$ for non-abelian symmetries.
\subsection{1D SPT entanglers with time reversal symmetry}\label{str}
%%%%%%%%%%%%%%%%%%%%%%%%%%%%%%%%%%%%%%%%%%%%%%%%%%%%%%%%%%%%%%%%%%%%%%%%%%%%%%%%%%%%
Time reversal symmetry is different from the unitary, on-site symmetries discussed in the previous sections because it cannot be restricted. This is because it is an antiunitary symmetry, taking the form $T=U_TK$, where $K$ is complex conjugation. While $U_T$ is a unitary operator that can be restricted, $K$ acts everywhere; there is no way to restrict complex conjugation.

However, we can restrict the SPT entangler $U$. In 1D, there is a $\mathbb{Z}_2$ classification of SPTs with time reversal symmetry. We will now show how to compute the corresponding $\mathbb{Z}_2$-valued invariant from the SPT entangler. Our invariant is closely related to the state-based concept of ``local Kramers degeneracy" described in Ref.~\onlinecite{levintr}.

First, we note that for an SPT entangler to be time reversal symmetric, it must satisfy 
\begin{equation}
T^{-1}U^\dagger T=U^\dagger\to T^{-1}U^\dagger TU=\mathbbm{1}.
\end{equation}

Now suppose that we truncate $U$ to $U_A$, which is fully supported in the interval $A$. Then 
\begin{equation}
T^{-1}U_A^\dagger TU_A=O_{A}^U=L_{A,T}^U\otimes R_{A,T}^U,
\end{equation}
where $L_{A,T}^U$ and $R_{A,T}^U$ are local operators at the left and right endpoints of $A$. Conjugating $O_{A}^U$ by $T$ gives
\begin{equation}
T^{-1}\left(T^{-1}U_A^{\dagger}TU_A\right)T=T^{-2}U_A^{\dagger}T^{-1}T^2U_AT.
\end{equation}

Now using $T^2=T^{-2}=\pm1$, we get
\begin{align}
\begin{split}
T^{-1}\left(T^{-1}U_A^{\dagger}TU_A\right)T&=U_A^{\dagger}T^{-1}U_AT\\
&=O_{A}^{U\dagger}.
\end{split}
\end{align}

This means that $T^{-1}O_{A}^UTO_{A}^{U}=\mathbbm{1}$. However, it may not be true that $T^{-1}L_{A,T}^UTL_{A,T}^U=\mathbbm{1}$ and $T^{-1}R_{A,T}^UTR_{A,T}^U=\mathbbm{1}$ individually. We claim that the following topological invariant classifies time reversal invariant SPT entanglers:
\begin{equation}\label{timerev}
\eta=\overline{\mathrm{Tr}}\left(T^{-1}R_{A,T}^UTR_{A,T}^{U}\right).
\end{equation}

In particular, $\eta=1$ for time reversal invariant FDQCs and $\eta=-1$ for a nontrivial time reversal SPT entangler. 

We will now show that $\eta$ is a topological invariant in that it is invariant under modification of $U\to U_rU$ or $U\to UU_r$ where $U_r$ is any time reversal symmetric local unitary. It is clear that if $U\to U_rU$ and $U_r$ is fully supported in $A$ or $\overline{A}$, then $U_r$ commutes through $T$ so $O_A^U$ is left unchanged: $O_A^U=O_A^{U_rU}$. Similarly, if $U\to UU_r$ where $U_r$ is fully supported deeper than $\xi$ within $A$ or $\overline{A}$, then $UU_rU^\dagger$ commutes through $T$ and $O_A^U$ is also left unchanged. Therefore, the only modifications of $U$ that might change $\eta$ are those near the endpoints of $A$. However, if we modify $U\to U'$ near the right endpoint of $A$ and then restrict, the resulting operator $O_A^U$ must still satisfy $T^{-1}O_A^{U'}TO_A^{U'}=\mathbbm{1}$, so
\begin{equation}\label{leftright}
\left(T^{-1}L_{A,T}^{U'}TL_{A,T}^{U'}\right)\left(T^{-1}R_{A,T}^{U'}TR_{A,T}^{U'}\right)=\mathbbm{1}.
\end{equation}

Since a modification near the right endpoint of $A$ does not change the first factor in (\ref{leftright}), it cannot change the second factor, and therefore cannot change $\eta$. A similar argument shows that $\eta$ is invariant under modifications of $U$ near the left endpoint of $A$, so $\eta$ is in fact invariant under $U\to U_rU$ or $UU_r$ where $U_r$ is a time-reversal symmetric local unitary anywhere in the system. This confirms that $\eta$ is a topological invariant for time reversal symmetric SPT entanglers.
% it is clear that $\eta$ is invariant under modification of $U_A\to U_AU_r$ where $U_r$ is a time reversal invariant local unitary anywhere in the 1D system, because $U_r$ commutes with $T$ and leaves $O_{A,g}^U$ unchanged. $\eta$ is invariant under modification of $U_A\to U_rU_A$ because if $U_r$ is supported deep within $A$ or $\overline{A}$, then we can write $U_rU_A=U_A\left(U_A^\dagger U_rU_A\right)$. Then, observing that $U_A^\dagger U_rU_A$ is time reversal symmetric (as long as $U_r$ is supported deep within $A$ or $\overline{A}$, we can commute it through $T$ as we did for $U_A\to U_AU_r$. If $U_r$ is supported near the left endpoint of $A$, then it cannot change $\eta$ because $U_r$ does not modify $\tilde{R}_{A,g}^U$. If $U_r$ is supported near the right endpoint of $A$, then it cannot change $\eta$ because the value of $\eta$ can also be computed from $\tilde{L}_g^U$: $\eta^{-1}=\overline{\mathrm{Tr}}\left(T\tilde{L}_{A,g}^UT^{-1}\tilde{L}_{A,g}^{U}\right)$. This comes from the fact that $TO_{A,g}^UT^{-1}O_{A,g}^U=\mathbbm{1}$. In conclusion, 

We can also show that $\eta=\pm1$, using the fact that $TR_{A,T}^U$ is an antiunitary operator and $\eta$ is a scalar with unit norm. To do this, we compute $\left(TR_{A,T}^U\right)^3$ and use associativity:
\begin{equation}\label{etareal}
\left(TR_{A,T}^U\right)^3=\eta T^2TR_{A,T}^U=TR_{A,T}^U T^2\eta^*.
\end{equation}

Here, $\eta$ is complex conjugated in the last term because it appears to the right of $T$, which is antiunitary. Canceling the $T^2=\pm 1$ in Eq.~(\ref{etareal}), we see that $\eta$ to be real. Combined with the fact that $\eta$ is a scalar with unit norm, this means that $\eta$ must equal $\pm1$.

A simple example of an SPT entangler with time reversal symmetry is the same LPU we used for $\mathbb{Z}_2\times\mathbb{Z}_2$, written in Eq.~(\ref{uz2z2}). We take time reversal to act as
\begin{equation}
T=\left(\prod_{r\in\Lambda}\sigma^x_r\right)K.
\end{equation}

Truncating $U$ defined in (\ref{uz2z2}) to $U_A=\prod_{r=L}^{R-1}e^{\frac{i\pi}{4}(-1)^r\sigma^z_r\sigma^z_{r+1}}$ gives $O_{A}^U=U_A^2=\sigma^z_L\sigma^z_R$. Therefore, $R_{A,T}^U=\sigma^z_R$. In this case, it is easy to compute $\eta=-1$.
%%%%%%%%%%%%%%%%%%%%%%%%%%%%%%%%%%%%%%%%%%%%%%%%%%%
\subsection{Higher dimensional SPT entanglers with 1D decorated domain walls}\label{sdecdomain}
%%%%%%%%%%%%%%%%%%%%%%%%%%%%%%%%%%%%%%%%%%%%%%%%%%%
We can use our invariants for 1D SPT entanglers to obtain closed form formulas for topological invariants of certain kinds of higher dimensional SPT entanglers. These SPT entanglers entangle SPTs of a symmetry group that is a product of two groups $G\times H$ (here, we assume that $G$ and $H$ are both unitary), where the anomaly corresponds to ``decorated domain walls." This means that the domain walls of one symmetry, say $G$, carry SPTs of another symmetry, say $H$. Physically, the action of the SPT entangler can be thought of as depositing $H$ SPTs onto $G$ domain walls.

\begin{figure}[tb]
   \centering
   \includegraphics[width=.9\columnwidth]{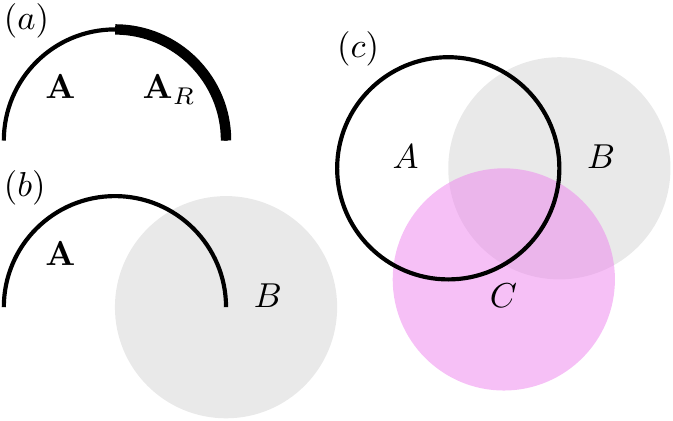} % requires the graphicx package
   \caption{Setup for calculation of 2D topological invariants described in Sec.~\ref{sabelian2d}. (a) To compute $e^{i\theta_{g_i}}$, we use $W_{\mathbf{A},g_i}^U$, which is supported on $\mathbf{A}$ is a subset of the 1D boundary of a 2D region $A$. $\mathbf{A}_R\subset\mathbf{A}$ is the right half of $\mathbf{A}$, and $W_{\mathbf{A},g_i}^U$ and $W_{\mathbf{A}_{R},g_i}^U$ are restricted in the same way on the right endpoint. (b) To compute $e^{i\theta_{g_i,g_j}}$, we use $W_{\mathbf{A},g_i}^U$ and $U_{B,g_j}$, where $B$ is a region that includes the right endpoint of $\mathbf{A}$ but not the left endpoint. (c) To compute $e^{i\theta_{g_i,g_j,g_k}}$, we use $W_{A,g_i}^U$ and symmetry operators for two other overlapping regions $B$ and $C$.}
   \label{fig:ABC}
\end{figure}

%For example, for a 2D system with $\mathbb{Z}_2\times H$ symmetry, the domain walls of the $\mathbb{Z}_2$ subgroup can carry 1D $H$ SPTs of order two.\footnote{By order two, we simply mean that a tensor product of two $H$ SPT states is trivial and can be disentangled by a symmetric FDQC.} For example, $H$ can be $\mathbb{Z}_2\times\mathbb{Z}_2$, which has an SPT of order two entangled by the SPT entangler written in Eq.~(\ref{uz2z2}). In this case, the flux insertion operator for the $\mathbb{Z}_2$ symmetry is equivalent to the 1D SPT entangler in Eq.~(\ref{uz2z2}). This means that we can simply use the flux insertion operator in place of $U$ in the formulas for the 1D topological invariants, given in (\ref{flowdiscrete}) for abelian $H$ and (\ref{flownonabelian}) for nonabelian $H$. 
More specifically, consider a 2D system with symmetry group $G\times H$, with three overlapping disks $A,B, $ and $C$ as illustrated in Fig.~\ref{fig:ABC}.c. Let $k$ be the generator of the $G$ symmetry and define $W_{C,k}^U$ as usual:
\begin{equation}
W_{C,k}^U=U_{C,k}UU_{C_{\mathrm{in}},k}^\dagger U^{\dagger}.
\end{equation}

Any $W_{C,k}^U$ equivalent to $W_{C,k}^{\mathbbm{1}}$ would be an $H$ symmetric FDQC of order $|k|$. Our topological invariants in this case describe obstructions to making $W_{C,k}^U$ an $H$-symmetric FDQC of order $|k|$. $W_{C,k}^U$ is supported near the 1D boundary of $C$, and the only constraints on $W_{C,k}^U$ is that it is a $G\times H$ symmetric LPU, and has order $|k|$. If $W_{C,k}^U$ entangles an SPT of the symmetry $H$, then by definition, it cannot be an $H$ symmetric FDQC, and is therefore an anomalous representation of the symmetry. If $H$ is abelian, we can simply plug in $W_{C,k}^U$ in (\ref{flowdiscrete}) in place of $U$:
\begin{equation}\label{decdomaininv}
c(g,h;k)=\overline{\mathrm{Tr}}\left(W_{C,k}^{U\dagger}U_{A,g} W_{C,k}^{U}U_{B,h}W_{C,k}^{U\dagger}U_{A,g}^\dagger W_{C,k}^{U}U_{B,h}^\dagger\right),
\end{equation}
where $g,h,\in H$. Notice that from Fig.~\ref{fig:ABC}.c, the overlap of $A$ and $B$ with the support of $W_{C,k}^U$ naturally gives two overlapping 1D intervals. We can write $U_{A,g}=U_{A_{\partial C},g}U_{A_{\overline{\partial C}},g}$ and $U_{B,h}=U_{B_{\partial C},h}U_{B_{\overline{\partial C}},h}$, where $A_{\partial C}$ and $B_{\partial C}$ are the overlaps of $A$ and $B$ respectively with the support of $W_{C,k}^U$. The trace in (\ref{decdomaininv}) then splits into a product of two traces: one over the support of $W_{C,k}^U$ and one over the rest of the Hilbert space of the (finite) 2D lattice. The former trace evaluates to $c(g,h)$ while the latter trace evaluates to
\begin{equation}
\overline{\mathrm{Tr}}\left(U_{A_{\overline{\partial C}},g}U_{B_{\overline{\partial C}},h}U_{A_{\overline{\partial C}},g}^\dagger U_{B_{\overline{\partial C}},h}^\dagger \right)=1,
\end{equation}
because $H$ is abelian. 

Eq.~(\ref{decdomaininv}) gives a completely closed form formula for a topological invariant $c(g,h;k)$ labeling the $G\times H$-symmetric SPT entangler. If $H$ is non-abelian, then we can similarly plug in $W_{C,k}^U$ in place of $U$ to compute the characters $\{e^{i\phi(\gamma)}\}$ defined in (\ref{flownonabelian}). 

It is easy to check that $c(g,h;k)$ is invariant under modification of $U$ by any $G\times H$-symmetric local unitary. It is invariant under modification of $U$ by $G\times H$ symmetric local unitaries deep in $C$ or $\overline{C}$ because these do not change $W_{C,k}^U$. Any other local, symmetric unitary changes $W_{C,k}^U\to U_r^\dagger W_{C,k}^UU_r$, where $U_r$ is a symmetric local unitary supported near the boundary of $C$. But according to the arguments in Sec.~\ref{sabelian}, $U_r$ can be removed by commuting it through $U_{A,g}$ or $U_{B,h}$, because it is supported either deep in $A$ or $\overline{A}$, or deep in $B$ or $\overline{B}$. 

This procedure can be easily generalized to higher dimensions by using $(d+1)$ overlapping $d$-balls, and continuing to substitute flux insertion operators for SPT entanglers. This gives completely closed form formulas for topological invariants of SPT entanglers related to decoration with 1D SPTs.

%%%%%%%%%%%%%%%%%%%%%%%%%%%%%%%%%%%%%%%%%%%%%%%%%%%
\section{2D SPT entanglers with discrete, abelian, unitary, on-site symmetries}\label{sabelian2d}
%%%%%%%%%%%%%%%%%%%%%%%%%%%%%%%%%%%%%%%%%%%%%%%%%%%
We now consider SPT entanglers in 2D. Like in 1D, we can also expect that topological invariants of 2D SPT entanglers correspond to gauge invariant quantities. For entanglers of in-cohomology SPTs, these gauge invariant quantities should completely specify $\omega(g,h,k)\in H^3(G,U(1))$. For a general cocycle, such a set of gauge invariant quantities may be complicated. However, there is a rather simple set of such gauge invariant quantities when $G$ is a discrete, abelian, unitary, on-site symmetry, i.e. a product of cyclic groups: $G=\prod_{i=1}^M\mathbb{Z}_{N_i}$. When $G$ is of this form, all of the anomalies are encoded in the generators of the cyclic groups. 

Let us denote the generators of $G$ by $g_1,g_2,\cdots g_{M}$. There are three different kinds of invariants $e^{i\theta_{g_i}},e^{i\theta_{g_i,g_j}},$ and $e^{i\theta_{g_i,g_j,g_k}}$ associated with these generators, that completely specify $\omega(g,h,k)\in H^3(G,U(1))$. These three kinds of invariants specify anomalies of type I, type II, and type III cocycles respectively\cite{topologicalinvariants,zaleteldetect, wang2015, wang2015spt,natinvariants}. Since SPTs described by type III cocycles are just particular examples of decorated domain wall SPTs, the topological invariant for these kinds of entanglers can be identified with Eq.~(\ref{decdomaininv}): $e^{i\theta_{g_i,g_j,g_k}}=c(g_i,g_j;g_k)$. The three invariants describe three different obstructions to making $\{W_{A,g}^U\}\sim\{W_{A,g}^{\mathbbm{1}}\}$:
\begin{itemize}
\item{$e^{i\theta_{g_i}}$ is an obstruction to making the restricted flux insertion operator $W_{\mathbf{A},g_i}^U$ on an open interval an FDQC of order $N_i$.}
\item{$e^{i\theta_{g_i,g_j}}$ is an obstruction to making $W_{\mathbf{A},g_i}^U$ a $\mathbb{Z}_{N_j}$ symmetric FDQC of order $N_{ij}$, where $N_{ij}$ is the least common multiple of $N_i$ and $N_j$.}
\item{$e^{i\theta_{g_i,g_j,g_k}}$ is an obstruction to making $W_{A,g_k}^U$ a $\mathbb{Z}_{N_i}\times\mathbb{Z}_{N_j}$ symmetric FDQC. This is a special case of the decorated domain wall invariants described in Sec.~\ref{sdecdomain}.}
\end{itemize}

We will now derive Eqs.~(\ref{thetag1def}) and (\ref{thetagigj}), which compute $e^{i\theta_{g_i}}$ and $e^{i\theta_{g_i,g_j}}$ from the SPT entangler. Since $e^{i\theta_{g_i,g_j,g_k}}$ is a decorated domain wall invariant, we will not repeat the derivation here.

Notice that a non-anomalous representation $\{W_{A,g_i}^{\mathbbm{1}}\}$ would not have any of the above obstructions. In this section, we will derive the invariants from the perspective of the above obstructions. We will directly connect these invariants to gauge invariant combinations of cocycles in Appendix~\ref{sproofs2d}.

%%%%%%%%%%%%%%%%%%%%%%%%%%%%%%%%%%%%%%%%%%%%%%%%%%%
\subsection{Type I invariant: $e^{i\theta_{g_i}}$}\label{sthetag}
%%%%%%%%%%%%%%%%%%%%%%%%%%%%%%%%%%%%%%%%%%%%%%%%%%%
The type I invariant, $e^{i\theta_{g_i}}$, involves only a single cyclic group, generated by $g_i$. A prototypical example of an SPT labeled a nontrivial $e^{i\theta_{g_i}}$ is the Levin-Gu $\mathbb{Z}_2$ SPT\cite{levingu}. We will discuss this example in more detail in Sec.~\ref{s2Dexamples}. Note that this invariant is closely related to results from Ref.~\onlinecite{symmedge}, as evident in Appendix~\ref{sfluxfusion}.

Since $\{W_{A,g}^U\}$ forms a representation of the cyclic group $\mathbb{Z}_{N_i}$, $\left(W_{A,g_i}^U\right)^{N_i}=\mathbbm{1}$. This means that $W_{A,g_i}^U$ is a 1D FDQC of order $N_i$. Although $W_{A,g_i}^U$ has order $N_i$, the restricted flux insertion operator $W_{\mathbf{A},g_i}^U$ on an open 1D interval $\mathbf{A}$ does not necessarily have order $N_i$. In general, $W_{\mathbf{A},g_i}^U$ satisfies
\begin{equation}\label{truncW}
\left(W_{\mathbf{A},g_i}^U\right)^{N_i}=L_{\mathbf{A},g_i}^{U}\otimes R_{\mathbf{A},g_i}^{U},
\end{equation}
where $L_{\mathbf{A},g_i}^U$ and $R_{\mathbf{A},g_i}^U$ are local operators supported near the left and right endpoints of $\mathbf{A}$ respectively (note that these operators are \emph{not} related to $L_{A,g}^U$ and $R_{A,g}^U$ discussed in Sec.~\ref{s1d}). While $W_{A,g_i}^U$ is a 1D representation of $\mathbb{Z}_{N_i}$, it may not be possible to make the restricted operator $W_{\mathbf{A},g_i}^U$ into a representation of $\mathbb{Z}_{N_i}$, satisfying $\left(W_{\mathbf{A},g_i}^U\right)^{N_i}=\mathbbm{1}$, for any choice of restriction of $W_{A,g_i}^U$. In particular, there do not exist any local unitary operators $O_L$ and $O_R$ near the left and right endpoints of $\mathbf{A}$ such that $\left(W_{\mathbf{A},g_i}^UO_LO_R\right)^{N_1}=\mathbbm{1}$. 

This presents an obstruction to $W_{A,g_i}^U$ being equivalent to the trivial representation. Specifically, we cannot have $W_{A,g_i}^U=V^\dagger W_{A,g_i}^{\mathbbm{1}}V$ where $V$ is a $G$ symmetric FDQC, because $V^\dagger W_{A,g_i}^{\mathbbm{1}}V^\dagger$ can always be truncated as $V_{A}W_{\mathbf{A},g_i}^{\mathbbm{1}}V_A^\dagger$. This satisfies $\left(V_{A}W_{\mathbf{A},g_i}^{\mathbbm{1}}V_A^\dagger\right)^{N_i}=\mathbbm{1}$ because $\left(W_{\mathbf{A},g_i}^{\mathbbm{1}}\right)^{N_i}=\mathbbm{1}$. 

The obstruction to making $W_{\mathbf{A},g_i}^{U}$ a representation of $\mathbb{Z}_{N_i}$ is encoded in $e^{i\theta_{g_i}}$, which is given by
\begin{equation}\label{thetagiR}
e^{i\theta_{g_i}}=\overline{\mathrm{Tr}}\left(W_{\mathbf{A},g_i}^{U\dagger}R_{\mathbf{A},g_i}^UW_{\mathbf{A},g_i}^UR_{\mathbf{A},g_i}^{U\dagger}\right)
\end{equation}

An alternative way to write $e^{i\theta_{g_i}}$, without restricting $\left(W_{\mathbf{A},g_i}^U\right)^{N_i}$ to $R_{\mathbf{A},g_i}^U$ is given by
\begin{equation}\label{thetag1def}
e^{i\theta_{g_i}}=\overline{\mathrm{Tr}}\left[W_{\mathbf{A}_R,g_i}^{U\dagger}\left(W_{\mathbf{A},g_i}^U\right)^{N_i}W_{\mathbf{A}_R,g_i}^U\left(W_{\mathbf{A},g_i}^{U\dagger}\right)^{N_1}\right]
\end{equation}
where $\mathbf{A}_R\subset\mathbf{A}$ is the right half of $\mathbf{A}$, as shown in Fig.~\ref{fig:ABC}.a. It is important that $W_{\mathbf{A}_R,g_i}^U$ has the same truncation as $W_{\mathbf{A},g_i}^U$ near the right endpoint of $\mathbf{A}_R$, and that $\mathbf{A}_R$ does not contain the support of $L_{\mathbf{A},g_i}^U$. The latter point ensures that $W_{\mathbf{A}_R,g_i}^U$ commutes with $L_{\mathbf{A},g_i}^U$.

We now confirm that $e^{i\theta_{g_i}}$ defined above is invariant under modification of $U$ by any local, symmetric unitary. Since $R_{\mathbf{A},g_i}^U$ is only sensitive to modifications of $U$ near the right endpoint of $\mathbf{A}$, we only need to check that $e^{i\theta_{g_i}}$ is invariant under modification of $U$ near the right endpoint of $\mathbf{A}$. %However, $W_{\mathbf{A},g_i}^U$ must commute with $L_{\mathbf{A},g_i}^U\otimes R_{\mathbf{A},g_i}^U$ (because it commutes with itself), and since a modification of $U$ near the right endpoint of $\mathbf{A}$ would not affect $L_{\mathbf{A},g_i}^U$, $e^{i\theta_{g_i}}$ as defined in Eq.~(\ref{thetagiR}) cannot be affected by such a local modification. 
%
%More precisely Local, symmetric unitaries fully supported deep in $A$ or $\overline{A}$ leave $W_{A,g_i}^U$ and therefore $W_{\mathbf{A},g_i}^{U}$ invariant. Local, symmetric unitaries along the interval $\mathbf{A}$ do not affect $R_{\mathbf{A},g_i}^U$ or its commutator with $W_{\mathbf{A},g_i}^U$. In particular, they modify $W_{\mathbf{A},g_i}^U\to U_r^\dagger W_{\mathbf{A},g_i}^UU_r$, and 
%\begin{equation}
%\left(W_{\mathbf{A},g_i}^U\right)^{N_i}=\left(U_r^\dagger W_{\mathbf{A},g_i}^UU_r\right)^{N_i}
%\end{equation}
%if $U_r$ is supported far from the endpoints of $\mathbf{A}$. Finally, local unitaries at the right endpoint of $\mathbf{A}$ cannot change the commutator of $R_{\mathbf{A},g_i}^U$ with $W_{\mathbf{A},g_i}^{U}$ because they do not affect $L_{\mathbf{A},g_i}^U$. 
This fact comes from the observation that the commutator of $W_{\mathbf{A},g_i}^U$ and $L_{\mathbf{A},g_i}^U\otimes R_{\mathbf{A},g_i}^U$ must vanish, because $W_{\mathbf{A},g_i}^U$ commutes with itself. This means that
\begin{align}
\begin{split}
\overline{\mathrm{Tr}}&\left(W_{\mathbf{A},g_i}^{U\dagger}R_{\mathbf{A},g_i}^UW_{\mathbf{A},g_i}^UR_{\mathbf{A},g_i}^{U\dagger}\right)\\
&=\overline{\mathrm{Tr}}\left(W_{\mathbf{A},g_i}^{U\dagger}L_{\mathbf{A},g_i}^{U}W_{\mathbf{A},g_i}^UL_{\mathbf{A},g_i}^{U\dagger}\right)^*.
\end{split}
\end{align}

Due to this constraint, $e^{i\theta_{g_i}}$ is insensitive to modifications of $U$ by local, symmetric unitaries fully supported near only the left endpoint or the right endpoint of $\mathbf{A}$. 

%%%%%%%%%%%%%%%%%%%%%%%%%%%%%%%%%%%%%%%%%%%%%%%%%%%
\subsection{Type II invariant: $e^{i\theta_{g_i,g_j}}$}\label{sthetaij}
%%%%%%%%%%%%%%%%%%%%%%%%%%%%%%%%%%%%%%%%%%%%%%%%%%%
The second invariant, $e^{i\theta_{g_i,g_j}}$ involves two cyclic groups $\mathbb{Z}_{N_i}$ and $\mathbb{Z}_{N_j}$, with generators $g_i$ and $g_j$ respectively.\footnote{We can also compute $e^{i\theta_{g_i,g_i}}$, but this is equal to $e^{2i\theta_{g_i}}$, so it is not an independent invariant.} It detects when the symmetries have a mixed anomaly. Physically, this means that in the edge theory, domain walls of one symmetry carry fractional charge of the other symmetry and vice versa\cite{wang2015spt}. In the 2D bulk, flux of one symmetry binds fractional charge of the other symmetry and vice versa\cite{zaleteldetect,wang2015}. 

$W_{A,g_i}^U$ again gives a representation of $\mathbb{Z}_{N_i}$, and hence is a FDQC of order $N_i$. However, now $W_{A,g_i}^U$ has an additional constraint. Since $g_i$ and $g_j$ commute, $W_{A,g_i}^U$ must commute with the global symmetry operator for $g_j$ according to Property 3: 
\begin{equation}
[W_{A,g_i}^U,U_{g_j}]=0.
\end{equation}

In fact, as we show in Appendix~\ref{spropinvariants}, $W_{A,g_i}^U$ can always be written as a $\mathbb{Z}_{N_i}\times\mathbb{Z}_{N_j}$ symmetric FDQC. Therefore, $W_{A,g_i}^U$ is a 1D $\mathbb{Z}_{N_i}\times\mathbb{Z}_{N_j}$ symmetric FDQC of order $N_i$ on a closed loop.

Again, we detect the anomaly using the restricted flux insertion operator $W_{\mathbf{A},g_i}^U$. In order to apply our invariant, we require that this restriction satisfies $[W_{\mathbf{A},g_i}^U,U_{g_j}]=0$. This symmetric restriction is always possible because, as mentioned in the previous paragraph, $W_{A,g_i}^U$ can always be written as a $\mathbb{Z}_{N_i}\times\mathbb{Z}_{N_j}$ symmetric FDQC. Naively, we can consider the operator $\left(W_{\mathbf{A},g_i}^U\right)^{N_i}$ as we did in the last section, which would be a product of local operators at the left and right endpoints of $\mathbf{A}$. While $\left(W_{\mathbf{A},g_i}^U\right)^{N_i}$ commutes with $U_{g_j}$, the operators on the left and right endpoints of $\mathbf{A}$ may not individually commute with $U_{g_j}$. 

However, there is an important subtlety in that the restriction is ambiguous in the following way: we can choose a different restriction that differs from $W_{\mathbf{A},g_i}^U$ by opposite charges under $U_{g_i}$ or $U_{g_j}$ at the endpoints of $\mathbf{A}$, and this new restricted flux insertion operator would still commute with $U_{g_j}$. More precisely,
\begin{equation}
W_{\mathbf{A},g_i}^U\sim W_{\mathbf{A},g_i}^UO_lO_r,
\end{equation}
where 
\begin{align}
\begin{split}
U_{g_i}^\dagger O_rU_{g_i}&=O_re^{2\pi in_i/N_i}\\
U_{g_j}^\dagger O_rU_{g_j}&=O_re^{2\pi in_j/N_j},
\end{split}
\end{align}
where $n_i,n_j\in\mathbb{Z}$. As long as $U_{g_i}$ and $U_{g_j}$ commute with the product $O_lO_r$, $W_{\mathbf{A},g_i}^U$ and $W_{\mathbf{A},g_i}^UO_lO_r$ are both equally valid restrictions. 

Alternatively, we can think of $O_lO_r$ as arising from truncating an equivalent $W_{A,g_i}^U$, of the form
\begin{equation}
W_{A,g_i}^{U'}=V^\dagger W_{A,g_i}^UV=W_{A,g_i}^UU_{n_i,n_j},
\end{equation}
where $U_{n_i,n_j}$ is a 1D FDQC (or more specifically, a 1D Floquet unitary\cite{cohomology}) that, when restricted to $\mathbf{A}$, pumps $n_i$ units of $\mathbb{Z}_{N_i}$ charge and $n_j$ units of $\mathbb{Z}_{N_j}$ charge (which are 0D SPTs) to the right endpoint of $\mathbf{A}$ and opposite charge to the left endpoint of $\mathbf{A}$. 

To remove this ambiguity, we instead consider the operator $\left(W_{\mathbf{A},g_i}^{U}\right)^{N_{ij}}$, where $N_{ij}$ is the least common multiple of $N_i$ and $N_j$. This operator is insensitive to the choice of restriction, because charges added to the endpoints of $W_{\mathbf{A},g_i}^U$ from a particular choice of restriction are neutralized upon taking $W_{\mathbf{A},g_i}^U$ to the power of $N_{ij}$. Physically, while the restriction causes an ambiguity of ``integer" charge, $e^{i\theta_{g_i,g_j}}$ measures ``fractional" charge attached to the endpoints of $W_{\mathbf{A},g_i}^U$. 

In general, $\left(W_{\mathbf{A},g_i}^{U}\right)^{N_{ij}}$ is a product of operators near the left and right endpoints of $\mathbf{A}$:
\begin{equation}\label{WNij}
\left(W_{\mathbf{A},g_i}^{U}\right)^{N_{ij}}=\left(L_{\mathbf{A},g_i}^U\right)^{N_{ij}/N_i}\otimes \left(R_{\mathbf{A},g_i}^U\right)^{N_{ij}/N_i},
\end{equation}
where $L_{\mathbf{A},g_i}^U$ and $R_{\mathbf{A},g_i}^U$ are defined in (\ref{truncW}). While $\left(L_{\mathbf{A},g_i}^U\right)^{N_{ij}/N_i}\otimes \left(R_{\mathbf{A},g_i}^U\right)^{N_{ij}/N_i}$ commutes with $U_{g_j}$ because $W_{\mathbf{A},g_i}^U$ is a $\mathbb{Z}_{N_j}$ symmetric FDQC, each operator individually may not commute with $U_{g_j}$, for any choice of restriction. In other words, there does not exist any local, symmetric unitary operators $O_L$ and $O_R$ near the left and right endpoints of $\mathbf{A}$ such that $\left(W_{\mathbf{A},g_i}^{U}O_LO_R\right)^{N_{ij}}$ is a product of two symmetric operators on the left and right endpoints of $\mathbf{A}$. Again, one can check that this presents an obstruction to $W_{A,g_i}^U$ being of the form $V^\dagger W_{A,g_i}^{\mathbbm{1}}V$, where $V$ is a $G$ symmetric FDQC supported near the boundary of $A$. The obstruction is given explicitly by
\begin{equation}\label{thetagigj}
e^{i\theta_{g_i,g_j}}=\overline{\mathrm{Tr}}\left[U_{B,g_j}^{\dagger}\left(W_{\mathbf{A},g_i}^{U\dagger}\right)^{N_{ij}}U_{B,g_j}\left(W_{\mathbf{A},g_i}^{U}\right)^{N_{ij}}\right],
\end{equation}
where $B$ is a disk that encloses the support of $R_{\mathbf{A},g_i}^U$ but not the support of $L_{\mathbf{A},g_i}$, as shown in Fig.~\ref{fig:ABC}.b. 

To check that $e^{i\theta_{g_i,g_j}}$ given in (\ref{thetagigj}) is invariant under modification of $U$ by local, symmetric (under $\mathbb{Z}_{N_i}\times\mathbb{Z}_{N_j}$) unitaries anywhere in $\Lambda$, we can use a similar argument as in the previous section. First, $R_{\mathbf{A},g_i}^U$ is again only sensitive to modification of $U$ by local unitaries near the right endpoint of $\mathbf{A}$. Furthermore, since $U_{g_j}$ commutes with $\left(L_{\mathbf{A},g_i}^U\right)^{N_{ij}/N_i}\otimes \left(R_{\mathbf{A},g_i}^U\right)^{N_{ij}/N_i}$, $e^{i\theta_{g_i,g_j}}$ cannot be affected by modifications of $U$ near the right endpoint of $\mathbf{A}$, which are far away from the support of $L_{\mathbf{A},g_i}^U$.%. It is invariant under local symmetric unitaries in $A$ or $\overline{A}$ because these do not change $W_{\mathbf{A},g_i}^{U}$. It is invariant under local symmetric unitaries along the interval $\mathbf{A}$ because these do not affect $\left(W_{\mathbf{A},g_i}^{U}\right)^{N_{ij}}$, which acts as identity along the bulk of the interval $\mathbf{A}$. Finally, it is invariant under local symmetric unitaries at the endpoints of $\mathbf{A}$ because, as long as we choose a restriction $W_{\mathbf{A},g_i}^U$ that commutes with the global symmetry operator $U_{g_j}$, such unitaries can only modify the endpoints of $W_{\mathbf{A},g_i}^U$ by $\mathbb{Z}_{N_i}$ and $\mathbb{Z}_{N_j}$ charges, as discussed earlier in this section. These charges are removed because we take $W_{\mathbf{A},g_i}^U$ to the $N_{ij}$th power. This confirms that $e^{i\theta_{g_i,g_j}}$ as defined in (\ref{thetagigj}) is indeed an invariant of $G$ symmetric SPT entanglers. 

%$e^{i\theta_{g_i,g_j}}$ has the following physical interpretations. On the boundary, it measures the fractional charge of one symmetry attached to the domain walls of the other symmetry. In the bulk, it measures fractional charge of one symmetry bound to flux of the other symmetry. It can also be understood from a bulk domain wall decoration perspective: $e^{i\theta_{g_i,g_j}}$ detects 0D SPTs decorating junctions between bulk domain walls. In this sense, we can actually interpret $\left(W_{A,g_i}^U\right)^{N_{ij}}=\mathbbm{1}$ as a 1D Floquet system on a closed manifold. On an open manifold, $\left(W_{\mathbf{A},g_i}^U\right)^{N_{ij}}$ is described by a boundary unitary as written on the right hand side of (\ref{WNij}). A 1D $\mathbb{Z}_{N_j}$ symmetric Floquet system can pump 0D $\mathbb{Z}_{N_j}$ SPTs (charges) to a physical boundary (endpoints of $\mathbf{A}$) every period. $e^{i\theta_{g_i,g_j}}$ can be interpreted as an edge invariant for this 1D Floquet system.

In Appendix~\ref{sproofs2d}, we review how to show that $e^{2i\theta_{g_i}}=e^{i\theta_{g_i,g_i}}$. This relation allows us to compute $e^{2i\theta_{g_i}}$ in a somewhat more closed-form way than in Eq.~(\ref{thetag1def}). According to Eq.~(\ref{thetagigj}),
\begin{equation}\label{thetagigi}
e^{2i\theta_{g_i}}=\overline{\mathrm{Tr}}\left[U_{B,g_i}^\dagger \left(W_{\mathbf{A},g_i}^{U\dagger}\right)^{N_i}U_{B,g_i}\left(W_{\mathbf{A},g_i}^{U}\right)^{N_i}\right]
\end{equation}
where, as in Eq.~(\ref{thetagigj}), $B$ is a disk that contains the full support of $R_{\mathbf{A},g_i}^U$ but not the support of $L_{\mathbf{A},g_i}^U$. Notice that (\ref{thetagigi}) \emph{completely} specifies $e^{i\theta_{g_i}}$ when $N_i$ is odd, but does not completely specify $e^{i\theta_{g_i}}$ when $N_i$ is even. In particular, it does not distinguish between $e^{i\theta_{g_i}}=\pm 1$.

\subsection{Example of a 2D SPT entangler with $\mathbb{Z}_2$ symmetry}\label{s2Dexamples}
%%%%%%%%%%%%%%%%%%%%%%%%%%%%%%%%%%%%%%%%%%%%%%%%%%%%%%%%%%%%%%%%%%%%%%%%%%%%%%%%%%%%
\begin{figure}[tb]
   \centering
   \includegraphics[width=.9\columnwidth]{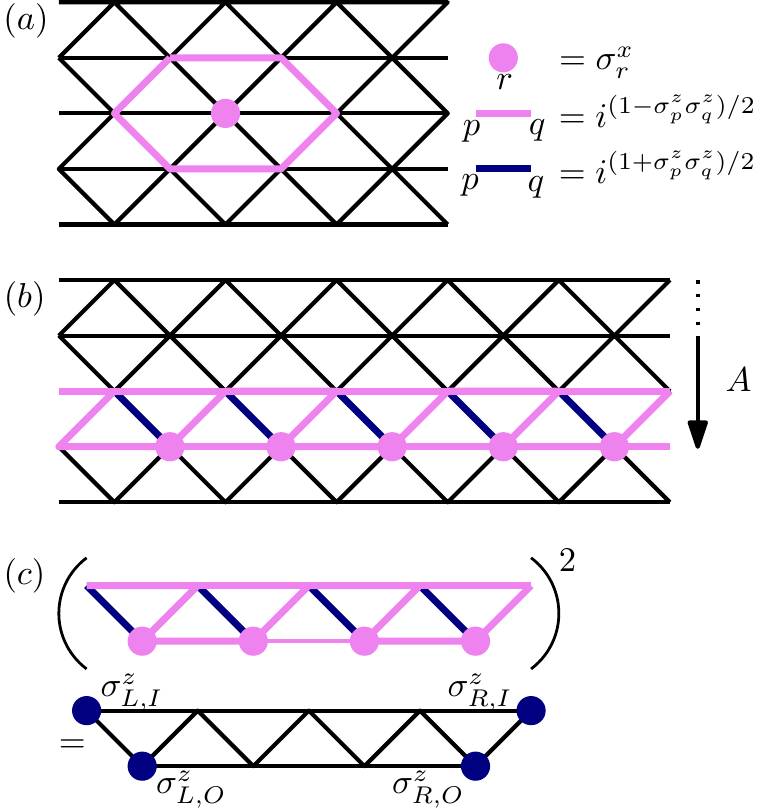} % requires the graphicx package
   \caption{$(a)$ The SPT entangler $U$ for the Levin-Gu SPT transforms $\sigma^x_r$ for a single site into a plaquette operator $B_p$ supported on 7 sites. $(b)$ The operator $W_{A,g_1}^U$ is a product of $\sigma^x_r$ operators and link operators, as illustrated. $A$ is the upper region extending down to the bottom pink links and $A_{\mathrm{in}}$ is the upper region extending down to the top pink links. $(c)$ Truncating $W_{A,g_1}^U$ to $W_{\mathbf{A},g_1}^U$ and squaring it gives a product of four $\sigma^z$ operators as indicated.}
   \label{fig:z2example}
\end{figure}
In this section, we present a more in-depth discussion of an example of a 2D SPT entangler with $\mathbb{Z}_2$ symmetry. This SPT entangler has a nontrivial $e^{i\theta_{g_i}}$, and entangles the Levin-Gu SPT\cite{levingu}. 

We consider a 2D triangular lattice, with a spin-1/2 on each vertex $r$. The original on-site symmetry is given by
\begin{equation}
U_{g_1}=\prod_{r}\sigma^x_r.
\end{equation}

An example of a symmetric, gapped Hamiltonian with a symmetric ground state is given, as in Sec.~\ref{sexample}, by $H_0=-\sum_r\sigma^x_r$. The ground state of $H_0$ is simply a product state, with each spin-1/2 in the $+1$ eigenstate of $\sigma^x_r$. A $\mathbb{Z}_2$ symmetric SPT entangler is given by
\begin{equation}
U=\prod_{\langle pqr\rangle}e^{\frac{i\pi}{24}\left(3\sigma^z_p\sigma^z_q\sigma^z_r-\sigma^z_p-\sigma^z_q-\sigma^z_r\right)},
\end{equation}
where the product runs over all triangles $\langle pqr\rangle$. One can check that
\begin{equation}
U^\dagger\sigma^x_rU=-\sum_pB_p,
\end{equation}
where 
\begin{equation}
B_p=-\sigma^x_r\prod_{\langle pq\rangle\in\hexagon_r}i^{(1-\sigma^z_p\sigma^z_q)/2},
\end{equation}
where $p,q$ are neighboring sites on the 6 links of the hexagon surrounding $r$, as illustrated in Fig.~\ref{fig:z2example}.a. This is precisely the Hamiltonian in Ref.~\onlinecite{levingu} for the Levin-Gu SPT.

We will now evaluate $e^{i\theta_{g_i}}$ for this SPT entangler. The first step is to compute $W_{A,g_1}^U$. 
%\begin{equation}
%U \sigma^x_rU^\dagger=-\sigma^x_r\prod_{\langle pq\rangle\in\hexagon_r}(-i)^{(1-\sigma^z_p\sigma^z_q)/2},
%\end{equation}
The action of $U$ on $U_{A_{\mathrm{in}},g_1}$ is given by
\begin{align}
\begin{split}\label{conjg}
UU_{A_{\mathrm{in}},g_1}^\dagger U^\dagger&=\prod_{r\in A_{\mathrm{in}}}U\sigma^x_rU^\dagger\\
&=(-1)^{|A_{\mathrm{in}}|}\prod_{r\in A_{\mathrm{in}}}\left[\sigma^x_r\prod_{\langle pq\rangle\in\hexagon_r}(-i)^{(1-\sigma^z_p\sigma^z_q)/2}\right].
\end{split}
\end{align}

In the bulk of $A_{\mathrm{in}}$, all links have two factors of $(-i)^{(1-\sigma^z_p\sigma^z_q)/2}$, which multiply to $\sigma^z_p\sigma^z_q$. This means that all vertices have a factor of $\left(\sigma^z_p\right)^6=\mathbbm{1}$, so as expected, $U$ leaves $U_{A_{\mathrm{in}},g_1}$ invariant deep in the bulk of $A_{\mathrm{in}}$. Let us bring all the $\sigma^x_r$ operators to the left in Eq.~(\ref{conjg}). For $r$ near the boundary of $A_{\mathrm{in}}$, we need to be careful about commuting the $\sigma^x_r$ operators through the $(-i)^{(1-\sigma^z_p\sigma^z_q)/2}$ operators. Let us assume that $|A_{\mathrm{in}}|$ is even. Then
\begin{align}
\begin{split}\label{UZ2}
&UU_{A_{\mathrm{in}},g_1}^\dagger U^\dagger\\
&=\prod_{r\in A_{\mathrm{in}}}\sigma^x_r\prod_{\langle pq\rangle\in \mathrm{pink}}(-i)^{(1-\sigma^z_p\sigma^z_q)/2}\prod_{\langle pq\rangle\in\mathrm{blue}}(-i)^{(1+\sigma^z_p\sigma^z_q)/2},
\end{split}
\end{align}
where the pink links and blue links lie in $\partial A$ and are indicated in Fig.~\ref{fig:z2example}.b. Specifically, the pink links include horizontal links and diagonal links of orientation ``$\slash$", while the blue links are diagonal links of orientation ``$\backslash$". We now easily obtain
\begin{align}
\begin{split}
&U_{A,g_1}UU_{A_{\mathrm{in}},g_1}^\dagger U^\dagger\\
&=\prod_{r\in A\setminus A_{\mathrm{in}}}\sigma^x_r\prod_{\langle pq\rangle\in \mathrm{pink}}(-i)^{(1-\sigma^z_p\sigma^z_q)/2}\prod_{\langle pq\rangle\in\mathrm{blue}}(-i)^{(1+\sigma^z_p\sigma^z_q)/2}.
\end{split}
\end{align}

We can choose the restriction illustrated in Fig.~\ref{fig:z2example}.c. It is easy to check that
\begin{equation}
\left(W_{A,g_1}^U\right)^2=\sigma^z_{L,I}\sigma^z_{L,O}\sigma^z_{R,I}\sigma^z_{R,O},
\end{equation}
where $\sigma^z_{L,I},\sigma^z_{L,O},\sigma^z_{R,I},$ and $\sigma^z_{R,O}$ are indicated in Fig.~\ref{fig:z2example}.c. It follows that $R_{\mathbf{A},g_1}^U=\sigma^z_{R,I}\sigma^z_{R,O}$ where $\sigma^z_{R,I}$ acts inside $A_{\mathrm{in}}$ and $\sigma^z_{R,O}$ acts outside $A_{\mathrm{in}}$. $R_{\mathbf{A},g_1}^U$ anticommutes with $W_{\mathbf{A},g_1}^U$, so from (\ref{thetag1def}) $e^{i\theta_{g_1}}=-1$. While we evaluated $e^{i\theta_{g_1}}$ using a particular restriction, the same answer holds for \emph{any} restriction.
%%%%%%%%%%%%%%%%%%%%%%%%%%%%%%%%%%%%%%%%%%%%%%%%%%%%%%%%%%%%%%%%%%%%%%%%%%%%%%%%%%%%
\subsection{Higher dimensional SPT entanglers with 2D decorated domain walls}\label{s3D}
%%%%%%%%%%%%%%%%%%%%%%%%%%%%%%%%%%%%%%%%%%%%%%%%%%%%%%%%%%%%%%%%%%%%%%%%%%%%%%%%%%%%
%Like for 2D SPTs, the complete set of gauge invariant quantities defining a 3D bosonic SPT for a general group $G$ is not known. However, it is known when $G$ is again a product of cyclic groups. In this section, we will show how to compute these quantities from the SPT entangler.
Like in Sec.~\ref{sdecdomain}, we can leverage our invariants for 2D SPT entanglers to obtain invariants for higher dimensional SPT entanglers that entangle SPTs with decorated domain walls. In fact, \emph{all} 3D SPT entanglers with discrete, abelian, on-site symmetries are of this form\cite{propitius,topologicalinvariants}, so we can completely classify 3D SPT entanglers with such symmetries using our 2D invariants. 

Specifically, SPT phases in 3D are classified by $\omega(g,h,k,l)\in H^4(G,U(1))$. For discrete, abelian, on-site symmetries, of the form $\prod_i\mathbb{Z}_{N_i}$, the gauge invariant quantities defining $\omega(g,h,k,l)$ was given in Ref.~\onlinecite{topologicalinvariants}. Like in 2D, there are three kinds of gauge invariant quantities: $e^{i\theta_{g_i;g_l}}$, $e^{i\theta_{g_i,g_j;g_l}}$, and $e^{i\theta_{g_i,g_j,g_k;g_l}}$, where $g_i,g_j,g_k,$ and $g_l$ are generators of different cyclic groups. These invariants are related to decorating domain walls of the $\mathbb{Z}_{N_{l}}$ symmetry with 2D SPTs with $\mathbb{Z}_{N_i}$ symmetry, $\mathbb{Z}_{N_i}\times\mathbb{Z}_{N_j}$ symmetry, and $\mathbb{Z}_{N_i}\times\mathbb{Z}_{N_j}\times\mathbb{Z}_{N_k}$ symmetry respectively\cite{topologicalinvariants}.

This means that we can obtain formulas for $e^{i\theta_{g_i;g_l}}$, $e^{i\theta_{g_i,g_j;g_l}}$, and $e^{i\theta_{g_i,g_j,g_k;g_l}}$ by simply replacing $U$ in the formulas for $e^{i\theta_{g_i}},e^{i\theta_{g_i,g_j}},$ and $e^{i\theta_{g_i,g_j,g_k}}$ by a flux insertion operator $W_{D,g_l}^U$, as we did in Sec.~\ref{sdecdomain} for 2D decorated domain wall SPT entanglers. 

We must also specify the geometry of the regions. In 3D, $A,B,C,$ and $D$ are overlapping balls, which we choose to be centered at the corners of a tetrahedron. This geometry can be thought of as extending three overlapping disks in Fig.~\ref{fig:ABC}.(c) into 3D overlapping balls. There are two points at which the boundaries of $A,B,$ and $C$ all intersect. We then add a fourth ball $D$ that overlaps with all three balls and contains only one of these two points. The closed flux insertion operator $W_{D,g_l}^U$ is supported near the closed 2D surface of $D$. 
 
These three 3D invariants are also related to different obstructions to making $W_{D,g_l}^U\sim W_{D,g_l}^{\mathbbm{1}}$. For example, $e^{i\theta_{g_i;g_l}}$ is an obstruction to making $W_{D,g_l}^U$ a $\mathbb{Z}_{N_i}$ symmetric FDQC of order $N_l$. It is not hard to check that all three are insensitive to modifications of $U$ by any symmetric, local unitaries, and therefore are invariant under composition of $U$ by any $G$ symmetric FDQC.
\section{Fermionic systems}\label{sfermionic}
%%%%%%%%%%%%%%%%%%%%%%%%%%%%%%%%%%%%%%%%%%%%%%%%%%%%%%%%%%%%%%%%%%%%%%%%%%%%%%%%%%%%%
We expect that it would not be difficult to generalize our framework to obtain topological invariants for broad classes of fermionic SPT entanglers. Here we will present some results about two particularly interesting fermionic systems: the Kitaev wire and the generator of the 2D $\mathbb{Z}_2\times\mathbb{Z}_2^f$ SPT. The latter phase is characterized by the property that the $\mathbb{Z}_2$ domain walls are decorated by Kitaev wires. 

First, it is well-known that the Kitaev wire is entangled by a QCA rather than an FDQC\cite{huangcomplexity,fermionic}. Since our definition of $W_{A,g}^U$ in Eq.~(\ref{WAgU}) does not require us to truncate the entangler, we can also compute $\{W_{A,g}^U\}$ when $U$ is a QCA, and use it to obtain a $\mathbb{Z}_2$-valued topological invariant for the Kitaev wire entangler. Second, a symmetric FDQC that entangles the generator of the 2D $\mathbb{Z}_2\times\mathbb{Z}_2^f$ SPT has not yet been found\cite{tarantino2016,nat2018}. Even if this phase cannot be entangled by an FDQC, it might be entangled by a symmetric QCA. We will show that this phase actually cannot be entangled by any symmetric QCA.

%%%%%%%%%%%%%%%%%%%%%%%%%%%%%%%%%%%%%%%%%%%%%%%%%%%%%%%%%%%%%%%%%%%%%%%%%%%%%%%%%%%%
\subsection{SPT entangler for the Kitaev wire}\label{sfermionic1d}
%%%%%%%%%%%%%%%%%%%%%%%%%%%%%%%%%%%%%%%%%%%%%%%%%%%%%%%%%%%%%%%%%%%%%%%%%%%%%%%%%%%%
In this section, we present a formula that gives a $\mathbb{Z}_2$ index $\zeta$ that completely classifies 1D fermionic QCA with no other symmetry besides fermion parity (modulo bosonic translations). Nontrivial QCA of this kind entangle the Kitaev wire\cite{fermionic}.

First, we note that for systems that include fermionic degrees of freedom, $U$ must conserve fermion parity in order to maintain locality. This means that $U$ commutes with the total fermion parity operator $\Gamma=\prod_{r\in\Lambda}\Gamma_r$. $\Gamma_r$ is the fermion parity operator of site $r$, with eigenvalues $\pm1$ describing the fermion parity of the states in the local Hilbert space on $r$. $\Gamma$ can be restricted to $\Gamma_A$, which measures the fermion parity in region $A$.

$U$ is classified by $W_{A,\Gamma}^U$, which is defined in the usual way:
\begin{equation}\label{wgamma}
W_{A,\Gamma}^U=\Gamma_AU\Gamma_{A_{\mathrm{in}}}U^\dagger=L_{A,\Gamma}^U\otimes R_{A,\Gamma}^U,
\end{equation}
where we used $\Gamma_{A_{\mathrm{in}}}^\dagger =\Gamma_{A_{\mathrm{in}}}$. While $L_{A,\Gamma}^U\otimes R_{A,\Gamma}^U$ is fermion parity even, $L_{A,\Gamma}^U$ and $R_{A,\Gamma}^U$ may not individually be fermion parity even. When $L_{A,\Gamma}^U$ and $R_{A,\Gamma}^U$ are fermion parity odd, $W_{A,\Gamma}^U$ is not equivalent to $W_{A,\Gamma}^{\mathbbm{1}}$. This means that $U$ is a nontrivial fermionic QCA and $\eta=-1$. Therefore, to obtain a formula for $\eta$, we simply compute the fermion parity of $R_{A,\Gamma}^U$. To do this without restricting $W_{A,\Gamma}^U$ to $R_{A,\Gamma}^U$, we use $\Gamma_B$, which includes the full support of $R_{A,\Gamma}^U$ but does not contain the support of $L_{A,\Gamma}^U$:
\begin{equation}\label{etagrgr}
\eta=\overline{\mathrm{Tr}}\left(\Gamma_B W_{A,\Gamma}^U\Gamma_B W_{A,\Gamma}^U\right),
\end{equation}
where we used the fact that $\Gamma_B$ and $W_{A,\Gamma}^U$ are both hermitian. Like in Sec.~\ref{s1d}, we can simplify Eq.~(\ref{etagrgr}) using the explicit expression for $W_{A,\Gamma}^U$ in (\ref{wgamma}). Further simplifying using the hermiticity of $\Gamma_A$ and $\Gamma_B$, we get
%\begin{align}
%\begin{split}
%\eta&=\overline{\mathrm{Tr}}\left[\Gamma_B \left(\Gamma_AU\Gamma_{A_{\mathrm{in}}}U^\dagger\right)\Gamma_B \left(\Gamma_AU\Gamma_{A_{\mathrm{in}}}U^\dagger\right)\right]\\
%&=\overline{\mathrm{Tr}}\left(\Gamma_B U\Gamma_{A_{\mathrm{in}}}U^\dagger\Gamma_B U\Gamma_{A_{\mathrm{in}}}U^\dagger\right)
%\end{split}
%\end{align}
%We can then relabel $A_{\mathrm{in}}$ as $A$ to obtain
\begin{equation}\label{flowfermions}
\zeta=\overline{\mathrm{Tr}}\left[\left(\Gamma_BU\Gamma_AU^\dagger\right)^2\right].
\end{equation}

It is easy to check that an FDQC composed of gates that are all fermion parity even gives $\zeta(U)=1$. On the other hand, the Kitaev wire entangler gives $\zeta=-1$. To check this, we can compute $\eta$ for the Majorana translation, which entangles the Kitaev wire. To define the Majorana translation, we consider a chain with a single spinless fermion on each site and we define each physical fermion in terms of two Majorana fermions in the usual way:
\begin{align}
\begin{split}
a_{n}&=\frac{1}{2}(\gamma_{2n-1}+i\gamma_{2n})\\ 
a_{n}^\dagger&=\frac{1}{2}(\gamma_{2n-1}-i\gamma_{2n}).
\end{split}
\end{align}

Let $A=[a_l,a_r]$ and $B=[b_l,b_r]$, with $a_l<b_l<a_r<b_r$. The fermion parity operators $\Gamma_A$ is given by
\begin{align}
\begin{split}
\Gamma_A&=(i\gamma_{2a_l-1}\gamma_{2a_l})(i\gamma_{2a_l+1}\gamma_{2a_l+2})\cdots(i\gamma_{2a_r-1}\gamma_{2a_r})\\
&=i^{a_r-a_l+1}\gamma_{2a_l-1}\cdots\gamma_{2a_r}.
\end{split}
\end{align}

$\Gamma_B$ is defined in a similar way. A Majorana translation taking $\gamma_r\to\gamma_{r+1}$ gives
\begin{equation}
U\Gamma_AU^\dagger=i^{a_r-a_l+1}\gamma_{2a_l}\cdots\gamma_{2a_r+1}.
\end{equation}

Evaluating $\Gamma_BU\Gamma_AU^\dagger$, we get
\begin{align}
\begin{split}\label{fermionex}
&\Gamma_BU\Gamma_AU^\dagger\\
&=i^{b_r-b_l+a_r-a_l+2}\left(\gamma_{2a_l}\cdots\gamma_{2b_l-2}\right)\cdot\left(\gamma_{2a_r+2}\cdots\gamma_{2b_r}\right).
\end{split}
\end{align}
The two hermitian factors on the right side of (\ref{fermionex}) are both fermion parity odd, so they anticommute. It follows that $\zeta=-1$ as expected.

%%%%%%%%%%%%%%%%%%%%%%%%%%%%%%%%%%%%%%%%%%%%%%%%%%%%%%%%%%%%%%%%%%%%%%%%%%%%%%%%%%%%
\subsection{No symmetric entangler for the 2D $\mathbb{Z}_2\times\mathbb{Z}_2^f$ SPT}\label{sfermionic2d}
%%%%%%%%%%%%%%%%%%%%%%%%%%%%%%%%%%%%%%%%%%%%%%%%%%%%%%%%%%%%%%%%%%%%%%%%%%%%%%%%%%%%
While symmetric entanglers have been obtained for many fermionic SPTs\cite{nat2018}, entanglers for beyond super-cohomology phases are still lacking. One example of such a beyond super-cohomology phase is the generator of the 2D $\mathbb{Z}_2\times\mathbb{Z}_2^f$ SPT\cite{tarantino2016}. Even if the entangler is not an FDQC, one may ask if there exists a 2D symmetric QCA that entangles the phase. In this section, we will present a simple argument for why there cannot exist a symmetric QCA that entangles this particular SPT. 

This SPT phase is characterized by a domain-wall decoration structure: domain walls of the $\mathbb{Z}_2$ symmetry are decorated with Kitaev wires. This means that $W_{A,g_1}^U$, where $g_1$ is the generator of the $\mathbb{Z}_2$ symmetry, is a Kitaev wire entangler. Because $\{W_{A,g}^U\}$ forms a representation of $\mathbb{Z}_2\times\mathbb{Z}_2^f$, $W_{A,g_1}^U$ must satisfy $\left(W_{A,g_1}^U\right)^2$. On the other hand, according to the well-established classification of 1D fermionic QCA\cite{fermionic}, there does not exist any QCA that entangles the Kitaev wire and has order two. In particular, in order for $W_{A,g_1}^U$ to entangle the Kitaev wire, it must have order $\sim|\partial A|$. Therefore, there does not exist any QCA that entangles the aforementioned 2D SPT.
%%%%%%%%%%%%%%%%%%%%%%%%%%%%%%%%%%%%%%%%%%%%%%%%%%%%%
\section{Discussion}\label{sdiscussion}
%%%%%%%%%%%%%%%%%%%%%%%%%%%%%%%%%%%%%%%%%%%%%%%%%%%%%%%%%%%%%%%%%%%%%%%%%%%%%%%%%%%%

In this work, we presented a general framework for classifying (strict) LPUs with symmetry, based on anomalies computed from explicit flux insertion operators $\{W_{A,g}^U\}$ defined in Eq.~(\ref{WAgdef}). We then applied this framework to obtain explicit formulas for topological invariants for various kinds of SPT entanglers. We conclude by highlighting interesting directions for extending our results and some relations between insights related to this framework and other topics of research.

First, we expect our framework to generalize naturally to broad classes of fermionic SPT phases, namely those classified by cohomology and supercohomology. Symmetric entanglers have already been obtained for these phases\cite{nat2018,ellisondisentangle}, and we expect that studying trivialization obstructions to $\{W_{A,g}^U\}$ obtained by these entanglers can be used to obtain topological invariants like the ones presented here.

Another direction for future work is making our invariants for entanglers of SPTs with anti-unitary symmetries more explicit. This is difficult because anti-unitary symmetries cannot be truncated, as explained in Sec.~(\ref{str}), so we cannot study how the entangler transforms restricted symmetry operators. Along similar lines, it would be interesting to see if our invariants for entanglers of higher dimensional SPTs (beyond those described by decorated 1D domain walls) can be made more explicit. For example, our formulas for $e^{i\theta_{g_i}}$ (\ref{thetag1def}) and $e^{i\theta_{g_i,g_j}}$ (\ref{thetagigj}) are not completely closed form because they involve truncating the flux insertion operators. It may be possible to obtain more explicit formulas for these quantities, that only require restriction of the original global symmetry operators. If this is not possible, it would be interesting to understand more precisely why it is not possible. 

In 2D and 3D we only obtained gauge invariant topological invariants for symmetries of the form $G=\prod_{i=1}^M\mathbb{Z}_{N_i}$ by studying different obstructions to making $\{W_{A,g}^U\}\sim\{W_{A,g}^{\mathbbm{1}}\}$. We show that these gauge invariant quantities correspond to known quantities defining the cocycle labeling the SPT phase in Appendix~\ref{sproofs2d} and \ref{sproofs3d}. For non-abelian symmetries, analogous gauge invariant topological invariants are not known; there is no easy generalization of  the method used for 1D SPTs with non-abelian symmetries discussed in Sec.~\ref{snonabelian}. While the particular kinds of obstructions we studied are for abelian groups, the framework of studying anomalous representations of $G$ given by $\{W_{A,g}^U\}$ is applicable to \emph{any} group. Specifically, Property 3 applies for non-abelian groups as well. It may be possible to obtain topological invariants for SPT entanglers with nonabelian symmetries in higher dimensions by considering more general obstructions to making $\{W_{A,g}^U\}$ a non-anomalous representation of $G$.

One particularly difficult entangler to study, with which we cannot apply our framework, is the QCA that entangles the beyond-cohomology bosonic SPT in 3D protected by time-reversal symmetry\cite{haahnontrivial,haahclifford,moreqca}. 1D SPTs with time-reversal symmetry can be entangled by FDQCs, so even though we cannot restrict time-reversal symmetry to compute $W_{A,T}^U$, we can still restrict the SPT entangler and obtain a topological invariant using the restricted entangler. The 3D time reversal SPT, however, can only be entangled by a QCA. In this case, we cannot restrict the symmetry (which is related to computing flux insertion operators) or the entangler (which is related to computing the boundary representation of the symmetry). It would be interesting to see if the framework presented here can be extended to obtain an invariant for this SPT entangler. One interesting direction to pursue would be to consider the higher-form SPT formulation of these phases, and use our framework to detect anomalies of the flux insertion operators for these higher form symmetries\cite{highercup,hsinexotic}.

As mentioned in the introduction, although the obstructions that we discuss in this work are all related to SPT invariants, not all nontrivial symmetric LPUs are SPT entanglers. There are also obstructions that are not related to SPT invariants, such as the obstruction classifying $U(1)$ symmetric LPUs in 1D characterized by chiral charge transport\cite{u1floquet}. It would be interesting to study more generally what differentiates SPT obstructions for other kinds of obstructions.

Interestingly, our framework adds a different perspective to the classification of certain kinds of Floquet systems. An MBL Floquet system is described by a path of unitaries parameterized by $t\in[0,T)$, with the constraint that $U(T)$ satisfies the MBL condition: $U(T)=\prod_rU_r$, where $\{U_r\}$ are mutually commuting local unitaries (possibly with exponentially decaying tails)\cite{chiralbosons}. Another interesting kind of Floquet circuit is one where $U(T)$ does not satisfy the MBL condition, but $U(NT)=U(T)^N$ does, where $N$ is a finite integer. For example, in the ``radical" Floquet circuit studied in Ref.~\onlinecite{radical}, $U(T)$ does not satisfy the MBL condition, but $U(T)^2$ does. These kinds of circuits are related to $\{W_{A,g}^U\}$, because for finite groups, $\left(W_{A,g}^U\right)^{|g|}=\mathbbm{1}$ on a closed manifold, so $\left(W_{A,g}^U\right)^{|g|}$ satisfies the MBL condition. Therefore, the study of different anomalous representations $\{W_{A,g}^U\}$ is related to the study of circuits that, roughly speaking, are the $|g|$th root of an MBL Floquet system. For example, there may be a $\mathbb{Z}_2$ symmetric LPU in 3D with $W_{A,g}^U$ (where $g$ generates the $\mathbb{Z}_2$ symmetry) equivalent to the radical Floquet circuit. Such an LPU would not be an SPT entangler, because the classification of $\mathbb{Z}_2$-symmetric bosonic SPTs in 3D is trivial.
%The classification of MBL Floquet systems is closely related to the classification of ``unitary loops" which are paths of unitaries $\{U(t):t\in[0,T)\}$ that satisfy $U(0)=U(T)=\mathbbm{1}$ on a closed manifold. 

\acknowledgments

C.Z. thanks Michael Levin for many helpful conversations, especially related to Sec.~\ref{sabelian2d}, and for comments on the drafts of this paper. C.Z. also thanks Yu-An Chen and Tyler Ellison for helpful discussions related to the heptagon equations and 3D QCA. C.Z. acknowledges the support of the Kadanoff Center for Theoretical Physics at the University of Chicago, the Simons Collaboration on Ultra-Quantum Matter (651440, M.L.), and the National Science Foundation Graduate Research Fellowship under Grant No. 1746045.
%%%%%%%%%%%%%%%%%%%%%%%%%%%%%%%%%%%%%%%%%%%%%%%%%%%
\appendix
%%%%%%%%%%%%%%%%%%%%%%%%%%%%%%%%%%%%%%%%%%%%%%%%%%%
%%%%%%%%%%%%%%%%%%%%%%%%%%%%%%%%%%%%%%%%%%%%%%%%%%%%%%%%%%%%%%%%%%%%%%%%%%%%%%%%%%%%
\section{Group cohomology}\label{sgroupcohomology}
%%%%%%%%%%%%%%%%%%%%%%%%%%%%%%%%%%%%%%%%%%%%%%%%%%%%%%%%%%%%%%%%%%%%%%%%%%%%%%%%%%%%
Many of the SPTs entanglers we discuss are related to bosonic in-cohomology SPTs. Here, we will briefly review the aspects of group cohomology relevant to the study of these SPTs. 

An $n$-cochain $\omega(g_1,\cdots g_n)$ is a map from $n$ group elements to $U(1)$:
\begin{equation}
\omega(g_1,\cdots g_n):G\times G\times\cdots \times G\to U(1),
\end{equation}
where $G$ is repeated $n$ times. The collection of $n$-cochains forms an abelian group $\mathcal{C}^n$ with group multiplication given by 
\begin{equation}
(\omega_1\cdot\omega_2)(g_1,\cdots g_n)=\omega_1(g_1,\cdots g_n)\cdot \omega_2(g_1,\cdots g_n).
\end{equation}

The coboundary operator $\delta$ is a map $\mathcal{C}^n\to\mathcal{C}^{n+1}$, defined by
\begin{align}
\begin{split}\label{coboundary}
\delta& \omega(g_1,\cdots, g_{n+1})\\
&=\omega(g_2,\cdots, g_{n+1})\omega(g_1,\cdots, g_n)^{(-1)^{n+1}}\\
&\times\prod_{i=1}^n\left[\omega(g_1,\cdots,g_ig_{i+1},\cdots,g_{n+1})\right]^{(-1)^i}.
\end{split}
\end{align}

One can check that $\delta(\omega_1\cdot\omega_2)=\delta\omega_1\cdot\delta\omega_2$ and $\delta^2=1$. The coboundary operator allows us to define $n$-cocycles and $n$-coboundaries, which are particular kinds of $n$-cochains. An $n$-cocycle is an $n$-cochain that satisfies $\delta\omega=1$. For example, from Eq.~(\ref{coboundary}), $3-$cocycles satisfy
\begin{equation}
\frac{\omega(g_2,g_3,g_4)\omega(g_1,g_2g_3,g_4)\omega(g_1,g_2,g_3)}{\omega(g_1g_2,g_3,g_4)\omega(g_1,g_2,g_3g_4)}=1,
\end{equation}
and $4-$cocycles satisfy
\begin{equation}
\frac{\omega(g_2,g_3,g_4,g_4)\omega(g_1,g_2g_3,g_4,g_5)\omega(g_1,g_2,g_3,g_4g_5)}{\omega(g_1g_2,g_3,g_4,g_5)\omega(g_1,g_2,g_3g_4,g_5)\omega(g_1,g_2,g_3,g_4)}=1.
\end{equation}

An $n$-coboundary is an $n$-cocycle that can be written as $\nu=\delta\omega$ where $\omega\in\mathcal{C}^{n-1}$. Because $\delta^2=1$, an $n$-coboundary must be an $n$-cocycle. We call two $n$-cocycles equivalent if they differ by a $n$-coboundary: 
\begin{equation}
\omega_1\sim\omega_2:\omega_1=\omega_2\cdot\delta\omega,
\end{equation}
where $\omega\in\mathcal{C}^{n-1}$. The equivalence classes of $n$-cocycles form an Abelian group $H^n(G,U(1))$, which classify many bosonic SPTs.

%%%%%%%%%%%%%%%%%%%%%%%%%%%%%%%%%%%%%%%%%%%%%%%%%%%%%%%%%%%%%%%%%%%%%%%%%%%%%%%%%%%%
\section{Proof that $\{W_{A,g}^{U'}\}\sim\{W_{A,g}^U\}$ iff $U'\sim U$}\label{sproofbd}
%%%%%%%%%%%%%%%%%%%%%%%%%%%%%%%%%%%%%%%%%%%%%%%%%%%%%%%%%%%%%%%%%%%%%%%%%%%%%%%%%%%%
In this appendix, we provide the more precise version of the ``only if" direction of Property 4 in Sec.~\ref{sboundaryrep}, as well as a proof of the ``if" direction. 

To prove the ``only if" direction, we will show that if $U=YU$ where $Y$ is a $G$ symmetric FDQC, then $W_{A,g}^{U'}=V^\dagger W_{A,g}^UV$ for every $g\in G$, where $V$ is as defined earlier. By definition, $W_{A,g}^{U'}$ is given by
\begin{equation}
W_{A,g}^{U'}=U_{A,g}YUU_{A_{\mathrm{in}},g}^\dagger U^\dagger Y^\dagger.
\end{equation}

Here, $U$ has an operator spreading length $\xi$, so $U'$ has an operator spreading length $\xi'=\xi+2n\lambda$. Let us write $Y=Y_NY_{N-1}\cdots Y_1$, where $Y_n=\prod_rY_{n,r}$ for every $n\in[1,N]$. Because $Y$ is a $G$ symmetric FDQC, every local gate $Y_{n,r}$ is symmetric. $U$ only modifies $U_{A_{\mathrm{in}},g}^{\dagger}$ within $\xi$ of the boundary of $A_{\mathrm{in}}$, so we can remove all the gates in $Y_1$ fully supported outside of $\partial_{\xi}A_{\mathrm{in}}$ by commuting them through $UU_{A_{\mathrm{in}},g}^{\dagger}U^\dagger$. Let us denote the remaining gates in $Y_1$ by $\tilde{Y}_1$. We can then remove all the gates in $Y_2$ fully supported outside of $\partial_{\xi+2\lambda}A_{\mathrm{in}}$ by commuting them through $\tilde{Y}_1UU_{A_{\mathrm{in}},g}U^\dagger \tilde{Y}_1^\dagger$. Continuing in this way, we get $\tilde{Y}=\tilde{Y}_N\tilde{Y}_{N+1}\cdots\tilde{Y}_1$. This operator is fully supported within $\xi+2n\lambda=\xi'$ of the boundary of $A_{\mathrm{in}}$. Because it is fully supported inside $A$ and contains only $G$ symmetric gates, it commutes with $U_{A,g}$. Therefore, we have
\begin{equation}
W_{A,g}^{U'}=\tilde{Y}U_{A,g}UU_{A_{\mathrm{in}},g}^\dagger U^\dagger\tilde{Y}^\dagger.
\end{equation}

Identifying $\tilde{Y}=V^\dagger$, we obtain the desired result. 

We will now prove the ``if" direction. Our proof in 1D uses methods similar to those in Sec. VII.C of Ref.~\onlinecite{u1floquet}, and our proof for higher dimensions is a generalization of the same line of argument. 

Because our invariants are multiplicative under stacking and composition, we only need to consider the case where $U=\mathbbm{1}$. Relabeling $U'\to U$, we will show that if $\{W_{A,g}^U\}=\{V^\dagger W_{A,g}^{\mathbbm{1}}V\}$ where $V$ is a $G$ symmetric FDQC supported within $\xi$ of the boundary of $A_{\mathrm{in}}$, then $U$ is a $G$ symmetric FDQC (up to products of $(d-1)$ dimensional $G$ symmetric LPUs). We will first prove this in 1D, and then we generalize to higher dimensions. Our strategy is the following: we will show that if $\{W_{A,g}^U\}=\{V^\dagger W_{A,g}^{\mathbbm{1}} V\}$, and we assume that $U$ is an FDQC, then we can modify the individual gates in $U$ (without changing $U$ as a whole) so that each gate is symmetric. Note that since, by this method, we can already find a symmetric FDQC giving $\{W_{A,g}^U\}$, we do not need to consider if $U$ is a QCA. %For simplicity, we will choose $U_{g,\mathrm{in}}^A=\mathbbm{1}$.

We begin with the proof in 1D. Without loss of generality, we can cluster the sites in a 1D bosonic spin chain into ``supersites" so that $U$ takes the form of a depth two FDQC where each layer consists of disjoint gates supported over two supersites\cite{GNVW}. In terms of supersites, $\lambda=1$ and $\xi=2$. Let us write $U=U_2U_1$ where
\begin{equation}\label{u2u1}
U_1=\prod_{r}U_{2r,2r+1}\qquad U_2=\prod_rU_{2r-1,2r}.
\end{equation}

$U$ is $G$-symmetric, but $U_2$ and $U_1$ are not necessarily individually $G$-symmetric. Since $\xi=2$, we choose $A_{\mathrm{in}}=[-1,2]$ and $A=[-3,4]$. This gives
\begin{equation}
W_{A,g}^{\mathbbm{1}}=U_{-3,g}U_{-2,g}U_{3,g}U_{4,g}.
\end{equation}

A 0D FDQC is simply a local unitary operator. We can write $V=V_LV_R$ where $V_L$ is supported on $[-3,0]$ and $V_R$ is supported on $[1,4]$. Then we have, as our assumption,
\begin{equation}
W_{A,g}^U=V_L^\dagger\left(U_{-3,g}U_{-2,g}\right)V_L\otimes V_R^\dagger\left(U_{3,g}U_{4,g}\right)V_R.
\end{equation}

Now we substitute $W_{A,g}^U=U_{A,g}UU_{A_{\mathrm{in}},g}^\dagger U^\dagger$ on the left hand side. Because $V_L$ and $V_R$ are both $G$-symmetric, we can multiply both sides by $U_{A,g}^\dagger$ to obtain
\begin{equation}
UU_{A_{\mathrm{in}},g}^\dagger U^\dagger=V_L^\dagger\left(U_{-1,g}^\dagger U_{0,g}^\dagger\right)V_L\otimes V_R^\dagger\left(U_{1,g}^\dagger U_{2,g}^\dagger\right)V_R.
\end{equation}
Using the explicit form of $U$, and conjugating both sides by $U_2^\dagger$, we get
\begin{align}
\begin{split}\label{proof1}
U_{-2,-1}&U_{0,1}U_{2,3}U_{A_{\mathrm{in}},g}^\dagger U_{2,3}^\dagger U_{0,1}^\dagger U_{-2,1}^\dagger\\
&=\tilde{V}_L^\dagger\left(U_{-1,g}^\dagger U_{0,g}^\dagger\right)\tilde{V}_L\otimes \tilde{V}_R^\dagger\left(U_{1,g}^\dagger U_{2,g}^\dagger\right)\tilde{V}_R,
\end{split}
\end{align}
where $\tilde{V}_L=V_LU_{-3,-2}U_{-1,0}$ and $\tilde{V}_R=V_RU_{1,2}U_{3,4}$. 

Notice that the first line of (\ref{proof1}) is fully supported on $[-2,3]$ and breaks into a tensor product of three disjointly supported operators:
\begin{align}
\begin{split}\label{proof2}
U_{-2,-1}&U_{0,1}U_{2,3}U_{A_{\mathrm{in}},g}^\dagger U_{2,3}^\dagger U_{0,1}^\dagger U_{-2,1}^\dagger\\
&=U_{[-2,-1],g}^\dagger\otimes U_{[0,1],g}^\dagger\otimes U_{[2,3],g}^\dagger.
\end{split}
\end{align}

On the other hand, the second line of (\ref{proof1}) can be written as a tensor product of two disjointly supported operators. In order for it to have the same support as the first line of $(\ref{proof1})$, $\tilde{V}_L^\dagger\left(U_{-1,g}U_{0,g}\right)\tilde{V}_L$ must be fully supported on $[-2,0]$ and $\tilde{V}_R^\dagger\left(U_{1,g}U_{2,g}\right)\tilde{V}_R$ must be fully supported on $[1,3]$. Because (\ref{proof2}) is a product of disjoint operators on $[-2,1],[0,1],$ and $[2,3]$, we must have
\begin{align}
\begin{split}
\tilde{V}_L^\dagger\left(U_{-1,g}^\dagger U_{0,g}^\dagger \right)\tilde{V}_L&=U_{[-2,-1],g}^\dagger\otimes \tilde{U}_{0,g}^\dagger\\
\tilde{V}_R^\dagger\left(U_{1,g}^\dagger U_{2,g}^\dagger \right)\tilde{V}_R&=\tilde{U}_{1,g}^\dagger\otimes U_{[2,3],g}^\dagger,
\end{split}
\end{align}
and $U_{[0,1],g}=\tilde{U}_{0,g}\otimes\tilde{U}_{1,g}$. Since the spectrum of $U_{[-2,-1].g}$ matches that of $U_{-1,g}$ from (\ref{proof2}), the spectrum of $\tilde{U}_{0,g}$ and $\tilde{U}_{1,g}$ must match those of $U_{0,g}$ and $U_{1,g}$ respectively. This means that there exists on-site operators $R_r$ such that
\begin{equation}
R_0\tilde{U}_{0,g}^\dagger R_{0}^\dagger=U_{0,g}^\dagger\qquad R_1\tilde{U}_{1,g}^\dagger R_{1}^\dagger=U_{1,g}^\dagger.
\end{equation}

We can repeat the exercise with other choices of $A_{\mathrm{in}}$ in order to get all the on-site operators $\{R_r\}$. Using these on-site operators, we can define $\tilde{U}_1=\prod_r\tilde{U}_{2r-1,2r}$ and $\tilde{U}_2=\prod_r\tilde{U}_{2r,2r+1}$, where
\begin{align}
\begin{split}\label{redefU}
\tilde{U}_{2r-1,2r}&=R_{2r-1}R_{2r}U_{2r-1,2r}\\
\tilde{U}_{2r,2r+1}&=U_{2r,2r+1}R_{2r}^\dagger R_{2r+1}^\dagger.
\end{split}
\end{align}

It is easy to check that $\tilde{U}_{2}\tilde{U}_1=U_2U_1=U$ and $\tilde{U}_2$ and $\tilde{U}_1$ consist of gates that are all $G$-symmetric. This concludes the proof for 1D systems. 

Now we proceed to higher dimensions. In higher dimensions, WLOG we can write any FDQC $U$ as $U=U_2U_1$ where $U_1$ and $U_2$ each consist of commuting $(d-1)$ dimensional FDQCs. In other words, we replace the local two-site unitaries in Eq.~(\ref{u2u1}) by $(d-1)$ dimensional FDQC. For example, in 2D, we can divide the plane into vertical strips, with a 1D vertical FDQC for each two-site interval in $\hat{x}$. In this case, $U_{2r,2r+1}$ is not a local unitary, but rather a 1D FDQC. We now consider $A$ and $A_{\mathrm{in}}$ to be infinite strips, with finite extent in $\hat{x}$. Using the notation $U_{x=r,g}=\prod_yU_{(r,y),g}$, we have
\begin{equation}
W_{A,g}^{\mathbbm{1}}=U_{x=-3,g}U_{x=-2,g}U_{x=3,g}U_{x=4,g}.
\end{equation}

Proceeding in the same way as for 1D, we have
\begin{align}
\begin{split}\label{proof3}
&U_{-2,-1}U_{0,1}U_{2,3}U_{A_{\mathrm{in}},g}^\dagger U_{2,3}^\dagger U_{0,1}^\dagger U_{-2,1}^\dagger\\
&=\tilde{V}_L^\dagger\left(U_{x=-1,g}^\dagger U_{x=0,g}^\dagger\right)\tilde{V}_L\otimes \tilde{V}_R^\dagger\left(U_{x=1,g}^\dagger U_{x=2,g}^\dagger\right)\tilde{V}_R,
\end{split}
\end{align}
where $\tilde{V}_L=V_LU_{-3,-2}U_{-1,0}$ and $\tilde{V}_R=V_RU_{1,2}U_{3,4}$. Here, $V_L$ is a 1D FDQC supported on $x\in[-3,0]$ and $V_R$ is a 1D FDQC supported on $x\in[1,4]$. From the same arguments as for 1D, we can write 
\begin{align}
\begin{split}
\tilde{V}_L^\dagger\left(U_{x=-1,g}^\dagger U_{x=0,g}^\dagger \right)\tilde{V}_L&=U_{x=[-2,-1],g}^\dagger\otimes \tilde{U}_{x=0,g}^\dagger\\
\tilde{V}_R^\dagger\left(U_{x=1,g}^\dagger U_{x=2,g}^\dagger \right)\tilde{V}_R&=\tilde{U}_{x=1,g}^\dagger\otimes U_{x=[2,3],g}^\dagger,
\end{split}
\end{align}
where $U_{x=[-2,-1],g},U_{x=[0,1],g},$ and $U_{x=[1,2],g}$ are defined similarly to how they are in (\ref{proof2}) and $U_{x=[0,1],g}=\tilde{U}_{x=0,g}^\dagger\otimes \tilde{U}_{x=1,g}^\dagger$. From the same arguments as for 1D, $\tilde{U}_{x=0,g}$ and $\tilde{U}_{x=1,g}$ must have the same spectra as $U_{x=0,g}$ and $U_{x=1,g}$ respectively. This means that there exist 1D FDQCs (that are not necessarily $G$-symmetric) $R_{x=0}$ and $R_{x=1}$ rotating these operators into each other:
\begin{align}
\begin{split}
R_{x=0}\tilde{U}_{x=0,g}^\dagger R_{x=0}^\dagger&=U_{x=0,g}^\dagger\\
 R_{x=1}\tilde{U}_{x=1,g}^\dagger R_{x=1}^\dagger&=U_{x=1,g}^\dagger.
\end{split}
\end{align}

We can again repeat the exercise for other choices of $A_{\mathrm{in}}$ to obtain a set of 1D FDQCs $\{R_{x=r}\}$. Using this set, we can again define $\tilde{U}_1$ and $\tilde{U}_2$ as in (\ref{redefU}). $\tilde{U}_2\tilde{U}_1=U_2U_1=U$, and each consist of 1D FDQCs $\tilde{U}_{r,r+1}$, which are individually $G$-symmetric.
 
This means that $U$ can only differ from $\mathbbm{1}$ by, at most, a product of $(d-1)$ dimensional $G$ symmetric LPUs along $\hat{y}$. However, by the same method with $A$ and $A_{\mathrm{in}}$ chosen to be infinite strips with finite extent along $\hat{y}$, we can show that $U\sim\mathbbm{1}$ up to a product of $(d-1)$ dimensional $G$-symmetric LPUs along $\hat{x}$. We conjecture that this means that $U$ can be written as a $G$ symmetric FDQC, because it cannot be a product of $G$-symmetric LPUs along $\hat{x}$ or $\hat{y}$. The same method of proof applies to higher dimensions. 

Notice that in 1D, $\{R_r\}$ were simply on-site unitary operators. In 2D, we must specify that $\{R_r\}$ are 1D FDQCs, not 1D QCAs. This ensures that $\{R_r\}$ provides a smooth map from $U_1$ and $U_2$ to $\tilde{U}_1$ and $\tilde{U}_2$. The fact that $V$ is an FDQC guarantees that $\{R_r\}$ are FDQCs. 

%%%%%%%%%%%%%%%%%%%%%%%%%%%%%%%%%%%%%%%%%%%%%%%%%%%%%%%%%%%%%%%%%%%%%%%%%%%%%%%%%%%%%
%\section{Relations between SPT entangler invariants and cocycles}\label{sproofs1D}
%%%%%%%%%%%%%%%%%%%%%%%%%%%%%%%%%%%%%%%%%%%%%%%%%%%%%%%%%%%%%%%%%%%%%%%%%%%%%%%%%%%%%
%In this section, we will show that all the topological invariants we presented in the previous sections are directly related to gauge invariant quantities defining the SPT phase entangled by the unitary. In particular, they are gauge invariant combinations of cocycles. 
%%%%%%%%%%%%%%%%%%%%%%%%%%%%%%%%%%%%%%%%%%%%%%%%%%%%%%%%%%%%%%%%%%%%%%%%%%%%%%%%%%%%
\section{Relation between $\{W_{A,g}^U\}$ and boundary representation of the symmetry}\label{sboundaryrep}
%%%%%%%%%%%%%%%%%%%%%%%%%%%%%%%%%%%%%%%%%%%%%%%%%%%%%%%%%%%%%%%%%%%%%%%%%%%%%%%%%%%%
An SPT in $d$ dimensions can also characterized by how the symmetry is realized anomalously on the $(d-1)$ dimensional boundary. On the lattice, we say that the representation on the boundary cannot be made ``on-site." In this appendix, we will review the method described in Ref.~\onlinecite{symmedge} for computing the boundary representation of the symmetry $\{\tilde{W}_{A,g}^U\}$ using a symmetric SPT entangler. We will show in this appendix that $\{\tilde{W}_{A,g}^U\}$ and $\{W_{A,g}^U\}$ are representations of $G$ that carry the same anomaly. Physically, this recovers the fact that boundary domain walls and bulk symmetry fluxes have the same fusion properties (i.e. same $F$-symbol). 

A system with a boundary to the vacuum has a boundary Hilbert space spanned by all the states with excitations within $\xi$ of the boundary. Specifically, consider a state of the form $|\psi_i\rangle=|\psi_{i,A_{\mathrm{bd}}}\rangle\otimes|\psi_{0,\left(\Lambda\setminus A_{\mathrm{bd}}\right)}\rangle$, where $|\psi_{0,\left(\Lambda\setminus A_{\mathrm{bd}}\right)}\rangle$ is a symmetric product state. The state $|\psi_i\rangle$ has an excitation in $A_{\mathrm{bd}}=A\setminus A_{\mathrm{in}}$, which is a strip of width $\xi$. The set of all such states $\{|\psi_i\rangle\}$ spans the boundary Hilbert space of a system supported within $A$. If $U$ is an FDQC, we can truncate $U$ to $U_A$, which is fully supported in $A$. Then $\{U_A|\psi_{i}\rangle\}$ describe a set of states that look like the SPT deep inside of $A$ but remain in the trivial product state outside of $A$. The action of the global symmetry operator $U_g$ on the SPT state $U_A|\psi_{i}\rangle$ is given by the boundary representation of the symmetry on the edge Hilbert space $\tilde{W}_{A,g}^U$:
\begin{equation}\label{bdrep1}
U_g\left(U_A|\psi_{i}\rangle\right)=U_A\tilde{W}_{A,g}^U|\psi_{i}\rangle.
\end{equation}

Note that $|\psi_{0,\left(\Lambda\setminus A_{\mathrm{bd}}\right)}\rangle$ is invariant under the action of $U_g$, so $|\psi_i\rangle=U_{\left(\Lambda\setminus A_{\mathrm{bd}}\right),g}|\psi_i\rangle$. This means that  
\begin{equation}\label{bdrep2}
U_g\left(U_A|\psi_{i}\rangle\right)=U_A\tilde{W}_{A,g}^UU_{\left(\Lambda\setminus A_{\mathrm{bd}}\right),g}|\psi_{i}\rangle.
\end{equation}

Putting together (\ref{bdrep1}) and (\ref{bdrep2}), we see that one explicit definition of $\tilde{W}_{A,g}^U$ is given by
\begin{equation}\label{bdrep}
\tilde{W}_{A,g}^U=U_A^\dagger U_gU_AU_{\left(\Lambda\setminus A_{\mathrm{bd}}\right),g}^\dagger.
\end{equation}

Truncating $\tilde{W}_{A,g}^U$ to an open $(d-1)$ dimensional manifold gives a boundary domain wall operator, which creates domain walls in a symmetry broken boundary theory. These are the boundary analogues of the bulk symmetry fluxes, and their fusion properties encode the anomaly of the SPT.

To relate $\tilde{W}_{A,g}^U$ to $W_{A,g}^U$, we begin by writing $U_{\left(A\setminus A_{\mathrm{bd}}\right),g}$ as $U_{\left(A\setminus A_{\mathrm{bd}}\right),g}=U_{\left(\Lambda\setminus A\right),g}\otimes U_{A_{\mathrm{in}},g}$. The first factor commutes with $U_A$ because it is supported outside of $A$, so we have
\begin{equation}
\tilde{W}_{A,g}^U=U_A^\dagger U_{A,g}U_AU_{A_{\mathrm{in}},g}^\dagger.
\end{equation}

Comparing this with (\ref{WAgdef}) and using $U_AU_{A_{\mathrm{in}},g}^\dagger U_A^\dagger=UU_{A_{\mathrm{in}},g}^\dagger U^\dagger$, we see that
\begin{equation}
W_{A,g}^U=U_A\tilde{W}_{A,g}^UU_A^\dagger.
\end{equation}

The fact that the two $(d-1)$ dimensional representations differ by conjugation by a QCA means that they carry the same anomaly. In particular, if we compute a cocycle from restrictions of $\{W_{A,g}^U\}$ using the method presented in Ref.~\onlinecite{symmedge}, it would match with the cocycle computed using $\{\tilde{W}_{A,g}^U\}$, if we simply define the restriction of $W_{A,g}^U$ as $W_{\mathbf{A},g}^U=U_A\tilde{W}_{\mathbf{A},g}^UU_A^\dagger$. Note that since $\{\tilde{W}_{A,g}^U\}$ is defined differently from $\{W_{A,g}^U\}$, it does not necessarily have to satisfy (\ref{wequiv}) in order to carry the same anomaly as $\{W_{A,g}^U\}$. 
%%%%%%%%%%%%%%%%%%%%%%%%%%%%%%%%%%%%%%%%%%%%%%%%%%%%%%%%%%%%%%%%%%%%%%%%%%%%%%%%%%%%
\section{Relation between 1D invariants and $\omega(g,h)$}\label{sproofs1D}
%%%%%%%%%%%%%%%%%%%%%%%%%%%%%%%%%%%%%%%%%%%%%%%%%%%%%%%%%%%%%%%%%%%%%%%%%%%%%%%%%%%%
We will prove that our invariants for 1D SPT entanglers correspond to gauge invariant quantities that completely define the cocycle $\omega(g,h)\in H^2(G,U(1))$ labeling the SPT entangled by $U$. We will first prove this for when $G$ is abelian. The generalization to non-abelian groups is straightforward.
%%%%%%%%%%%%%%%%%%%%%%%%%%%%%%%%%%%%%%%%%%%%%%%%%%%%%
\subsection{Abelian symmetries}\label{sproofabelian}
%%%%%%%%%%%%%%%%%%%%%%%%%%%%%%%%%%%%%%%%%%%%%%%%%%%%%%%%%%%%%%%%%%%%%%%%%%%%%%%%%%%%
Our invariant for 1D SPT entanglers with abelian, unitary, discrete symmetries is given by
\begin{equation}\label{cgh}
c(g,h)=\overline{\mathrm{Tr}}\left(U^\dagger U_{A,g} U U_{B,h}U^\dagger U_{A,g}^\dagger U U_{B,h}^{\dagger}\right).
\end{equation}

We must show that the right hand side indeed computes $c(g,h)=\frac{\omega(g,h)}{\omega(h,g)}$, which defines the SPT entangled by $U$. Specifically, $c(g,h)$ is given by
\begin{align}
\begin{split}
c(g,h)&=\overline{\mathrm{Tr}}\left(R_{A,g}^UR_{A.h}^UR_{A,g}^{U\dagger}R_{A,h}^{U\dagger}\right)\\
&=\overline{\mathrm{Tr}}\left(L_{A,g}^UL_{A,h}^UL_{A,g}^{U\dagger}L_{A,h}^{U\dagger}\right)^*,
\end{split}
\end{align}
where $L_{A,g}^U\otimes R_{A,g}^U=W_{A,g}^U$. $L_{A,g}^U$ and $R_{A,g}^U$ form opposite projective representations of $G$, localized near the left and right endpoints of $A$. As mentioned in Sec.~\ref{sabelian}, it will be convenient to use instead an equivalent representation $\{\mathcal{W}_{A,g}^U\}$, defined by 
\begin{align}
\begin{split}
\mathcal{W}_{A,g}^U=\mathcal{L}_{A,g}^U\otimes\mathcal{R}_{A,g}^U&=U^\dagger W_{A,g}^UU\\
&=U_{A_{\mathrm{in}},g}^{\dagger}U^{\dagger}U_{A,g}U,
\end{split}
\end{align}
where in the last line we inserted the definition of $W_{A,g}^U$ and used the fact that $U_{A_{\mathrm{in}},g}^{\dagger}$ commutes with $U^\dagger U_{A,g}U$. Since $\{\mathcal{W}_{A,g}^U\}$ forms an equivalent representation as $\{W_{A,g}^U\}$, $\mathcal{R}_{A,g}^U$ forms the same projective representation as $R_{A,g}^U$. We therefore have
\begin{align}
\begin{split}
c(g,h)&=\overline{\mathrm{Tr}}\left(\mathcal{R}_{A,g}^U\mathcal{R}_{A.h}^U\mathcal{R}_{A,g}^{U\dagger}\mathcal{R}_{A,h}^{U\dagger}\right)\\
&=\overline{\mathrm{Tr}}\left(\mathcal{L}_{A,g}^U\mathcal{L}_{A,h}^U\mathcal{L}_{A,g}^{U\dagger}\mathcal{L}_{A,h}^{U\dagger}\right)^*,
\end{split}
\end{align}
\begin{figure}[tb]
   \centering
   \includegraphics[width=.9\columnwidth]{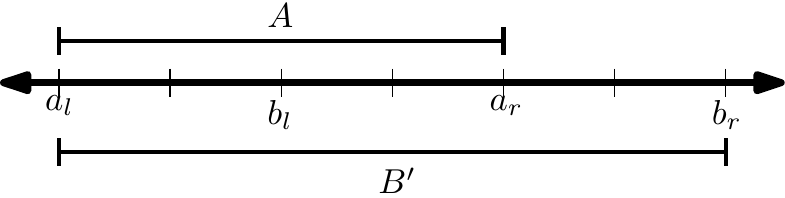} % requires the graphicx package
   \caption{The setup for proving (\ref{cgh}). Regions $A=[a_l,a_r]$ and $B'=[a_l,b_r]$ are aligned at the left endpoint only, and we define $B$ in (\ref{cgh}) to be $[b_l,b_r]$. The distances between labeled points in the figure must all be greater than or equal to $2\xi$.}
   \label{fig:cohomology}
\end{figure}

Let us specify $A=[a_l,a_r]$ and $B=[b_l,b_r]$; these are overlapping intervals in the 1D chain, as shown in Fig.~\ref{fig:ABintervals}. We will prove (\ref{cgh}) by first considering a slightly different setup, with an interval $B'=[a_l,b_r]$ that aligns with the left endpoint of $A$ and the right endpoint of $B$, as illustrated in Fig.~\ref{fig:cohomology}. Assuming that $G$ is abelian, we have
\begin{equation}\label{ghhg}
\left(U^\dagger U_{A,g}U\right)\left(U^\dagger U_{B',h}U\right)\left(U^\dagger U_{A,h}^\dagger U\right)\left(U^\dagger U_{B',h}^\dagger U\right)=\mathbbm{1}.
\end{equation}

We will show that the left hand side of the above equation splits into two contributions. One contribution is localized near $a_l$, and corresponds to $\mathcal{L}_g^U\mathcal{L}_h^U\mathcal{L}_g^{U\dagger}\mathcal{L}_h^{U\dagger}=c(g,h)^{-1}\mathbbm{1}$. The other contribution is precisely the operator inside the trace in the right hand side of Eq.~(\ref{cgh}). Since these two factors multiply to $\mathbbm{1}$, this proves Eq.~(\ref{cgh}).

The first step is to use
\begin{align}
\begin{split}\label{UAU}
U^\dagger U_{A,g}U&=U_{A_{\mathrm{in}},g}\mathcal{L}_{A,g}^U\mathcal{R}_{A,g}^U\\
&=\left(U_{A_{\mathrm{in},L},g}\mathcal{L}_{A,g}^U\right)\left(U_{A_{\mathrm{in},R},g}\mathcal{R}_{A,g}^U\right),
\end{split}
\end{align}
where $A_{\mathrm{in},L}$ and $A_{\mathrm{in},R}$ are the left and right halves of $A_{\mathrm{in}}=[a_l+\xi,a_r-\xi]$. Specifically, $A_{\mathrm{in},L}=[a_l+\xi,b_l-1]$ and $A_{\mathrm{in},R}=[b_l,a_r-\xi]$. Similarly,
\begin{equation}\label{UBU}
U^\dagger U_{B',h}U=\left(U_{B_{\mathrm{in},L}',h}\mathcal{L}_{B',h}^U\right)U_{B_{\mathrm{in},M}',h}\left(U_{B_{\mathrm{in},R}',h}\mathcal{R}_{B',h}^U\right),
\end{equation}
where $B_{\mathrm{in},L}'=A_{\mathrm{in},L}$, $B_{\mathrm{in},M}'=[b_l,a_r-1]$, and $B_{\mathrm{in},R}'=[a_r,b_r-\xi]$. 

The purpose for making the above partitions is to split the regions of the chain into a ``left" region $[a_l-\xi,b_l-1]\supset A_{\mathrm{in},L}=B_{\mathrm{in},L}'$, a ``middle" region $B_{\mathrm{in},M}'\supset A_{\mathrm{in},R}$, and a ``right" region $[a_r,b_r+\xi]\supset B_{\mathrm{in},L}'$. Expanding out Eq.~(\ref{ghhg}) using (\ref{UAU}) and (\ref{UBU}), we can commute all the operators fully supported on the left region past the operators fully supported in the other two regions. Using the fact that all the operators fully supported on $A_{\mathrm{in},L}$ commute with $\mathcal{L}_{A,g}^U$ and $\mathcal{L}_{A,h}^U$, since they are supported on disjoint spaces, we get
\begin{align}
\begin{split}
&\left(\mathcal{L}_{A,g}^U\mathcal{L}_{A,h}^U\mathcal{L}_{A,g}^{U\dagger}\mathcal{L}_{A,h}^{U\dagger}\right)\left(U_{A_{\mathrm{in},R},g}\mathcal{R}_{A,g}^U\right)\\
&\times\left(U_{B_{\mathrm{in},M}',h}U_{B_{\mathrm{in},R}',h}\mathcal{R}_{B',h}^U\right)\left(U_{A_{\mathrm{in},R},g}\mathcal{R}_{A,g}^U\right)^\dagger\\
&\times\left(U_{B_{\mathrm{in},M}',h}U_{B_{\mathrm{in},R}',h}\mathcal{R}_{B',h}^U\right)^\dagger=\mathbbm{1},
\end{split}
\end{align}
where we used $\mathcal{L}_{A,h}^U=\mathcal{L}_{B',h}^U$ and $U_{A_{\mathrm{in},L},g}U_{B_{\mathrm{in},L}',h}U_{A_{\mathrm{in},L},g}^\dagger U_{B_{\mathrm{in},L}',h}^\dagger=1$. We can identify the first term with $c(g,h)$. Since $R_{B',h}^U$ is supported far away from $U_{A_{\mathrm{in},R},g}\mathcal{R}_{A,g}^U$, we can commute it through $\left(U_{A_{\mathrm{in},R},g}\mathcal{R}_{A,g}^U\right)^\dagger$ to cancel with $\mathcal{R}_{B',h}^{U\dagger}$. Then replacing $B_{\mathrm{in},M}'\cup B_{\mathrm{in},R}'$ by $B$, we get
\begin{equation}
\left(U_{A_{\mathrm{in},R},g}\mathcal{R}_{A,g}^U\right)U_{B,h}\left(U_{A_{\mathrm{in},R},g}\mathcal{R}_{A,g}^U\right)^\dagger U_{B,h}^\dagger=c(g,h)\mathbbm{1}.
\end{equation}

Finally, we can multiply the left hand side by $\left(U_{A_{\mathrm{in},L},g}\mathcal{L}_{A,g}^U\right)\left(U_{A_{\mathrm{in},L},g}\mathcal{L}_{A,g}^U\right)^\dagger$. Commuting $\left(U_{A_{\mathrm{in},L},g}\mathcal{L}_{A,g}^U\right)^\dagger$ through $U_{B,h}$ and then using Eq.~(\ref{UAU}), we get
\begin{equation}
U^\dagger U_{A,g}UU_{B,h}U^\dagger U_{A,g}^\dagger UU_{B,h}^\dagger=c(g,h)\mathbbm{1}.
\end{equation}

After taking the normalized trace of both sides, we obtain (\ref{cgh}).

%%%%%%%%%%%%%%%%%%%%%%%%%%%%%%%%%%%%%%%%%%%%%%%%%%%%%%%%%%%%%%%%%%%%%%%%%%%%%%%%%%%%
\subsection{Non-abelian symmetries}\label{sproofnonabelian}
%%%%%%%%%%%%%%%%%%%%%%%%%%%%%%%%%%%%%%%%%%%%%%%%%%%%%%%%%%%%%%%%%%%%%%%%%%%%%%%%%%%%
Our proof for the non-abelian invariant (\ref{flownonabelian}) follows the same steps as our proof for the abelian invariant. For non-abelian symmetries, as with abelian symmetries, we must compute all the gauge invariant phases in order to determine the projective representation. However in this case, the set of gauge invariant phases is not given by $\{c(g,h)\}$, but rather $\{e^{i\phi(\gamma_n)}\}$ (see Sec.~\ref{snonabelian}). Consider $\gamma_n$ given by 
\begin{equation}
\gamma_n=g_1h_1g_1^{-1}h_1^{-1}g_2h_2g_2^{-1}h_2^{-1}.
\end{equation}

By definition, multiplying the elements on the right hand side in the group gives identity:
\begin{equation}
U_{A,g_1}U_{B',h_1}U_{A,g_1}^{-1}U_{B',h_1}^{-1}U_{A,g_2}U_{B',h_2}U_{A,g_2}^{-1}U_{B',h_2}^{-1}=1,
\end{equation}
where $A$ and $B'$ have the same definition as in Sec.~\ref{sproofabelian}. We can conjugate each element on the left hand side by $U$, to obtain an equation like (\ref{ghhg}). Crucially, all the conjugated symmetry operators $U^\dagger U_{A,g_1}U,U^\dagger U_{B',h_1}U,U^\dagger U_{A,g_2}U$, and $U^\dagger U_{A,h_2}U$ all again separate into operators supported in the left region, the middle region, and the right region, just as in the abelian case. Using the same steps as in Sec.~\ref{sproofabelian} to rearrange the operators by commuting them appropriately, we get 

\begin{align}
\begin{split}
&\mathcal{L}_{A,g_1}^U\mathcal{L}_{A,h_1}^U\mathcal{L}_{A,g_1}^{U\dagger}\mathcal{L}_{A,h_1}^{U\dagger}\mathcal{L}_{A,g_2}^U\mathcal{L}_{A,h_2}^U\mathcal{L}_{A,g_2}^{U\dagger}\mathcal{L}_{A,h_2}^{U\dagger}\\
&\times\left(U_{B_M',g_1}\mathcal{R}_{A,g_1}^U\right)\left(U_{B_{\mathrm{in},M}',h_1}U_{B_{\mathrm{in},R}',h_1}\mathcal{R}_{B',h_1}^U\right)\\
&\times\left(U_{B_M',g_1}\mathcal{R}_{A,g_1}^U\right)^\dagger\left(U_{B_{\mathrm{in},M}',h_1}U_{B_{\mathrm{in},R}',h_1}\mathcal{R}_{B',h_1}^U\right)^\dagger \\
&\times\left(U_{B_M',g_2}\mathcal{R}_{A,g_2}^U\right)\left(U_{B_{\mathrm{in},M}',h_2}U_{B_{\mathrm{in},R}',h_2}\mathcal{R}_{B',h_2}^U\right)\\
&\times\left(U_{B_M',g_2}\mathcal{R}_{A,g_2}^U\right)^\dagger\left(U_{B_{\mathrm{in},M}',h_2}U_{B_{\mathrm{in},R}',h_2}\mathcal{R}_{B',h_2}^U\right)^\dagger=\mathbbm{1}.
\end{split}
\end{align}
The first line gives $e^{i\phi(\gamma_n)}$, and we can then use the same steps as in Sec.~\ref{sproofabelian} to (1) remove $\mathcal{R}_{B',h_1}^U$ and $\mathcal{R}_{B',h_2}^U$, (2) replace $B_{\mathrm{in},M}'\cup B_{\mathrm{in},R}'\to B$, and (3) multiply by $\left(U_{A_{\mathrm{in},L},g_1}\mathcal{L}_{A,g_1}^U\right)\left(U_{A_{\mathrm{in},L},g_1}\mathcal{L}_{A,g_1}^U\right)^\dagger\left(U_{A_{\mathrm{in},L},g_2}\mathcal{L}_{A,g_2}^U\right)\\ \times\left(U_{A_{\mathrm{in},L},g_2}\mathcal{L}_{A,g_2}^U\right)^\dagger$. This gives
\begin{align}
\begin{split}
e^{i\phi(\gamma_n)}\mathbbm{1}&=U^\dagger U_{A,g_1}U U_{B,h_1}U^\dagger U_{A,g_1}^\dagger U U_{B,h_1}^\dagger \\
&\times U^\dagger U_{A,g_2}U U_{B,h_2}U^\dagger U_{B,g_2}^\dagger U U_{B,h_2}^\dagger.
\end{split}
\end{align}

We can obtain all other gauge invariant phases by a similar construction. In general, we have
\begin{align}
\begin{split}
e^{i\phi(\gamma_n)}&=\overline{\mathrm{Tr}}\left[\prod_i\left(\mathcal{R}_{A,g_i}^U\mathcal{R}_{A,h_i}^U\mathcal{R}_{A,g_i}^{U\dagger} \mathcal{R}_{A,h_i}^{U\dagger}\right)\right]\\
&=\overline{\mathrm{Tr}}\left[\prod_i\left(U^\dagger U_{A,g_i}U U_{B,h_i}U^\dagger U_{A,g_i}^\dagger U U_{B,h_i}\right)\right].
\end{split}
\end{align}

%%%%%%%%%%%%%%%%%%%%%%%%%%%%%%%%%%%%%%%%%%%%%%%%%%%%%%%%
\section{Manipulating symmetry fluxes}\label{sdefect}
%%%%%%%%%%%%%%%%%%%%%%%%%%%%%%%%%%%%%%%%%%%%%%%%%%%%%%%%%%%%%%%%%%%%%%%%%%%%%%%%%%%%%
\begin{figure}[tb]
   \centering
   \includegraphics[width=\columnwidth]{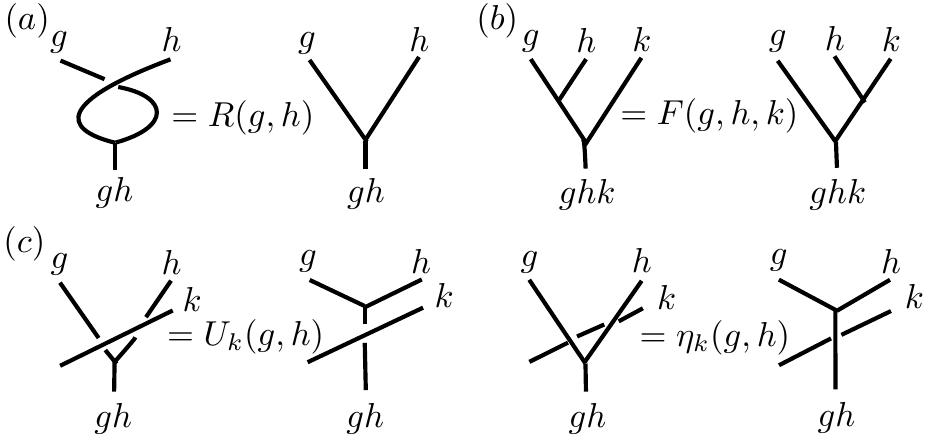} % requires the graphicx package
   \caption{Symmetry fluxes of a discrete, unitary, abelian group $G$ are described by the following processes. $(a)$ Braiding two symmetry fluxes $g$ and $h$ produces a $U(1)$ phase $R(g,h)$. $(b)$ Fusing symmetry fluxes in two different ways produces a $U(1)$ phase $F(g,h,k)$. $(c)$ Sliding a symmetry flux line over a fusion vertex produces a $U(1)$ phase $U_k(g,h)$, while sliding it under a fusion vertex produces a $U(1)$ phase $\eta_k(g,h)$.}
   \label{fig:RFUeta}
\end{figure}
In order to relate our 2D SPT entangler invariants to expressions in terms of cocycles (see Appendix~\ref{sproofs2d}), we need to first review various manipulations of symmetry fluxes in 2D SPT phases. In 2D, $W_{A,g}^U$ is supported on a closed loop. As we showed Sec.~\ref{sanomalous}, the restricted flux insertion operators $W_{\mathbf{A},g}^U$ insert symmetry flux at the endpoints of the interval $\mathbf{A}$. Here, we will first describe the more general framework of fusing, braiding, and sliding symmetry fluxes. Then, we will show how to perform these manipulations using concrete operators. This allows us to relate topological invariants in terms of various symmetry flux processes to our invariants for $\{W_{A,g}^U\}$.

A 2D bosonic SPT with symmetry $G$ can be understood as a $G$-crossed braided tensor category, where the original braided tensor category enriched by the $G$ symmetry is simply the trivial category, containing only trivial bosonic excitations\cite{symmfrac}. Because the original category is trivial, we will not need much of the more complex structures in $G$-crossed braided tensor categories; we simply describe it in this way in order to more easily relate fusion, braiding, and ``sliding". Enriching the trivial category with $G$ symmetry results a $G$-crossed braided tensor category $\mathcal{C}_G^{\times}$, which contains elements labeled by group elements $g\in G$, with fusion given by group multiplication: $g\times h=gh$.

The SPT corresponding to a given $\mathcal{C}_G^{\times}$ is defined by its fluxes (labeled by group elements) and their fusion, braiding, and sliding. In the following, we will use symmetry flux, symmetry defect, and group element interchangeably. Fusion of symmetry fluxes is described by the $F$ symbol, braiding of fluxes is described by the $R$ symbol, and sliding is described by $U$ and $\eta$. These three processes are illustrated in Fig.~\ref{fig:RFUeta}. In general, these quantities are all tensors, but because the fusion of the symmetry fluxes is just group multiplication, which is abelian, they are all $U(1)$ phases. 

The $F$ symbol takes as input three group elements $g,h,$ and $k$, and produces a $U(1)$ phase describing the difference between fusing three domain walls or symmetry fluxes in two different ways, as illustrated in Fig.~\ref{fig:RFUeta}.a. It can be computed in the bulk or the boundary of the SPT. For 2D bosonic SPTs, we can always choose a gauge in which $F(g,h,k)=\omega(g,h,k)\in H^3(G,U(1))$. 

In the bulk, symmetry fluxes can also be braided. The $R$ symbol $R(g,h)$ takes as input symmetry fluxes for $g,h\in G$ and describes the transformation associated with exchanging $g$ and $h$, as illustrated in Fig.~\ref{fig:RFUeta}.b. Therefore, $R(g,h)R(h,g)$ describes a full braid. Since we only consider abelian fusion, we will label the $R$ symbol by only two indices $g$ and $h$; their product $gh$ is fully determined by $g$ and $h$.

Finally, in order for fusion and braiding to be compatible, we must add ``sliding," which is shown in Fig.~\ref{fig:RFUeta}.c. $U_k(g,h)$ is the phase picked up from sliding a $k$ flux insertion line over a $g,h$ fusion vertex, while $\eta_k(g,h)$ is the phase picked up from sliding a $k$ flux insertion line under a $g,h$ fusion vertex. 

\begin{figure}[tb]
   \centering
   \captionsetup{width=\columnwidth}
   \includegraphics[width=0.92\columnwidth]{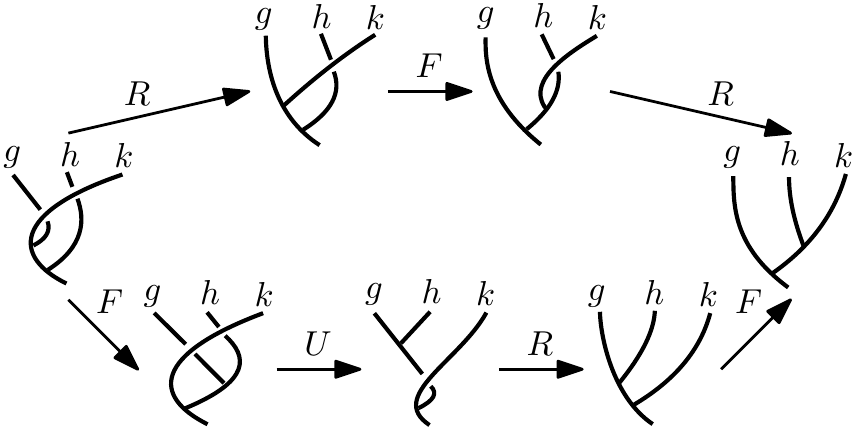} % requires the graphicx package
   \caption{The heptagon equations ensure consistency of symmetry flux fusion ($F$), braiding ($R$), and sliding ($U$ and $\eta$). Here, we specify to the case where the underlying topological order is trivial. In this case, fusion is simply given by group multiplication, so the label for bottom leg of each of the diagrams above is $ghk$. We only illustrate one of the heptagon equations here; the other (which involves $\eta$) takes a similar form\cite{symmfrac}.}
   \label{fig:heptagon}
\end{figure}

Fusion, braiding, and sliding satisfy two consistency equations, known as the heptagon equations. We will only need to use the first heptagon equation, which is illustrated in Fig.~\ref{fig:heptagon}. This equation tells us that
\begin{align}
\begin{split}\label{heptagon1}
R&(g,k)F(g,k,h)R(h,k)\\
&=F(k,g,h)U_k(g,h)R(gh,k)F(g,h,k)\\
%R^{-1}&(k,g)F(g,k,h)R^{-1}(k,h)\\
%&=F(k,g,h)\eta_k(g,h)R^{-1}(k,gh)F(g,h,k).
\end{split}
\end{align}

Rearranging (\ref{heptagon1}), we obtain
\begin{equation}\label{FandR}
\frac{F(k,g,h)F(g,h,k)}{F(g,k,h)}=\frac{R(g,k)R(h,k)}{R(gh,k)}U_k(g,h)^{-1}.
\end{equation}

Using $F(g,h,k)=\omega(g,h,k)$, we get
\begin{equation}\label{chiR}
\frac{\omega(k,g,h)\omega(g,h,k)}{\omega(g,k,h)}=\frac{R(g,k)R(h,k)}{R(gh,k)}U_k(g,h)^{-1}.
\end{equation}

We will see that for SPTs described by type I and type II cocycles (with nontrivial $e^{i\theta_{g_i}}$ or $e^{i\theta_{g_i,g_j}}$), we can always choose either $U_k(g,h)=1$ or $\eta_k(g,h)=1$ (but not both). For SPTs with III cocycles, which have a nontrivial $e^{i\theta_{g_i,g_j,g_k}}$, we cannot choose $U_k(g,h)=1$ or $\eta_k(g,h)=1$. In Appendix~\ref{sproofs2d}, we will show that $e^{i\theta_{g_i}}$, $e^{i\theta_{g_i,g_j}}$, and $e^{i\theta_{g_i,g_j,g_k}}$ can each be written as a product of $F$-symbols, which describes a fusion process. We will then use (\ref{heptagon1}) to translate these fusion processes into braiding and sliding processes.
 
We will now describe how to fuse and braid symmetry fluxes using concrete operators acting on an SPT state. We will not need to use sliding for this work, because we can instead prove our equation for $e^{i\theta_{g_i,g_j,g_k}}$ using dimensional reduction.

%%%%%%%%%%%%%%%%%%%%%%%%%%%%%%%%%%%%%%%%%%%%%%%%%%%%%%%%%%%%%%%%%%%%%%%%%%%%%%%%%%%%
\subsection{Symmetry flux fusion}\label{sfusion}
%%%%%%%%%%%%%%%%%%%%%%%%%%%%%%%%%%%%%%%%%%%%%%%%%%%%%%%%%%%%%%%%%%%%%%%%%%%%%%%%%%%%
Ref.~\onlinecite{symmedge} describes how $\omega(g,h,k)$ defines an anomaly related to domain wall fusion. We showed in Appendix~\ref{sboundaryrep} that boundary symmetry representations and flux insertion operators carry the same anomaly, so $\omega(g,h,k)$ also defines an anomaly related to symmetry flux fusion. In particular, we can compute $\omega(g,h,k)$ using restricted boundary representations of the symmetry $\{\tilde{W}_{\mathbf{A},g}^U\}$, and we can also compute it using restricted flux insertion operators $\{W_{\mathbf{A},g}^U\}$. 

The method for extracting $\omega(g,h,k)$ from $\{W_{\mathbf{A},g}^U\}$ was explained in Ref.~\onlinecite{symmedge}; we will briefly review it here. The first step is to notice that restricted flux insertion operators compose up to local operators near the boundary of the restriction:
\begin{equation}
W_{\mathbf{A},g}^UW_{\mathbf{A},h}^U=\Omega(g,h)W_{\mathbf{A},gh}^U.
\end{equation}

In particular, in two spatial dimensions, $W_{\mathbf{A},g}^U$ is supported on an open 1D interval, which has a boundary consisting of two points. Therefore, $\Omega(g,h)=\Omega_l(g,h)\otimes\Omega_r(g,h)$ where $\Omega_l(g,h)$ and $\Omega_r(g,h)$ are supported near the left and right endpoints of $\mathbf{A}$ respectively. $\Omega(g,h)$ is ambiguous in that it depends on the particular truncation. However, the expression for $\omega(g,h,k)$ does not depend on the choice of truncation. 

To obtain $\omega(g,h,k)$, we use the fact that multiplication of $W_{\mathbf{A},g}^U$ is associative. By considering the product $W_{\mathbf{A},g}^UW_{\mathbf{A},h}^UW_{\mathbf{A},k}^U$ and comparing the result from multiplying the first two operators first or multiplying the latter two operators first, we get
\begin{equation}
\Omega(g,h)\Omega(gh,k)=W_{\mathbf{A},g}^U\Omega(h,k)W_{\mathbf{A},g}^{U\dagger}\Omega(g,hk).
\end{equation}

However, the left and right endpoints of the restriction only need to be associative up to a phase. This phase is precisely $\omega(g,h,k)$:
\begin{align}
\begin{split}
&\Omega_r(g,h)\Omega_r(gh,k)\\
&=\omega(g,h,k)W_{\mathbf{A},g}^U\Omega_r(h,k)W_{\mathbf{A},g}^{U\dagger}\Omega_r(g,hk)\\
&=F(g,h,k)W_{\mathbf{A},g}^U\Omega_r(h,k)W_{\mathbf{A},g}^{U\dagger}\Omega_r(g,hk).
\end{split}
\end{align}
%%%%%%%%%%%%%%%%%%%%%%%%%%%%%%%%%%%%%%%%%%%%%%%%%%%%%%%%%%%%%%%%%%%%%%%%%%%%%%%%%%%%
\subsection{Symmetry flux braiding}\label{sbraiding}
%%%%%%%%%%%%%%%%%%%%%%%%%%%%%%%%%%%%%%%%%%%%%%%%%%%%%%%%%%%%%%%%%%%%%%%%%%%%%%%%%%%%

We can braid symmetry fluxes by using flux insertion operators to move symmetry fluxes around, but we must take care to only apply flux insertion operators on the SPT state. Here, as an example, we will demonstrate how to obtain an operator that braids symmetry fluxes $g$ and $h$ once. We will study more complex braiding processes later.

\begin{figure}[tb]
   \centering
   \includegraphics[width=0.9\columnwidth]{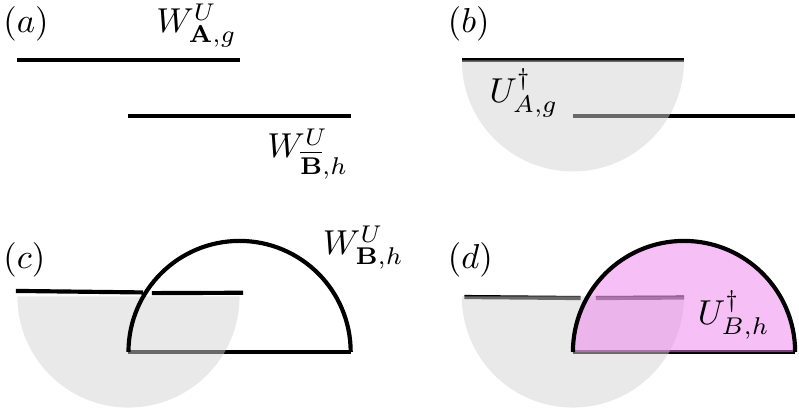} % requires the graphicx package
   \caption{The braiding phase $R(h,g)R(g,h)$ can be computed by comparing the above process, where we first create $g$ and $h$ fluxes and then move an $h$ flux around a $g$ flux, to the process where we first create an $h$ flux and move it in a loop and then create a $g$ flux. $(a)$ We first create opposite $g$ fluxes and $h$ fluxes. $(b)$ In order to move the $h$ flux around the $g$ flux at the right endpoint of $\mathbf{A}$, we must first undo the symmetry transformation near $W_{\mathbf{A},g}^U$ by applying $U_{A,g}^\dagger$. $(c)$ We can now apply $W_{\mathbf{B},h}^U$ in the local ground state, which moves the $h$ flux around the $g$ flux. $(d)$ To finish, we undo the symmetry transformation in the $h$ flux loop by applying $U_{B,h}^\dagger$.}
   \label{fig:braidsteps}
\end{figure}

Braiding the symmetry flux $h$ around $g$ consists of several steps, which are illustrated in Fig.~\ref{fig:braidsteps}. The first step is to create opposite $g$ and $h$ fluxes by applying $W_{\mathbf{A},g}^U$ and $W_{\overline{\mathbf{B}},h}^U$ on $|\psi_{\mathrm{SPT}}\rangle$. Notice that applying a restricted flux insertion operator $W_{\mathbf{A},g}^U$ modifies the state by a gauge transformation near $\mathbf{A}$. One way to see this is to recall that $W_{\mathbf{A},g}^U$ acts as the symmetry defect operator $D_{\mathbf{A},g}^U$ on $|\psi_{\mathrm{SPT}}\rangle$. In order to move the $h$ flux around the $g$ flux, we must undo the gauge transformation near $\mathbf{A}$, by applying $U_{A,g}^\dagger$. We can then apply $W_{\mathbf{B},h}^U$ to move the $h$ flux completely around the $g$ flux. Finally, we undo the gauge transformation in $B$ by applying $U_{B,h}^\dagger$. The phase difference between the above process and first moving the $h$ flux in a loop and then moving a $g$ flux into the loop is given by $R(h,g)R(g,h)$:
\begin{align}
\begin{split}
&U_{B,h}^\dagger W_{\mathbf{B},h}^U U_{A,g}^\dagger W_{\overline{\mathbf{B}},h}^UW_{\mathbf{A},g}^U|\psi_{\mathrm{SPT}}\rangle\\
&=R(h,g)R(g,h)U_{A,g}^{\dagger}W_{\mathbf{A},g}^UU_{B,h}^\dagger W_{\mathbf{B},h}^UW_{\overline{\mathbf{B}},h}^U|\psi_{\mathrm{SPT}}\rangle.
\end{split}
\end{align}

We can simplify the above expression by using the fact that $W_{\mathbf{A},g}^U$ commutes with $W_{\mathbf{B},h}^U$ (since they have disjoint support). Then, recalling the definition $D_{\overline{\mathbf{A}},g}^U=W_{\mathbf{A},g}^{U\dagger}U_{A,g}$:
\begin{align}
\begin{split}
&W_{\overline{\mathbf{B}},h}^{U\dagger}D_{\overline{\mathbf{B}},h}^{U}D_{\overline{\mathbf{A}},g}^{U}D_{\overline{\mathbf{B}},h}^{U\dagger}D_{\overline{\mathbf{A}},g}^{U\dagger}W_{\overline{\mathbf{B}},h}^U|\psi_{\mathrm{SPT}}\rangle\\
&=R(h,g)R(g,h)|\psi_{\mathrm{SPT}}\rangle.
\end{split}
\end{align}

Notice that if $g=g_i$ and $h=g_j$ are generators of $\mathbb{Z}_{N_i}$ and $\mathbb{Z}_{N_j}$ respectively, then as we show in Appendix~\ref{spropinvariants}, we can always choose $W_{A,g_i}^U$ and $W_{A,g_j}^U$ to be $\mathbb{Z}_{N_i}\times\mathbb{Z}_{N_j}$ symmetric FDQC. For this particular termination, $D_{\overline{\mathbf{B}},h}^{U}D_{\overline{\mathbf{A}},g}^{U}D_{\overline{\mathbf{B}},h}^{U\dagger}D_{\overline{\mathbf{A}},g}^{U\dagger}$ as an operator is a pure phase. This is because $D_{\overline{\mathbf{A}},g}^U$ and $D_{\overline{\mathbf{B}},h}^U$ can only fail to commute near the intersection $\mathbf{A}$ and $\mathbf{B}$ (see Fig.~(\ref{fig:braidsteps})). However, modifying $D_{\overline{\mathbf{A}},g}^U\to W_{\overline{\mathbf{A}},g}^{U\dagger}D_{\overline{\mathbf{A}},g}^U=UU_{A_{\mathrm{in}},g}U^\dagger$ and $D_{\overline{\mathbf{B}},h}^U$ similarly make these two operators commute, and these modifications are far away from the intersection of $\mathbf{A}$ and $\mathbf{B}$. This means that $D_{\overline{\mathbf{A}},g}^U$ and $D_{\overline{\mathbf{B}},h}^U$ can only fail to commute up to a phase. As a result, we can write
\begin{equation}\label{rgigj}
R(g_j,g_i)R(g_i,g_j)=\mathrm{Tr}(D_{\overline{\mathbf{B}},g_j}^{U}D_{\overline{\mathbf{A}},g_i}^{U}D_{\overline{\mathbf{B}},g_j}^{U\dagger}D_{\overline{\mathbf{A}},g_i}^{U\dagger}).
\end{equation}

Note that (\ref{rgigj}) only holds for symmetric terminations of flux insertion operators.
%%%%%%%%%%%%%%%%%%%%%%%%%%%%%%%%%%%%%%%%%%%%%%%%%%%%%%%%%%%%%%%%%%%%%%%%%%%%%%%%%%%%
\section{Relation between 2D invariants and $\omega(g,h,k)$}\label{sproofs2d}
%%%%%%%%%%%%%%%%%%%%%%%%%%%%%%%%%%%%%%%%%%%%%%%%%%%%%%%%%%%%%%%%%%%%%%%%%%%%%%%%%%%%
Bosonic SPTs in 2D with discrete, abelian, unitary, on-site symmetries are classified by 3-cocycles $\omega(g,h,k)\in H^3(G,U(1))$. In this section, we will show that our topological invariants for 2D SPT entanglers correspond to gauge invariant quantities that completely specify the 3-cocycle. 

It is convenient to first define an operation called the \emph{slant product} $\chi$, which takes a group element $g$ together with an $(d+1)$-cocycle and produces an $d$-cocycle. Explicitly,
\begin{align}
\begin{split}\label{slant}
\chi_g&(h_1,\cdots, h_d)\\
&=\prod_{i=1}^{d+1}\omega_{d+1}^{(-1)^{d+1+i}}(h_1,\cdots,h_{i-1},g,h_i,\cdots,h_d).
\end{split}
\end{align}

The physical interpretation of the slant product is that it describes an SPT in $d$ dimensions compactified in one spatial dimension over a circle, with $g$ flux inserted\cite{topologicalinvariants, natinvariants}. It therefore corresponds to a dimensional reduction process, mapping an SPT in $d$ spatial dimensions to one in $(d-1)$ spatial dimensions. For example, for $d=2$, it describes taking a 2D SPT on a long cylinder, with $g$ flux inserted through the cylinder. In this particular case, the slant product takes the form
\begin{equation}\label{chighk}
\chi_g(h,k)=\frac{\omega(g,h,k)\omega(h,k,g)}{\omega(h,g,k)}.
\end{equation}

The three invariants $e^{i\theta_{g_i}},e^{i\theta_{g_i,g_j}}$, and $e^{i\theta_{g_i,g_j,g_k}}$ are defined in terms of slant products by\cite{topologicalinvariants}\footnote{Here we define $e^{i\theta_{g_i,g_j,g_k}}$ as the inverse of the quantity labeled $e^{i\Theta_{ijk}}$ in Ref.~\onlinecite{topologicalinvariants}.}
\begin{align}
\begin{split}\label{cocycles}
e^{i\theta_{g_i}}&=\prod_{n=0}^{N_i-1}\chi_{g_i}(g_i,g_i^n)\\
e^{i\theta_{g_i,g_j}}&=\prod_{n=0}^{N_{ij}-1}\chi_{g_i}(g_j,g_j^n)\chi_{g_j}(g_i,g_i^n)\\
e^{i\theta_{g_i,g_j,g_k}}&=\frac{\chi_{g_k}(g_i,g_j)}{\chi_{g_k}(g_j,g_i)}.
\end{split}
\end{align} 

One can check that for abelian groups, all of the above quantities are gauge invariant, meaning they are invariant under changing $\omega(g,h,k)$ by a 3-coboundary. They were discussed in depth in Ref.~\onlinecite{topologicalinvariants} in the context of the Dijkgraaf-Witten theories that one obtains from gauging the SPTs we study here. In the Dijkgraaf-Witten theory, the above invariants correspond to various anyon braiding processes. As we will show, in the SPT, these expressions describe certain symmetry flux fusion, braiding, and sliding processes. Our expression for $e^{i\theta_{g_i}}$ in Eq.~(\ref{thetag1def}) can be directly related to a symmetry flux fusion process, because it is written entirely in terms of restricted flux insertion operators. However, our expressions for $e^{i\theta_{g_i,g_j}}$ and $e^{i\theta_{g_i,g_j,g_k}}=c(g_i,g_j;g_k)$ in Eqs.~(\ref{thetagigj}) and (\ref{decdomaininv}) also contain restricted global symmetry operators such as $U_{B,h}$. Using the heptagon equations, we will show that these expressions are actually related to braiding and sliding processes respectively. 
%%%%%%%%%%%%%%%%%%%%%%%%%%%%%%%%%%%%%%%%%%%%%%%%%%%%%%%%
\subsection{Relation between $e^{i\theta_{g_i}}$ and symmetry flux fusion}\label{sfluxfusion}
%%%%%%%%%%%%%%%%%%%%%%%%%%%%%%%%%%%%%%%%%%%%%%%%%%%%%%%%%%%%%%%%%%%%%%%%%%%%%%%%%%%%%
To show how Eq.~(\ref{thetag1def}) produces the quantity $e^{i\theta_{g_i}}$ defined in Eq.~(\ref{cocycles}), we use the fact that from (\ref{chighk}), $\chi_{g_i}(g_i,g_i^n)=\omega(g_i,g_i^n,g_i)$. Plugging this into Eq.~(\ref{cocycles}), we have
\begin{equation}\label{thetagicocycle}
e^{i\theta_{g_i}}=\prod_{n=0}^{N_i-1}\omega(g_i,g_i^n,g_i).
\end{equation}

From Appendix~\ref{sfusion}, we have
\begin{align}
\begin{split}\label{omegaggng}
&\Omega_r(g_i,g_i^n)\Omega_r(g_i^{n+1},g_i)\\
&=\omega(g_i,g_i^n,g_i)W_{\mathbf{A},g_i}^U\Omega_r(g_i^n,g_i)W_{\mathbf{A},g_i}^{U\dagger}\Omega_r(g_i,g_i^{n+1}).
\end{split}
\end{align}

We will now turn back to our expression for $e^{i\theta_{g_i}}$ in Eq.~(\ref{thetag1def}). We will use (\ref{omegaggng}) to get from (\ref{thetag1def}) to (\ref{thetagicocycle}). 

Our expression for $e^{i\theta_{g_i}}$ uses the operator $\left(W_{\mathbf{A},g_i}^U\right)^{N_i}$. Notice that we can write this as
\begin{align}
\begin{split}
\left(W_{\mathbf{A},g_i}^U\right)^{N_i}&=\prod_{n=0}^{N_i-1}W_{\mathbf{A},g_i^n}^UW_{\mathbf{A},g_i}^UW_{\mathbf{A},g_i^{n+1}}^{U\dagger}\\
%&=W_{\mathbf{A},g_i}^{U}W_{\mathbf{A},g_i}^{U}W_{\mathbf{A},g_i^2}^{U\dagger}W_{\mathbf{A},g_i^2}^{U}W_{\mathbf{A},g_i}^{U}W_{\mathbf{A},g_i^3}^{U\dagger}\cdots\\
&=\prod_{n=0}^{N_i-1}\Omega(g_i^n,g_i).
\end{split}
\end{align}

It follows that $R_{\mathbf{A},g_i}^U=\prod_{n=0}^{N_i-1}\Omega_r(g_i^n,g_i)$, and our expression for $e^{i\theta_{g_i}}$ can be written as
\begin{widetext}
\begin{equation}
e^{i\theta_{g_i}}=\overline{\mathrm{Tr}}\left[W_{\mathbf{A},g_i}^{U\dagger}\left(\prod_{n=0}^{N_i-1}\Omega_r(g_i^n,g_i)\right)W_{\mathbf{A},g_i}^U\left(\prod_{n=0}^{N_i-1}\Omega_r(g_i^n,g_i)\right)^\dagger\right].
\end{equation}

If we replace the first two factors in $\left(\prod_{n=0}^{N_i-1}\Omega_r(g_i^n,g_i)\right)$ by the second line of (\ref{omegaggng}) with $n=0$, we get

\begin{equation}
e^{i\theta_{g_i}}=\omega(g_i,g_i^0,g_i)\times\overline{\mathrm{Tr}}\left[W_{\mathbf{A},g_i}^{U\dagger}\Omega_r(g_i,g_i)\left(\prod_{n=2}^{N_i-1}\Omega_r(g_i^n,g_i)\right)W_{\mathbf{A},g_i}^U\left(\prod_{n=1}^{N_i-1}\Omega_r(g_i^n,g_i)\right)^\dagger\right].
\end{equation}

Again replacing $\Omega_r(g_i,g_i)\Omega_r(g_i^2,g_i)$ with the second line of (\ref{omegaggng}), and cyclically permuting $\Omega_r(g_i,g_i)$ to cancel with $\Omega_r(g_i,g_i)^\dagger$ on the right side of the trace, we get
\begin{equation}
e^{i\theta_{g_i}}=\omega(g_i,g_i^0,g_i)\omega(g_i,g_i,g_i)\times\overline{\mathrm{Tr}}\left[W_{\mathbf{A},g_i}^{U\dagger}\Omega_r(g_i^2,g_i)\left(\prod_{n=3}^{N_i-1}\Omega_r(g_i^n,g_i)\right)W_{\mathbf{A},g_i}^U\left(\prod_{n=2}^{N_i-1}\Omega_r(g_i^n,g_i)\right)^\dagger\right].
\end{equation}
\end{widetext}

Continuing in this way, we eventually remove all the $\Omega_r(g_i^n,g_i)$ factors, and obtain (\ref{thetagicocycle}). This completes the proof that our invariant in Eq.~(\ref{thetag1def}) is the gauge invariant quantity given in Eq.~(\ref{cocycles}). %In the gauged theory, $\theta_{g_i}$ is the topological spin of an anyon with $\mathbb{Z}_{N_i}$ gauge flux.

%%%%%%%%%%%%%%%%%%%%%%%%%%%%%%%%%%%%%%%%%%%%%%%%%%%%%%%%
\subsection{Relation between $e^{i\theta_{g_i}}$ and $e^{i\theta_{g_i,g_j}}$ to symmetry flux braiding}\label{sflusbraid}
%%%%%%%%%%%%%%%%%%%%%%%%%%%%%%%%%%%%%%%%%%%%%%%%%%%%%%%%%%%%%%%%%%%%%%%%%%%%%%%%%%%%%
We will not need the braiding formulation of $e^{i\theta_{g_i}}$, but will present it here for completeness. Plugging $g=g_i=k$ and $h=g_i^n$ into Eq.~(\ref{chiR}) and choosing a gauge with $U_{g_i}(g_i^n,g_i)=1$, we have
\begin{equation}
\chi_{g_i}(g_i,g_i^n)=\frac{R(g_i,g_i)R(g_i^n,g_i)}{R(g_i^{n+1},g_i)}.
\end{equation}

Using this in the equation for $e^{i\theta_{g_i}}$ in Eq.~(\ref{cocycles}), we obtain
\begin{align}
\begin{split}
e^{i\theta_{g_i}}&=\prod_{n=0}^{N_i-1}\frac{R(g_i,g_i)R(g_i^n,g_i)}{R(g_i^{n+1},g_i)}\\
&=\left[R\left(g_i,g_i\right)\right]^{N_i}.
\end{split}
\end{align}

This means that $e^{i\theta_{g_i}}$ is the phase from exchanging two $g_i$ symmetry defects $N_i$ times, or braiding them around each other $N_i/2$ times. 

Similarly, plugging $g=g_j,h=g_j^n,$ and $k=g_i$ into Eq.~(\ref{chiR}) with $U_{g_i}(g_j^n,g_j)=1$, we get
\begin{equation}
\chi_{g_i}(g_j,g_j^n)=\frac{R(g_j,g_i)R(g_j^n,g_i)}{R(g_j^{n+1},g_i)}.
\end{equation}

Plugging this into Eq.~(\ref{cocycles}) gives
\begin{align}
\begin{split}
e^{i\theta_{g_i,g_j}}&=\prod_{n=0}^{N_{ij}-1}\frac{R(g_j,g_i)R(g_j^n,g_i)}{R(g_j^{n+1},g_i)}\\
&\times\frac{R(g_i,g_j)R(g_i^n,g_j)}{R(g_i^{n+1},g_j)}\\
&=\left[R\left(g_j,g_i\right)R\left(g_i,g_j\right)\right]^{N_{ij}}.
\end{split}
\end{align}

This means that $e^{i\theta_{g_i,g_j}}$ is the phase from braiding $g_i$ and $g_j$ fluxes around each other $N_{ij}$ times. As we will show, our expression for $e^{i\theta_{g_i,g_j}}$ in Eq.~(\ref{thetagigj}) describes precisely this process. 

\begin{widetext}
Using our braiding result in Eq.~\ref{rgigj} and cyclically permuting within the trace, we have
\begin{equation}\label{RNij}
\left[R(g_j,g_i)R(g_i,g_j)\right]^{N_{ij}}=\overline{\mathrm{Tr}}\left[\left(D_{\overline{\mathbf{A}},g_i}^UD_{\overline{\mathbf{B}},g_j}^{U\dagger}D_{\overline{\mathbf{A}},g_i}^{U\dagger}D_{\overline{\mathbf{B}},g_j}^U\right)^{N_{ij}}\right].
\end{equation}

Notice that $D_{\overline{\mathbf{B}},g_j}^{U\dagger}D_{\overline{\mathbf{A}},g_i}^{U\dagger}D_{\overline{\mathbf{B}},g_j}^U$ is just a pure phase ($R(g_j,g_i)R(g_i,g_j)$) times $D_{\overline{\mathbf{A}},g_i}^{U\dagger}$, so it commutes with $D_{\overline{\mathbf{A}},g_i}^{U}$. Suppose that $N_{ij}=2$. Then we have
\begin{align}
\begin{split}
e^{i\theta_{g_i,g_j}}&=\overline{\mathrm{Tr}}\left(D_{\overline{\mathbf{A}},g_i}^UD_{\overline{\mathbf{B}},g_j}^{U\dagger}D_{\overline{\mathbf{A}},g_i}^{U\dagger}D_{\overline{\mathbf{B}},g_j}^UD_{\overline{\mathbf{A}},g_i}^UD_{\overline{\mathbf{B}},g_j}^{U\dagger}D_{\overline{\mathbf{A}},g_i}^{U\dagger}D_{\overline{\mathbf{B}},g_j}^U\right)\\
&=\overline{\mathrm{Tr}}\left[\left(D_{\overline{\mathbf{A}},g_i}^U\right)^2 D_{\overline{\mathbf{B}},g_j}^{U\dagger}\left(D_{\overline{\mathbf{A}},g_i}^{U\dagger}\right)^2 D_{\overline{\mathbf{B}},g_j}^U\right].
\end{split}
\end{align}

In the same way, we can simplify Eq.~(\ref{RNij}) for general $N_{ij}$, by pulling all the $D_{\overline{\mathbf{A}},g_i}^U$ operators together by commuting them through $D_{\overline{\mathbf{B}},g_j}^{U\dagger}D_{\overline{\mathbf{A}},g_i}^{U\dagger}D_{\overline{\mathbf{B}},g_j}^U$. For general $N_{ij}$, we have
\begin{equation}
e^{i\theta_{g_i,g_j}}=\overline{\mathrm{Tr}}\left[\left(D_{\overline{\mathbf{A}},g_i}^U\right)^{N_{ij}} D_{\overline{\mathbf{B}},g_j}^{U\dagger}\left(D_{\overline{\mathbf{A}},g_i}^{U\dagger}\right)^{N_{ij}} D_{\overline{\mathbf{B}},g_j}^U\right].
\end{equation}

We can further simplify this expression by using the definition of $D_{\overline{\mathbf{A}},g_i}^U$:
\begin{equation}
\left(D_{\overline{\mathbf{A}},g_i}^U\right)^{N_{ij}}=\left(W_{\mathbf{A},g_i}^{U\dagger}U_{A,g}\right)^{N_{ij}}=\left(W_{\mathbf{A},g_i}^{U\dagger}\right)^{N_{ij}},
\end{equation}
where we used the fact that $W_{\mathbf{A},g_i}^{U\dagger}$ commutes with $U_{A,g}$ and $U_{A,g}^{N_{ij}}=\mathbbm{1}$. Finally, since $\left(W_{\mathbf{A},g_i}^{U\dagger}\right)^{N_{ij}}$ is supported only at the endpoints of $\mathbf{A}$, it commutes with $W_{\overline{\mathbf{B}},g_j}^U$. Putting this all together, we have
\begin{equation}
e^{i\theta_{g_i,g_j}}=\overline{\mathrm{Tr}}\left[\left(W_{\mathbf{A},g_i}^{U\dagger}\right)^{N_{ij}} U_{B,g_j}^{\dagger}\left(W_{\mathbf{A},g_i}^{U}\right)^{N_{ij}} U_{B,g_j}\right],
\end{equation}
which matches precisely with Eq.~(\ref{thetagigj}). By the same derivation, we obtain
\begin{equation}
e^{2i\theta_{g_i}}=\overline{\mathrm{Tr}}\left[\left(W_{\mathbf{A},g_i}^{U\dagger}\right)^{N_i} U_{B,g_i}^{\dagger}\left(W_{\mathbf{A},g_i}^{U}\right)^{N_i} U_{B,g_i}\right],
\end{equation}
which matches with Eq.~(\ref{thetagigi}).
\end{widetext}
%%%%%%%%%%%%%%%%%%%%%%%%%%%%%%%%%%%%%%%%%%%%%%%%%%%%%%%%
\subsection{Relation between $e^{i\theta_{g_i,g_j,g_k}}$ and symmetry flux sliding}\label{sfluxslide}
\begin{figure}[tb]
   \centering
   \includegraphics[width=0.9\columnwidth]{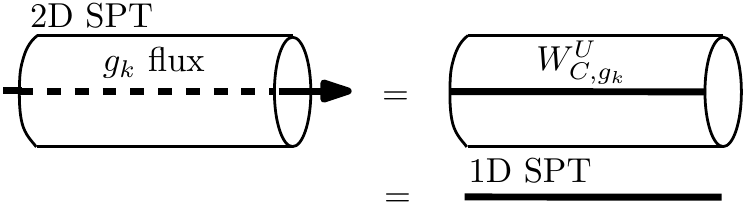} % requires the graphicx package
   \caption{Dimensional reduction from 2D to 1D: $\chi_{g_k}(g_i,g_j)$ corresponds to taking the 2D SPT described by $\omega(g_i,g_j,g_k)$ and wrapping it into a cylinder by compactifying one dimension, and threading through a $g_k$ flux. This is implemented by applying the flux insertion operator $W_{C,g_k}^U$ along the length of the cylinder. The resulting system describes a 1D system with cocycle $\chi_{g_k}(g_i,g_j)$. This means that $W_{C,g_k}^U$ must be a 1D SPT entangler for the SPT described by $\chi_{g_k}(g_i,g_j)$.}
   \label{fig:d2d1}
\end{figure}

%%%%%%%%%%%%%%%%%%%%%%%%%%%%%%%%%%%%%%%%%%%%%%%%%%%%%%%%%%%%%%%%%%%%%%%%%%%%%%%%%%%%%
We can already identify $e^{i\theta_{g_i,g_j,g_k}}$ with our formula in Eq.~(\ref{decdomaininv}) by dimensional reduction. This is because $\chi_{g_k}(g_i,g_j)$ describes a 2-cocycle $\omega(g_i,g_j)$ corresponding to a 1D SPT obtained from placing the 2D SPT on a cylinder with $g_k$ flux inserted (see Fig.~\ref{fig:d2d1}). This means that a 2D cylinder with a $g_k$ flux insertion line $W_{A,g_k}^U$ along the length describes a 1D SPT. Then clearly, $W_{A,g_k}$ must be the SPT entangler of the 1D SPT, which is precisely what is detected by Eq.~(\ref{decdomaininv}). 

Note that we can also derive our formula for $e^{i\theta_{g_i,g_j,g_k}}$ in Eq.~(\ref{decdomaininv}) using the heptagon equation, as we did with $e^{i\theta_{g_i,g_j}}$. We do not do this here because it is not necessary, as the dimensional reduction argument above already proves our desired result. However, we will sketch the heptagon equation method for completeness. In this case, Eq.~(\ref{chiR}) says that
\begin{align}
\begin{split}
\frac{\chi_{g_k}(g_i,g_j)}{\chi_{g_k}(g_j,g_i)}&=\frac{R(g_i,g_k)R(g_j,g_k)}{R(g_ig_j,g_k)}\frac{R(g_jg_i,g_k)}{R(g_j,g_k)R(g_i,g_k)}\\
&\times \frac{U_{g_k}(g_j,g_i)}{U_{g_k}(g_i,g_j)}.
\end{split}
\end{align}

The $R$ symbols cancel because $g_kg_j=g_jg_k$, so in this case, if we choose $F(g_i,g_j,g_k)=\omega(g_i,g_j,g_k)$, we cannot choose $U_{g_k}(g_j,g_i)=1$. In fact, $e^{i\theta_{g_i,g_j,g_k}}$ is given entirely by two sliding moves $U_{g_k}(g_j,g_i)$ and $U_{g_k}(g_i,g_j)$. 

\begin{equation}
e^{i\theta_{g_i,g_j,g_k}}=\frac{U_{g_k}(g_j,g_i)}{U_{g_k}(g_i,g_j)}.
\end{equation}

%%%%%%%%%%%%%%%%%%%%%%%%%%%%%%%%%%%%%%%%%%%%%%%%%%%%%%%%%%%%%%%%%%%%%%%%%%%%%%%%%%%%
\section{Relation between 3D invariants and $\omega(g,h,k,l)$}\label{sproofs3d}
%%%%%%%%%%%%%%%%%%%%%%%%%%%%%%%%%%%%%%%%%%%%%%%%%%%%%%%%%%%%%%%%%%%%%%%%%%%%%%%%%%%%
Bosonic SPTs in 3D with discrete, abelian, unitary, on-site symmetries are classified by $\omega(g,h,k,l)\in H^4(G,U(1))$. Each of the invariants we presented in Sec.~\ref{s3D} again corresponds to a different kind of gauge invariant quantity specifying $\omega(g,h,k,l)$. It can be shown that the three invariants $e^{i\theta_{g_i;g_l}},e^{i\theta_{g_i,g_j;g_l}}$, and $e^{i\theta_{g_i,g_j,g_k;g_l}}$ all dimensionally reduce to 2D invariants, like how $e^{i\theta_{g_i,g_j,g_k}}$ reduces to a 2-cocycle which is a 1D SPT invariant. This means that, writing the symmetry group as $G\times H$, they can all be thought of as obtained from decorating $G$ domain walls with $H$ SPTs. 

Specifically, we can write $e^{i\theta_{g_i;g_l}},e^{i\theta_{g_i,g_j;g_l}}$, and $e^{i\theta_{g_i,g_j,g_k;g_l}}$ in therms of slant products $\chi_g(h,k,l)$ which produce a 3-cocycle from 4-cocycles as given in (\ref{slant}) as
\begin{equation}
\chi_g(h,k,l)=\frac{\omega(h,g,k,l)\omega(h,k,l,g)}{\omega(g,h,k,l)\omega(h,k,g,l)}.
\end{equation}
\begin{figure}[tb]
   \centering
   \includegraphics[width=0.9\columnwidth]{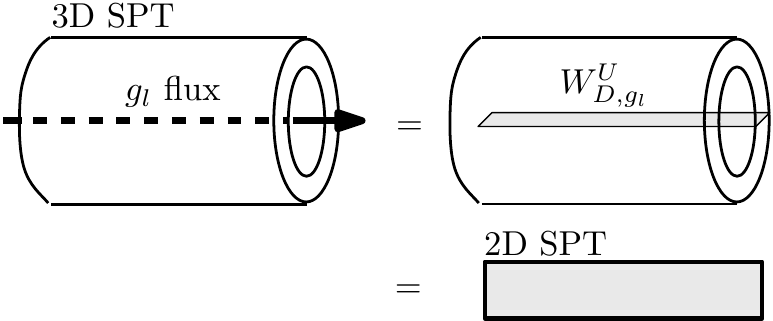} % requires the graphicx package
   \caption{Dimensional reduction from 3D to 2D. Here, we take a 3D SPT labeled by $\omega(g,h,k,g_l)$ on a thick slab, and wrap it into a thick cylinder with $g_l$ flux inserted. This is implemented by apply the flux insertion operator $W_{D,g_l}^U$ on a plane along the length of the thick cylinder. The resulting system is a 2D SPT with cocycle $\chi_{g_l}(g,h,k)$, so $W_{D,g_l}^U$ must entangle a 2D SPT described by $\chi_{g_l}(g,h,k)$.}
   \label{fig:d3d2}
\end{figure}

Like in the 2D case, $\chi_g(h,k,l)$ describes the 2D SPT resulting from taking a 3D SPT labeled by $\omega(h,g,k,l)$ on a thick slab, then wrapping it into a thick cylinder by compactifying one dimension, and threading through the thick cylinder a $g$ flux. This is illustrated in Fig.~\ref{fig:d3d2}. To relate our 3D invariants more explicitly to our 2D invariants, we also define the 2-cocycle from these 3-cocycles as in (\ref{chighk}):
\begin{equation}
\chi_{g,h}(k,l)=\frac{\chi_g(h,k,l)\chi_g(k,l,h)}{\chi_g(k,h,l)}.
\end{equation}

The three invariants are defined as
\begin{align}
\begin{split}\label{cocycles3D}
e^{i\theta_{g_i,g_l}}&=\prod_{n=0}^{N_i-1}\chi_{g_l,g_i}(g_i,g_i^n)\\
e^{i\theta_{g_i,g_j;g_l}}&=\prod_{n=0}^{N_{ij}-1}\chi_{g_l,g_i}(g_j,g_j^n)\chi_{g_l,g_j}(g_i,g_i^n)\\
e^{i\theta_{g_i,g_j,g_k;g_l}}&=\frac{\chi_{g_l,g_k}(g_i,g_j)}{\chi_{g_l,g_k}(g_j,g_i)}.
\end{split}
\end{align}

Since each of the invariants above can be written entirely as products of 3-cocycles $\chi_{g_l}$ indexed by $g_l$, all the invariants above correspond to 2D SPT invariants computed in a system with $W_{D,g_l}^U$ applied in a plane as in Fig.~\ref{fig:d3d2}. This means that they all correspond to 2D SPTs entangled by $W_{D,g_l}^U$. This justifies computing these three invariants by plugging in $W_{C,g_l}^U$ in place of the SPT entangler in the equations for the 2D invariants.

%%%%%%%%%%%%%%%%%%%%%%%%%%%%%%%%%%%%%%%%%%%%%%%%%%%%%%%%%%%%%%%%%%%%%%%%%%%%%%%%%%%%
\section{Properties of $e^{i\theta_{g_i}},e^{i\theta_{g_i,g_j}},$ and $e^{i\theta_{g_i,g_j,g_k}}$}\label{spropinvariants}
%%%%%%%%%%%%%%%%%%%%%%%%%%%%%%%%%%%%%%%%%%%%%%%%%%%%%%%%%%%%%%%%%%%%%%%%%%%%%%%%%%%%
The invariants in Eq.~(\ref{cocycles}) and (\ref{cocycles3D}) satisfy various properties. These include the following\cite{topologicalinvariants}
\begin{align}
\begin{split}\label{thetaproperties}
e^{2i\theta_{g_i}}&=e^{i\theta_{g_i,g_i}}\\
e^{i\theta_{g_i,g_j}}&=e^{i\theta_{g_j,g_i}}\\
e^{i\theta_{g_i,g_j,g_l}}&=e^{i\mathrm{sgn}(\hat{p})\theta_{\hat{p}(g_i,g_j,g_k)}},
\end{split}
\end{align}
where $\hat{p}(g_i,g_j,g_k)$ is a permutation of $(g_i,g_j,g_k)$ and $\mathrm{sgn}(\hat{p})=1$ if the permutation is cyclic and $-1$ otherwise. 

The third property in the list will be useful for showing some aspects of the flux insertion operators relevant for our invariants. These properties are:
\begin{itemize}
\item{We can always choose $W_{A,g_i}^U$ to be a $\mathbb{Z}_{N_i}$ symmetric FDQC. This ensures that we can always restrict $W_{A,g_i}^U$ to $W_{\mathbf{A},g_i}^U$ such that $W_{\mathbf{A},g_i}^U$ commutes with $U_{g_i}$. This kind of restriction is necessary to compute $e^{i\theta_{g_i,g_i}}=e^{2i\theta_{g_i}}$ using Eq.~(\ref{thetagigi}).}
\item{We can always choose $W_{A,g_i}^U$ and $W_{B,g_j}^U$ to be $\mathbb{Z}_{N_i}\times\mathbb{Z}_{N_j}$ symmetric FDQCs. This ensures that we can always restrict $W_{A,g_i}^U$ to $W_{\mathbf{A},g_i}^U$ and $W_{B,g_j}^U$ to $W_{\mathbf{B},g_j}^U$ such that $W_{\mathbf{A},g_i}^U$ and $W_{\mathbf{B},g_j}^U$ commute with $U_{g_i}$ and $U_{g_j}$.}
\end{itemize}

The first statement is a result of the observation that $e^{i\theta_{g_i,g_i,g_i}}=1$, so there is no obstruction to making $W_{A,g_i}^U$ a $\mathbb{Z}_{N_i}$ symmetric FDQC. Similarly, the second statement follows from the observation that $e^{i\theta_{g_i,g_j,g_k}}=1$ if any pair of indices are the same.

As a side note, while we simply quote (\ref{thetaproperties}) from previous work, it is actually possible to derive some of these properties using the explicit forms of these invariants in terms of operators. For example, it is easy to see that $c(g,h)=c(h,g)^*$ when $c(g,h)$ is defined explicitly in Eq.~(\ref{cgh}). This is because $c(h,g)$ is equal to $c(g,h)$ with $U$ replaced by $U^\dagger$, and $c(g,h)$ is multiplicative under composition of unitaries.

%%%%%%%%%%%%%%%%%%%%%%%%%%%%%%%%%%%%%%%%%%%%%%%%%%%%%%%%%%%%%%%%%%%%%%%%%%%%%%%%%%%%
\bibliography{flowsfloquetbib}
%\bibliographystyle{plain}
%%%%%%%%%%%%%%%%%%%%%%%%%%%%%%%%%%%%%%%%%%%%%%%%%%%%%%%%%%%%%%%%%%%%%%%%%%%%%%%%%%%%

\end{document}